 \documentclass[11pt]{article}

\usepackage[]{acl}

\usepackage{times}
\usepackage{latexsym}

\usepackage[T1]{fontenc}

\usepackage[utf8]{inputenc}
\usepackage[linesnumbered,ruled,vlined]{algorithm2e}

\usepackage{microtype}

\usepackage{inconsolata}

\usepackage{graphicx}

\usepackage{longtable}
\usepackage{amssymb}
\usepackage{tabularx} 

\usepackage{algorithm}
\usepackage{algorithmic}
\usepackage{array}      
\usepackage{makecell}   
\usepackage{booktabs}   

\usepackage[table]{xcolor}

\usepackage{amsmath} 
\usepackage{xcolor}
\definecolor{promptblue}{RGB}{33,150,243}
\definecolor{epsgreen}{RGB}{76,175,80}  
\definecolor{promptblue}{RGB}{33,150,243}
\definecolor{promptborder}{RGB}{30,136,229}
\definecolor{promptgray}{gray}{0.96}
\definecolor{tablesection}{RGB}{224,224,224} 

\newcommand{\EffectivePresentationScorer}{\emph{EffectivePresentationScorer}}

\definecolor{tableheader}{RGB}{200,200,200} 

\usepackage{graphicx}
\usepackage{booktabs}
\usepackage{multirow}

\title{\includegraphics[height=0.6cm]{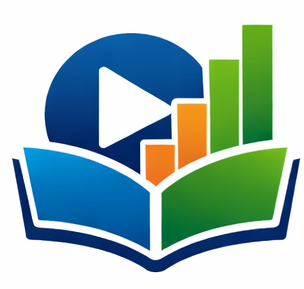}
\textbf{A Good Talk Doesn't Look Like a Summary, It Teaches You!}
\textbf{Measuring Takeaways from Paper-to-Video Talks}
}

\author{Ishani Mondal,$^{1}$\thanks{The work started when Ishani was a research intern at the Adobe Research Lab.}    
 \textbf{Aparna Garimella},$^{2}$ \\ 
 \textbf{Ananya Sai},$^{2}$ 
 \textbf{Pannaga Shivaswamy},$^{2}$ \textbf{Jordan Boyd-Graber}$^{1}$ \\ \\
  $^{1}$ University of Maryland, College Park, \hspace{0.1cm} 
  $^{2}$ Adobe Research \hspace{0.1cm}\\
  \texttt{{imondal@umd.edu}}\\
}

\usepackage[most]{tcolorbox}
\newcommand{\promptbox}[1]{%
  \begin{tcolorbox}[colback=yellow!5!white,colframe=yellow!80!black,title=Prompt]
  #1
  \end{tcolorbox}
}
\begin{document}
\maketitle
\begin{abstract}

Automatically generated videos from scientific papers are increasingly used for education and research dissemination. However, existing evaluation metrics mainly measure visual quality or whether key points from the paper appear in the video—without assessing whether the video actually helps viewers understand the ideas.
We introduce \EffectivePresentationScorer{}, a framework for evaluating the instructional quality of scientific presentation videos. It checks whether a video explains the main ideas clearly, introduces needed background concepts, and connects technical details to the paper’s main contribution.
When we apply  \EffectivePresentationScorer{} to the existing paper-to-video generation systems, we find that generated videos mention the correct topics and follow the structure of the paper but fail to explain prerequisite concepts or clarify why the method works. These failures are often ignored by existing video evaluation metrics, which focus on content presence rather than explanatory quality. 
\end{abstract}

\section{Beyond Summarization: Teaching through Videos at Conference}

Instructional videos generated from papers are a well-established medium for scientific education, shown to support conceptual understanding and long-term retention when explanations are structured and pedagogically coherent \cite{AYRES2025102077, ACKERMANS2025102137, Niekrenz2024-hb}. Unlike short-form or entertainment-oriented videos, instructional videos must respect conceptual prerequisites, present ideas in a logical order, and provide sufficient explanation for complex material \cite{Ramon-Arbues2025-ao}. These requirements are critical for videos generated 
from papers, where the goal is not merely to summarize, but to help viewers understand a paper’s core contributions.

Recent advances in paper-to-video generation pipelines have made it possible to convert scientific papers into narrated slide/video presentations with minimal human effort \cite{zhu2025paper2videoautomaticvideogeneration, zheng-etal-2025-pptagent}. However, a fundamental question remains largely unanswered: \emph{How effective/useful are these generated videos in helping one to comprehend core contributions of the paper?}


\begin{table}[t]
\centering
\scriptsize
\setlength{\tabcolsep}{3pt}
\renewcommand{\arraystretch}{1.1}

\resizebox{\columnwidth}{!}{
\begin{tabular}{p{3.4cm}cc}
\toprule
\textbf{Evaluation Metric} 
& \textbf{Video A} 
& \textbf{Video B} \\
\midrule

VideoScore \cite{he-etal-2024-videoscore}
& 0.85 
& 0.85 \\

PresentQuiz \cite{zhu2025paper2videoautomaticvideogeneration} 
& 0.87 
& 0.88 \\

\midrule

\rowcolor{gray!12}
\EffectivePresentationScorer{} 
&  
&  \\

\rowcolor{gray!12}
Conceptual Grounding 
& Mentions modules 
& Explains APE idea \\

\rowcolor{gray!12}
Prerequisite Satisfaction 
& Weak motivation 
& Problem $\rightarrow$ method \\

\rowcolor{gray!12}
Coherence 
& Slide reading 
& Causal flow \\

\rowcolor{gray!12}
Answerability 
& Hard to infer why 
& Reasoning clear \\

\rowcolor{gray!12}
EffectivePresentationScorer Utility 
& \textbf{Low} 
& \textbf{High} \\

\bottomrule
\end{tabular}
}

\caption{
Comparison of two videos generated from the same paper \citet{wu-etal-2021-training}. 
\textbf{Video A} narrates slides sequentially, describing components (APE, HasAnswer, scheduler) without explaining their causal relationship  (Narration in Table~\ref{tab:ape_frame_narration}).
\textbf{Video B} restructures the narration to build a reasoning chain: motivating retrieval inefficiency, explaining relevance prediction and adaptive scheduling, and clarifying why computation can be reduced without hurting accuracy (Narration in Table~\ref{tab:ape_frame_narration_teaching}). 
Although both videos receive similar scores from metrics such as \textsc{VideoScore} and \textsc{PresentQuiz}, \EffectivePresentationScorer{} reveals a large difference in instructional utility because Video B supports reasoning about method while Video A mainly reads slide content.
}
\label{tab:two_video_comparison}
\end{table}

Evaluating the true utility of such videos is inherently difficult, as it
hinges on downstream outcomes like comprehension, retention, and learning
gains relative to the effort invested, as discussed in 
\cite{McConnell2017InstructionalUA, reiter2025evaluaterealworldimpact}.
Yet most existing evaluations focus on surface-level properties. Benchmarks
like VBench \cite{huang2024vbenchcomprehensiveversatilebenchmark} and
EvalCrafter \cite{liu2024evalcrafterbenchmarkingevaluatinglarge} prioritize
visual realism, fluency, and text--video alignment, aligning with the
strengths of recent generative models such as Sora
\cite{liu2024sorareviewbackgroundtechnology} and Veo
\cite{wiedemer2025videomodelszeroshotlearners}, which typically produce
short (1--2 minute) visually appealing clips. However, such short-form
content is poorly suited for instructional or scientific communication,
which demands longer, logically structured, and pedagogically coherent
presentations \cite{Mayer2024ThePP}. Standard metrics
like Fr\'echet Video Distance (FVD) \cite{unterthiner2019accurategenerativemodelsvideo}
and CLIP-based similarity scores \cite{radford2021learningtransferablevisualmodels}
assess perceptual quality but overlook whether the video actually teaches.
Recent efforts use VLMs as Judge \cite{bansal2024videophyevaluatingphysicalcommonsense,
ku-etal-2024-viescore, he-etal-2024-videoscore}, but these too emphasize
summary-level fluency and alignment and cannot distinguish instructional
utility (Table~\ref{tab:two_video_comparison}). Crucially, even
document-to-video systems \cite{zheng-etal-2025-pptagent} are rarely
evaluated for their educational impact. \emph{We argue that existing
metrics, while necessary, leave a gap between benchmark performance (what
is being optimized) and instructional utility (what should be optimized).}

We argue that instructional videos should be evaluated by how well they support understanding rather than only surface-level quality, and therefore adopt the established notion of \emph{instructional utility} from education research \cite{McConnell2017InstructionalUA, cao2026developingauthenticsimulatedlearners, educsci15111507}, grounding our automated presentation-level evaluation framework in theories of multimedia learning and cognitive load \cite{Mayer2024ThePP, SWELLER1994295, SWELLER2024102423} (Section~\ref{sec:taskdescription}).
To measure this, we introduce \EffectivePresentationScorer{}, a framework that evaluates how effectively a video supports understanding (Section~\ref{sec:methodology}). We first construct a quiz from the source paper, guided by Bloom's Taxonomy~\cite{Adams2015-nv}, with questions ranging from simple recall to reasoning about the paper's methods and ideas. We then create pairs of videos that vary in content and delivery. Human participants watch each video and answer the same set of questions across variants, allowing us to identify which presentations lead to better understanding. Along with answer accuracy, we collect interaction signals such as pauses, replays, and self-reported difficulty---behavioral proxies for cognitive load. These signals form a utility-scored benchmark dataset (Section~\ref{sec:dataset}). Using them, \EffectivePresentationScorer{} analyzes whether a video explains ideas clearly, introduces the necessary background concepts, and presents information in a coherent order.

We validate our framework by comparing its scores against human
answerability and rationales, where it shows stronger correlation than
existing metrics. Finally, we apply \EffectivePresentationScorer{} to
both human conference videos and automatically generated videos
(Section~\ref{sec:results}). While current systems often produce visually
fluent summaries, they frequently fail to improve viewer
understanding. Beyond evaluation, \EffectivePresentationScorer{}
provides diagnostic feedback which can help generation systems improve
instructional videos by adding context and clarifying key ideas.

\begin{figure*}[t]
\centering
\includegraphics[width=0.98\textwidth]{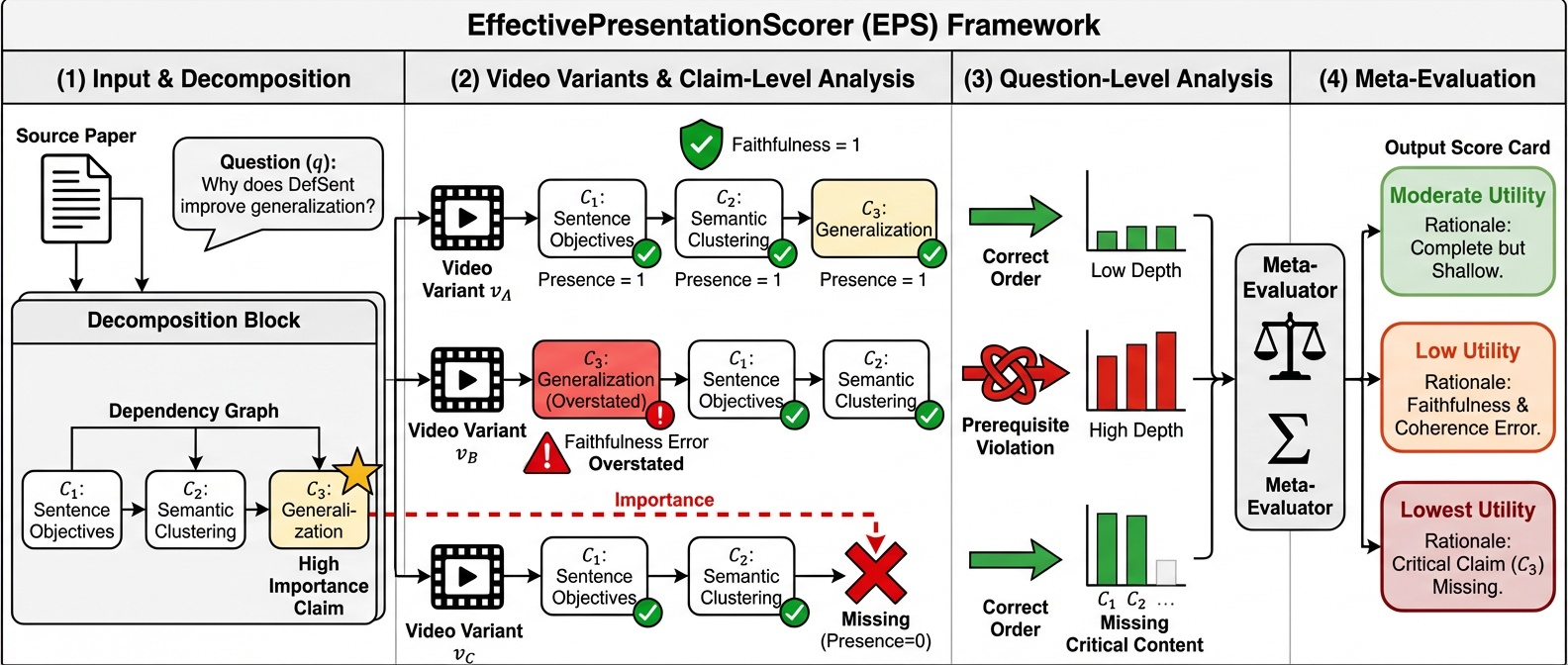}

\caption{Step-by-step illustration of \EffectivePresentationScorer{} using a running DefSent example. 
The pipeline starts from a source paper and an evaluation question (\emph{``Why does DefSent improve generalization?''}), which is decomposed into a dependency-structured set of explanatory claims ($C_1 \rightarrow C_2 \rightarrow C_3$). 
For each video variant ($v_A$, $v_B$, $v_C$) and for each claim, claim-wise agents assess \emph{Presence}, \emph{Faithfulness}, and \emph{Claim Importance}, revealing that $v_C$ omits the core causal claim and $v_B$ distorts it. 
Question-level agents then evaluate \emph{Coherence} by checking prerequisite order violations and \emph{Delivery} by measuring explanation depth and clarity across claims. 
Finally, a Meta-Evaluator aggregates claim-level and delivery-level diagnostics into an interpretable utility score per variant, explaining why superficially polished videos may receive lower educational utility due to missing, misordered, or weakly explained concepts.}
\label{fig:EffectiveScorer-diagram}
\end{figure*}

\section{Related Work}
Early text-to-video evaluation relied on perceptual and alignment metrics (e.g., IS/FID/FVD, CLIP), which assess visual quality and prompt relevance but provide limited insight into task success. Recent benchmarks—VBench \cite{huang2023vbench}, VBench++ \cite{huang2024vbenchcomprehensiveversatilebenchmark}, EvalCrafter \cite{liu2024evalcrafterbenchmarkingevaluatinglarge}, DEVIL \cite{liao2024evaluation}, and VideoGen-Eval \cite{yang2025videogen}—introduce multi-dimensional, structured, and dynamics-aware evaluations with improved alignment to human judgments.
However, these frameworks primarily measure perceptual and generative quality, not communicative or instructional success. They lack source-grounded evaluation of whether a video is faithful to the underlying document, respects conceptual prerequisites, presents ideas in a pedagogically correct order, and provides sufficient explanation for learning. 
Google’s NotebookLM \footnote{\url{https://notebooklm.google/}} generates videos grounded in user-provided documents, but offers limited support for goal-driven or user-adaptive generation. Consequently, success is measured mainly by fidelity to source content rather than by whether the video achieves an intended communicative goal.

\section{Define `Instructional Utility'}
\label{sec:taskdescription}

We adopt the established construct of \emph{instructional utility} from education research, defined by McConnell et al.~\cite{McConnell2017InstructionalUA} as the measurable learning gains an instructional activity produces relative to the time and effort invested---a framing also used in recent work on LLMs in learning contexts \cite{cao2026developingauthenticsimulatedlearners, educsci15111507}. Our contribution is not a new definition but an \emph{automated, presentation-level operationalization} of this construct for scientific videos, grounded in \citep{Mayer2024ThePP} and Cognitive Load Theory \cite{SWELLER1994295, SWELLER2024102423}. 
Specifically, Mayer argues that meaningful multimedia learning occurs when learners actively engage in selecting relevant words and images, organizing them into coherent verbal and pictorial representations, and integrating them with prior knowledge under limited cognitive capacity \cite{Mayer2024ThePP}. 
Consistent with this view, we operationalize instructional utility through paper-grounded question answering because it provides a scalable proxy for whether a learner has constructed an actionable understanding of document after watching the video. 
By using questions spanning multiple levels of Bloom's taxonomy (Section~3), our evaluation captures both factual recall and higher-order reasoning rather than surface familiarity alone. 
Beyond raw QA accuracy, our framework further incorporates cognitive-effort signals and explanation diagnostics, enabling a broader assessment of instructional effectiveness. Long-term learning outcomes such as retention and transfer across time remain out of scope and are discussed as future work in Section~\ref{sec:limitations}.

\section{EffectivePresentation-EvalBench}
\label{sec:dataset}

To evaluate whether scientific presentation videos help viewers understand a paper, we introduce a new dataset, \textbf{EffectivePresentation-EvalBench} which measures \emph{instructional utility}: whether viewers can correctly answer paper-specific questions after watching a presentation.

Existing resources either contain only real conference talks or evaluate generated videos using perceptual metrics such as visual quality or modality alignment. These metrics do not capture whether a video actually explains the scientific ideas. Our dataset instead evaluates learning outcomes by measuring how well viewers answer questions about the paper after watching different presentations of the same work (Statistics in Table~\ref{tab:dataset_stats} and Table~\ref{tab:topic_distribution}).



EffectivePresentation-EvalBench builds on 20 research papers from VISTA \cite{liu-etal-2025-talk}, each paired with multiple presentation videos explaining the same paper. For each paper we collect seven videos: six automatically generated variants and one human-authored conference presentation. The videos average 7.8 minutes in duration and contain roughly 28 slides each, including figures, equations, and textual explanations. In total the dataset contains \textbf{140 instructional videos} with human utility annotations.
This is designed as a controlled evaluation setting: each paper is paired with seven systematically perturbed video variants evaluated using the same question set, enabling within-paper analysis of instructional failures while isolating presentation quality from content and annotator effects. The underlying framework is modular and readily scalable to larger and more diverse domains.
To ensure that results reflect the quality of the presentation rather than differences in annotator background knowledge, we first construct a set of \emph{background screening questions} for each paper. These questions test prerequisite concepts required to understand the paper but explicitly exclude the paper’s novel contributions. Annotators must correctly answer at least seven out of ten screening questions to evaluate videos for that paper. This filtering ensures that annotators possess sufficient prior knowledge to judge whether a video explanation is clear.

Instructional utility is measured using a set of paper-grounded questions that capture the key ideas of the paper. To identify these ideas, we analyze citation contexts from papers that reference the target work, since citations often reflect the contributions research community associates with it. We extract these commonly cited contributions and convert them into evaluation questions. The questions span multiple levels of Bloom’s taxonomy, from factual recall to reasoning about the method and its implications, allowing us to assess whether a viewer understands the paper after watching the video.
A manual validation of these generated questions on an expanded subset is reported in Appendix~\ref{sec:question-validation}, where 88–91\% of questions are judged well-formed and pedagogically aligned. We emphasize that EffectivePresentationScorer is agnostic to question provenance: expert-authored or curriculum-aligned question sets can be used directly in place of LLM-generated ones.

To analyze how presentation choices affect learning, we generate multiple video variants for each paper that explain the same content but differ in presentation quality. These variants introduce controlled perturbations to either the \textbf{content} or the \textbf{delivery} of the presentation. 
Content perturbations modify what information is explained (e.g., omitting prerequisites or presenting concepts in an illogical order). Delivery perturbations modify how the explanation is presented (e.g., poor pacing, audio–visual misalignment, or overly dense slides). A final variant combines both types of perturbations. Because all variants explain the same paper and are evaluated with the same questions, differences in viewer understanding can be attributed to specific instructional failures (Details in Appendix~\ref{appendix:controlled_perturbations}).

Each video is evaluated by three independent annotators in a controlled A/B-style study. Annotators watch the video and answer the same set of paper-grounded questions. In addition to providing answers, annotators report perceived difficulty and viewing effort using Likert scales. These signals capture the intuition that a presentation may technically enable correct answers but still impose excessive cognitive burden.

We combine answer correctness with reported difficulty and effort to compute a \emph{utility score} for each video. Correct answers increase utility, while high difficulty or effort reduce it. Scores are averaged across annotators, questions, and papers to obtain a stable learner-centered estimate of instructional quality (Formally in Appendix~\ref{sec:human_utility}).

Finally, \textbf{EffectivePresentation-EvalBench} comprises, for each paper:
(i) a fixed set of paper-grounded background and takeaway questions with verified gold answers, and
(ii) seven presentation-style video variants—six automatically generated and systematically perturbed along content and presentation dimensions, and one human-authored conference presentation video serving as a high-quality reference and (iii) paper-video pair utility estimates.
This structure enables controlled, within-paper comparison of instructional utility across diverse presentations (Statistics and Paper Details in Table~\ref{tab:dataset_stats} and Table~\ref{tab:topic_distribution}).
This controlled structure enables systematic comparison of presentation strategies and provides a benchmark for evaluating whether automatically generated videos truly help viewers understand scientific papers.

\section{\EffectivePresentationScorer{}}
\label{sec:methodology}

Building on the dataset, we next introduce a framework  \EffectivePresentationScorer{} to estimate the \emph{educational utility} of videos generated from scientific papers (Algorithm~\ref{alg:eps}). 
Rather than relying on a single holistic score, our approach decomposes evaluation into content-level and delivery-level diagnostics that collectively measure whether a video explains a paper’s key ideas correctly (\emph{what}) and effectively (\emph{how}). Each stage produces structured outputs and rationales, enabling transparent aggregation.
Although we evaluate presentation-style scientific videos, the core abstractions of \EffectivePresentationScorer{}---claim decomposition, dependency-aware coverage, faithfulness, coherence, and delivery quality---are modality-agnostic and extend naturally to other educational formats such as interactive tutorials or code walkthroughs with format-specific input and delivery adaptations.

To illustrate the evaluation, we use a running example from \citet{tsukagoshi-etal-2021-defsent}  with the question: \emph{“Why does DefSent improve generalization?”} 
We compare three video variants: $v_A$, which briefly mentions the key ideas; $v_B$, which gives detailed but poorly structured explanations; and $v_C$, which is well-delivered but misses the core explanation. 
This example highlights our goal: whether the video correctly explains the concept and conveys it effectively?
(Figure~\ref{fig:EffectiveScorer-diagram}).

\paragraph{Multimodal Video Representation.}
To enable fine-grained instructional analysis, we convert each video $v$ into a structured multimodal representation
$\mathcal{V}=\{(s_i,t_i,a_i,d_i)\}_{i=1}^{M}$, where $s_i$ is the $i$-th slide image, $t_i$ its timestamp interval (indicates \emph{when} each explanation occurs in the video, allowing us to track how slides and narration jointly communicate the explanation over time),
$a_i$ the aligned narration, and $d_i$ a vision--language description of visual content.
This representation jointly captures \emph{what} is shown, \emph{what} is said, \emph{when} and \emph{how} they are delivered. Across the three DefSent video variants in Figure~\ref{fig:EffectiveScorer-diagram} (Narration ($a_i$) and Time ($t_i$) in Table~\ref{tab:defsent_narrations_textheavy}), we will observe how the consecutive slides and narration explain the answer.

\paragraph{Step 1: Identification of Reasoning Claims Required to Answer a Question.}
To ensure that claim decomposition reflects the paper’s actual explanations rather than the model’s prior knowledge, we ground decomposition directly in retrieved paper text.
Given a question $q$ and paper $p$, we preprocess the paper using the SciPDF parser\footnote{\url{https://github.com/titipata/scipdf_parser}}, embed sentence-level units from key explanatory sections (abstract, introduction, methods, results) using a Sentence-BERT encoder \cite{reimers-gurevych-2019-sentence}, and retrieve a compact, coherence-preserving context window via cosine similarity to $q$.
This retrieved context is provided verbatim to GPT-4o, which is instructed to derive claims only from the supplied text, ensuring that each claim corresponds to an explicit paper statement and enabling reliable downstream evaluation of claim presence, faithfulness, ordering, and delivery (Appendix~\ref{appendix:decomposition}).

For the DefSent example in Figure~\ref{fig:EffectiveScorer-diagram}, this yields three claims:
$c_1$: DefSent uses sentence-level objectives;
$c_2$: sentence-level representations improve semantic clustering; and
$c_3$: improved clustering leads to better generalization,
$c_1 \rightarrow c_2 \rightarrow c_3$

\paragraph{Step 2: Are All Claims Equally Important?}
Because not all claims contribute equally to answering the question, we assign each claim an \emph{importance score} using an Importance Agent.
This agent assigns a salience value $I(c)\in[0,1]$ based on the claim’s role in the paper’s explanatory structure (Details in Appendix~\ref{appendix:claim-importance-estimation}).
For DefSent, $c_3$ receives the highest importance because it directly answers why the method improves generalization, while $c_1$ serves as supporting context.
Importantly, these importance scores are used to weight how omissions or weak explanations affect downstream utility, whereas dependency depth is used only for ordering and coverage normalization.

\paragraph{Step 3: Are the Claims Correct and Present?}

\paragraph{Presence.}
Instructional videos may appear fluent and accurate while omitting a key explanatory step, particularly a downstream causal link.
To identify such omissions, we use a Presence Agent that explicitly checks whether each required claim is stated in the video.
The agent produces $\pi(c,v)\in\{0,1\}$ indicating whether claim $c$ is explicitly expressed (Appendix~\ref{appendix:presence}).
In $v_A$ and $v_B$, all three claims are present, whereas $v_C$ explains sentence-level objectives and clustering ($c_1$, $c_2$) but never states that clustering leads to generalization, yielding $\pi(c_3,v_C)=0$.
Claim coverage is aggregated as $\pi(q,v)=\sum_c \tilde d(c)\pi(c,v)/Z$, where $\tilde d(c)$ reflects dependency depth and $Z$ is a normalization constant.

\paragraph{Faithfulness.}
Even when a claim is stated, it may be incorrect, overstated, or unsupported by the paper.
The Faithfulness Agent verifies whether each expressed claim aligns with the source document, yielding $F(c,v,p)\in\{0,1\}$ (Appendix~\ref{appendix:faithfulness}).
In $v_A$ and $v_C$, all expressed claims are faithful, whereas $v_B$ overstates the causal strength of clustering, causing $F(c_3,v_B,p)=0$.
Aggregate faithfulness is computed as $F(q,v,p)=\sum_c \tilde d(c)\pi(c,v)F(c,v,p)/Z'$, ensuring that distorted core explanations reduce utility even when coverage is high.

\paragraph{Step 4: How Are Claims Explained?}

\paragraph{Coherence.}
Even when claims are present and faithful, instructional utility can suffer if they are introduced in an order that violates conceptual prerequisites.
For a question $q$ with claim set $\mathcal{C}(q)$, let $\tau_p(c)$ denote the position of claim $c$ in the paper-derived dependency order and $\tau_v(c)$ the point at which it is first introduced in the video.
We define coherence as
$C(q,v)=1-\frac{1}{|\mathcal{C}(q)|^2}\sum_{c_i,c_j\in\mathcal{C}(q)}\mathbb{I}[\tau_p(c_i)<\tau_p(c_j)\wedge\tau_v(c_i)>\tau_v(c_j)]$,
which penalizes prerequisite violations (Appendix~\ref{appendix:coherence}).
In $v_B$, generalization ($c_3$) precedes sentence-level objectives ($c_1$), lowering coherence, while $v_A$ and $v_C$ follow correct order.

\paragraph{Delivery Quality.}
Even when claims are correctly ordered, instructional utility depends on how clearly and thoroughly those claims are explained.
The Delivery Agent evaluates explanatory depth, time allocation, and audio--visual alignment, while explicitly weighting explanations by claim importance (Appendix~\ref{appendix:delivery}).

For a question $q$ and video $v$, delivery is computed as
$D_{\text{del}}(q,v)=\sum_{c\in\mathcal{C}(q)} I(c)\,\pi(c,v)\,\hat{T}(c,v)\,Q(c,v)\,A(c,v)$,
where $I(c)$ is the claim importance score, $\pi(c,v)$ indicates presence, $\hat{T}(c,v)$ is the normalized narration time devoted to explaining $c$, $Q(c,v)$ measures qualitative explanatory richness, and $A(c,v)\in[0,1]$ captures audio--visual alignment between narration and supporting visuals.
This formulation ensures that weak or rushed explanations of high-importance claims incur larger penalties than equally weak explanations of supporting claims.

For example, although $v_A$ mentions all claims, it allocates little time and weak visual grounding to $c_3$, resulting in a low delivery contribution for the most important claim.
Video $v_B$ provides richer explanations but suffers from ordering violations, while $v_C$ delivers strong explanations for $c_1$ and $c_2$ but contributes nothing for $c_3$ due to its absence.

\paragraph{Engagement.}
An Engagement Agent assigns an engagement score $E(q,v)\in[0,1]$ based on pacing, prosodic variation, and slide transitions (Appendix~\ref{appendix:engagement}).

\paragraph{Final Step: Meta-Evaluator Utility}
All agent outputs are collected into a diagnostic record and the overall score is a weighted linear combination of the individual components.
The Meta-Evaluator aggregates these signals using $U(q,v,p)=\lambda_1\pi(q,v)+\lambda_2F(q,v,p)+\lambda_3C(q,v)+\lambda_4D_{\text{del}}(q,v)+\lambda_5E(q,v)$, besides the aggregation is rationale-driven guided by the outputs of all the agents (Details in Appendix~\ref{appendix:meta-evaluation}). 
The Coverage, Faithfulness are assigned higher weights compared to coherence, delivery , followed by engagement.
As a result, $v_A$ receives moderate utility due to shallow delivery, $v_B$ is penalized primarily for coherence and faithfulness violations, and $v_C$ receives lowest utility despite strong delivery and engagement because it omits the core causal claim. 
The Meta-Evaluator explicitly reports these distinctions in its rationale, attributing score differences to missing, misordered, or weakly explained concepts rather than surface quality alone.

\paragraph{Paper--Video Utility.}
Given a paper $p$ with evaluation questions $\mathcal{Q}(p)$, we compute the paper--video utility as $U(p,v)=\frac{1}{|\mathcal{Q}(p)|}\sum_{q\in\mathcal{Q}(p)}U(q,v,p)$. 
Aggregating across questions allows the Meta-Evaluator to identify systematic patterns, such as consistent omission of causal explanations or recurring prerequisite violations, enabling reliable comparison between video variants while preserving interpretability (Appendix~\ref{appendix:paper-video-utility}).
 
\begin{table}[t]
\centering
\tiny
\setlength{\tabcolsep}{6pt}
\begin{tabular}{lcccc}
\toprule
\multirow{2}{*}{\textbf{Evaluation Method}} 
& \multicolumn{2}{c}{\textbf{RecallQ}} 
& \multicolumn{2}{c}{\textbf{Non-RecallQ}} \\
\cmidrule(lr){2-3} \cmidrule(lr){4-5}
& \textbf{K $\tau$ $\uparrow$} 
& \textbf{PA $\uparrow$}
& \textbf{K $\tau$ $\uparrow$} 
& \textbf{PA $\uparrow$} \\
\midrule

\rowcolor{epsgreen}
\textbf{\EffectivePresentationScorer{}} 
& \textbf{0.58} & \textbf{0.80} 
& \textbf{0.53} & \textbf{0.76} \\

Single-Agent LLM QA (F + T) 
& 0.49 & 0.72 
& 0.28 & 0.56 \\

Single-Agent LLM QA (T-only) 
& 0.47 & 0.70 
& 0.25 & 0.54 \\

Holistic LLM Utility Rating 
& 0.45 & 0.69 
& 0.26 & 0.55 \\

\midrule
VideoScore~\cite{he-etal-2024-videoscore} 
& 0.44 & 0.68 
& 0.20 & 0.52 \\

EvalCrafter~\cite{liu2024evalcrafterbenchmarkingevaluatinglarge} 
& 0.46 & 0.69 
& 0.22 & 0.53 \\

PresentQuiz~\cite{zhu2025paper2videoautomaticvideogeneration} 
& 0.46 & 0.70
& 0.31 & 0.59 \\

PPTEval~\cite{zheng-etal-2025-pptagent} 
& 0.36 & 0.67
& 0.32 & 0.63 \\

\bottomrule
\end{tabular}

\caption{Paper-level Kendall’s $\tau$ and pairwise ranking accuracy with respect to human-measured utility, separated into recall-only and non-recall (reasoning) questions. Rows highlighted in green correspond to \textbf{EffectivePresentationScorer}. Recall questions primarily test content coverage, where quiz- and similarity-based metrics remain competitive. For non-recall questions, which require multi-step reasoning and logical integration across scenes, all baselines degrade substantially.}
\label{tab:recall_vs_reasoning_kendall_pairwise}
\end{table}


\begin{table}[t]
\centering
\small
\setlength{\tabcolsep}{6pt}
\begin{tabular}{lccc}
\toprule
\textbf{Improvement Category}
& \textbf{P} $\uparrow$
& \textbf{R} $\uparrow$
& \textbf{F1} $\uparrow$ \\
\midrule
Missing background (BG) 
& 0.82 & 0.78 & 0.80 \\

Poor coherence (ORD) 
& 0.79 & 0.75 & 0.77 \\

Too dense / rushed (TIME) 
& 0.73 & 0.68 & 0.70 \\

Unfaithful content (FAITH) 
& 0.76 & 0.60 & 0.67 \\

Misaligned (AV)
& 0.71 & 0.57 & 0.63 \\

\midrule
\textbf{Micro-average} 
& 0.77 & 0.71 & 0.74 \\

\bottomrule
\end{tabular}

\caption{
Agreement between \EffectivePresentationScorer{} feedback and human feedback inferred from annotator comments.
}
\label{tab:dvue_human_feedback_agreement}
\end{table}

\begin{table}[!t]
\centering
\tiny
\setlength{\tabcolsep}{6pt}
\renewcommand{\arraystretch}{1.25}

\begin{tabular}{p{3cm}ccccc}
\toprule
\textbf{Paper-to-Video Generation}
& \multicolumn{2}{c}{\textbf{Recall Utility}}
& \multicolumn{3}{c}{\textbf{Reasoning Utility}}\\
\cmidrule(lr){2-3}
\cmidrule(lr){4-6}
& P
& F
& Pre.
& D
& O \\
\midrule

Paper2Video~\cite{zhu2025paper2videoautomaticvideogeneration}
& \cellcolor{green!25}0.82
& \cellcolor{green!25}0.74
& \cellcolor{red!20}0.41
& \cellcolor{red!20}0.38
& \cellcolor{red!20}0.43 \\

PPTAgent~\cite{zheng-etal-2025-pptagent}
& \cellcolor{green!25}0.81
& \cellcolor{green!25}0.76
& \cellcolor{yellow!25}0.52
& \cellcolor{red!20}0.45
& \cellcolor{yellow!25}0.51 \\

Doc2PPT~\cite{sun-etal-2021-d2s}
& \cellcolor{green!25}0.64
& \cellcolor{green!25}0.75
& \cellcolor{red!25}0.40
& \cellcolor{red!20}0.43
& \cellcolor{yellow!25}0.49 \\

Human (VISTA)
& \cellcolor{green!25}0.86
& \cellcolor{green!25}0.76
& \cellcolor{green!25}0.81
& \cellcolor{green!25}0.85
& \cellcolor{green!25}0.83 \\

\bottomrule
\end{tabular}

\caption{Recall vs. Reasoning utility under \EffectivePresentationScorer{}. While automated methods achieve strong recall utility (Presence (P), Faithfulness (F)), they exhibit substantial degradation on reasoning utility (Prerequisites (Pre.), Delivery (D), Overall (O)) due to weak prerequisite grounding, explanation quality, and illustrative use of visuals. More Evaluation in Appendix~\ref{app:leceval}.}
\label{tab:recall_reasoning_gap}
\end{table}

\section{Experimental Setup and Results}
\label{sec:results}
The primary goal of our experimental evaluation is to determine whether an automatic evaluator can reliably estimate the instructional utility of scientific presentation videos, where utility is defined as the extent to which a video enables a learner to understand a paper well enough to answer paper-grounded questions using the video alone.
Hence, our experiments are structured around the following research questions which we discuss below.

\paragraph{RQ1) Can \EffectivePresentationScorer{} reliably predict which scientific videos humans find more instructionally useful?} 
We compare \EffectivePresentationScorer{} on EffectivePresentation-EvalBench against a diverse set of baseline methods including single-agent QA and prior video evaluation metrics.
All the baselines produce a single scalar utility score for every (paper, video) pair.

\noindent
\underline{\textbf{Baselines.}} We consider the following single-agent LLM-based QA baselines as follows: (Details in Appendix~\ref{appendix:baselines}) \\
\textbf{[1] Single-Agent VLM QA (T-only))} In the transcript-only setting , we use GPT-4o to answer paper-derived questions using only the video transcript as input. The generated answer is compared against the paper-grounded gold answer using an LLM judge, yielding a correctness score in $\{0, 0.5, 1\}$. Scores are averaged across questions to obtain a paper-level utility estimate. \\
\textbf{[2] Single-Agent VLM QA (T + F)} We evaluate a multimodal variant in which GPT-4o is provided with both the transcript and uniformly sampled video keyframes, and is instructed to jointly reason over textual and visual content. Utility is computed identically as above.\\
\textbf{[3] Holistic VLM Utility Rating.} We prompt GPT-4o to directly rate the overall usefulness of a video for understanding the corresponding paper. \\
Besides, we include prior video evaluation metrics such as VideoScore \cite{he-etal-2024-videoscore}, EvalCrafter \cite{liu2024evalcrafterbenchmarkingevaluatinglarge} and PresentQuiz \cite{zhu2025paper2videoautomaticvideogeneration} and PPTEval from \cite{zheng-etal-2025-pptagent} using their official implementation (Details in Appendix~\ref{appendix:existing-metrics}).\\

\noindent
\underline{\textbf{Evaluation.}} For each paper, all video variants are ranked according to human judgments as well as the scores produced by each evaluation method. We then compute Kendall’s $\tau$ and pairwise accuracy to assess how well each method aligns with human.

\noindent
\underline{\textbf{Results.}}
Table~\ref{tab:recall_vs_reasoning_vertical_vlm} reports correlation separately for recall-only questions and non-recall questions, revealing a clear and systematic pattern.
For recall questions, which primarily test content coverage and factual presence, several baseline methods perform competitively, indicating that surface-level coverage is often sufficient to approximate human utility for recall-oriented understanding.
However, for non-recall questions, all baselines exhibit a substantial drop in correlation. 
This suggests that these approaches are insensitive to pedagogical structure, reasoning flow, and implicit explanation quality.

In contrast, \EffectivePresentationScorer{} maintains strong agreement with human judgments in both settings, achieving the highest Kendall’s $\tau$ and pairwise accuracy for reasoning questions. Bootstrap confidence intervals, permutation tests, and leave-one-out stability analyses confirming these gaps are statistically significant ($p<0.01$) are reported in Appendix~\ref{app:stats}
\emph{The smaller performance gap between recall and reasoning for \EffectivePresentationScorer{} indicates that it captures factors beyond mere content presence, such as explanation order, prerequisite coverage, and inferential support.}
An estimate of the per-pair API-call budget and practical adoptability of the multi-agent pipeline is provided in Appendix~\ref{app:cost}.

\begin{table}[t]
\centering
\small
\setlength{\tabcolsep}{6pt}
\begin{tabular}{lcc}
\toprule
\textbf{Ablation Variant} 
& $\Delta$ \textbf{Kendall $\tau$ $\downarrow$} 
& $\Delta$ \textbf{P-Acc $\downarrow$} \\
\midrule
\textbf{EPS} & 0.00 & 0.00 \\

\midrule
w/o Claim Decomposition
& $-$0.07 & $-$0.08 \\

w/o Faithfulness 
& $-$0.09 & $-$0.10 \\

w/o Coherence
& $-$0.05 & $-$0.06 \\

w/o Delivery 
& $-$0.04 & $-$0.05 \\

w/o Engagement
& $-$0.01 & $-$0.00 \\

\bottomrule
\end{tabular}
\caption{
Aggregate performance drop relative to the full multi-agent pipeline, measured using Kendall’s $\tau$ and pairwise accuracy across all questions.
Removing any single diagnostic component degrades performance, with the largest drops observed when claim decomposition, faithfulness, coherence.
}
\label{tab:delta_ablation_overall}
\end{table}

\paragraph{RQ2) To what extent does the actionable feedback from \EffectivePresentationScorer{} align with improvement needs inferred from human feedback?}

Beyond ranking agreement, an important desideratum for \EffectivePresentationScorer{} is that its diagnostic feedback aligns with the types of improvements that human evaluators implicitly desire. We examine whether the actionable feedback from  \EffectivePresentationScorer{} identifies the same instructional failure modes that humans surface when commenting on presentation videos.\\
\noindent
\underline{\textbf{Experimental Procedure.}}
For each paper–video variant pair, we collect two sources of feedback:
(i) free-form comments/remarks written by human annotators during the utility evaluation study, and
(ii) \EffectivePresentationScorer{} diagnostic report produced during utility evaluation.
To ensure a fair and symmetric comparison, we map both sources into a shared set of improvement categories using a LLM Prompt (Figure~\ref{fig:prompt-feedback-categorization}), which analyzes both human and \EffectivePresentationScorer{} feedback and assigns zero or more labels from the following taxonomy: \emph{Missing pre-requisities (BG)},  \emph{Poor coherence (ORD)},  \emph{Overly dense or rushed explanations (TIME)},  \emph{Unfaithful content (FAITH)}, and  \emph{Misaligned narration and visuals (AV)}. This procedure avoids manual labeling and ensures that both human and {\bf EffectivePresentationScorer} feedback are interpreted under identical criteria. Treating human-derived labels as the reference set and \EffectivePresentationScorer{}-derived labels as predictions, we compute precision, recall, and F1 per-category (Examples in Table~\ref{tab:feedback-mapping-examples}).

\noindent
\underline{\textbf{Results.}}
Table~\ref{tab:dvue_human_feedback_agreement} reports agreement between \EffectivePresentationScorer{} actionable feedback and human feedback.
Overall, \EffectivePresentationScorer{} exhibits strong alignment with human-identified improvement needs, indicating that its diagnostics capture not only which video variants are less useful, but also \emph{why} they fail instructionally. 
Agreement is strongest for missing background and poor ordering, reflecting the effectiveness of \EffectivePresentationScorer{} prerequisite planning and temporal coherence modeling.
\emph{These results suggest that the feedback from \EffectivePresentationScorer{} closely mirrors human instructional critique, while decomposing it into actionable signals.}

\paragraph{RQ3) How do state-of-the-art paper-to-video generation systems compare
to human-authored videos when evaluated for instructional utility?}
\label{sec:dvue_sota_video_analysis}

To answer this question, we compare videos generated for five NLP papers using multiple state-of-the-art automatic pipelines against human-authored presentation videos, all evaluated under identical settings using \EffectivePresentationScorer{}. The automatic systems include an end-to-end paper-to-video model, \citeauthor{zhu2025paper2videoautomaticvideogeneration}'s \textit{Paper2Video}~\cite{zhu2025paper2videoautomaticvideogeneration}, as well as document-to-slides pipelines based on \textit{PPTAgent}~\cite{zheng-etal-2025-pptagent} and \textit{Doc2PPT}~\cite{sun-etal-2021-d2s}. For the slide-based pipelines, we convert generated slide decks into narrated videos using a uniform MoviePy-based rendering process with text-to-speech narration, ensuring that differences in utility arise from content structuring rather than rendering artifacts. As a human upper bound, we use professionally authored presentation videos from VISTA dataset, which are created by humans to explain same papers.

Across all five papers, \EffectivePresentationScorer{} consistently assigns the highest instructional utility to human-authored videos. While the relative ordering of automatic methods varies by paper, all state-of-the-art pipelines exhibit a persistent utility gap compared to human videos, even when surface-level quality and content coverage are high (Table~\ref{tab:recall_reasoning_gap}). Diagnostic outputs reveal two recurring failure modes: insufficient motivation and causal linkage for prerequisite concepts, and passive treatment of figures that are shown but not explained to support visual reasoning.

These findings explain why quiz-based and similarity-driven metrics often overestimate video quality: structural correctness without explanatory depth yields low instructional utility. In contrast, human-authored videos progressively introduce prerequisites, motivate key concepts, and actively use figures as explanatory tools, resulting in higher grounding, coherence, and overall utility. \emph{Together, these results highlight that current paper-to-video systems organize content well but struggle to deliver grounded, learner-oriented explanations.}

\section{Further Analysis}

\paragraph{Generalizability across VLM Usage.}
We evaluate \EffectivePresentationScorer{} across three VLM backbones (GPT-4o, Gemini-3, Qwen-VL) and two question types (recall and non-recall), comparing against single-agent QA and holistic LLM ratings using Kendall’s $\tau$ and pairwise accuracy. 
Across all backbones, \EffectivePresentationScorer{} consistently achieves the strongest alignment with human utility judgments (Table~\ref{tab:recall_vs_reasoning_vertical_vlm}).
Beyond ranking alignment, we additionally measure cross-model agreement on the underlying per-claim decisions (presence, faithfulness, coherence) and analyze the qualitative distribution of disagreement cases; details and statistics are reported in Appendix~\ref{sec:llm-judge-robustness}.

\paragraph{Ablation of Components.}
In simple terms, Table~\ref{tab:delta_ablation_overall} shows what happens when we remove one part of our evaluation system at a time. The full \EffectivePresentationScorer{} (top row) performs best, and every ablated version performs worse. Removing claim decomposition or faithfulness causes the largest drop, meaning the system struggles most when it can no longer reason about what should be explained or whether the explanation is true to the paper. Removing delivery, coherence, or engagement also hurts performance, but to a smaller extent.
Overall, the results show that no single component is sufficient on its own: reliable utility estimation requires combining multiple diagnostic signals. This explains why simpler or single-agent evaluators are less aligned with human judgments than the full multi-agent framework.
\paragraph{Qualiative Analysis.}
We qualitatively investigate systematic instructional failures in generated videos that are not captured by surface-level metrics in Figure~\ref{tab:qualitative_case_study_three_regimes}). 
In the first two cases, key concepts are introduced without sufficient grounding or are presented out of logical order, leading to low coverage and coherence despite visually polished slides. 
While PresentQuiz \cite{zhu2025paper2videoautomaticvideogeneration} and VideoScore \cite{he2024videoscore} assign high utility based on content presence and visual quality, \EffectivePresentationScorer{} correctly penalizes these breakdowns by identifying missing prerequisites and ordering violations. 
In contrast, human-authored video provides explicit background, step-by-step explanation, and coherent narrative flow, resulting in high utility across all dimensions.
\section{Conclusion and Future Work}
\label{sec:conclusion}
We introduce \EffectivePresentationScorer{}, a diagnostic, interpretable, and goal-conditioned evaluation framework that measures instructional utility through learner-oriented dimensions.
Beyond evaluation, we envision \EffectivePresentationScorer{} as a feedback mechanism for paper-to-video generation systems. The category-level alignment between its diagnostics and human improvement needs (Table 3) suggests that its outputs can directly guide generation along controllable axes such as prerequisite coverage, claim ordering, and explanation depth—either as intermediate supervision signals, as reinforcement rewards, or as critique inputs in iterative revision loops. The present work intentionally scopes its contribution to establishing a reliable, interpretable evaluation framework; integrating EffectivePresentationScorer into a closed-loop generation pipeline, and quantifying the resulting improvements in instructional utility, is the focus of our ongoing work.
\section*{Limitations}
\label{sec:limitations}
While \EffectivePresentationScorer{} provides a structured and interpretable framework for evaluating instructional utility of paper-to-video presentations, we call out a few limitations of our approach.
\begin{enumerate}
    \item \textbf{Dependence on Paper-Grounded Question Design.}
Our evaluation operationalizes instructional utility through paper-grounded questions derived from citation context and Bloom’s taxonomy. While this ensures alignment with community-recognized contributions, it also means that the utility estimates are conditioned on the quality and coverage of the question set. Extending the framework to adaptive or learner-generated question sets remains an important direction for future work.
\item \textbf{LLM/VLM Reliance and Model Bias.}
Several agents in \EffectivePresentationScorer{} rely on large language and vision–language models for claim decomposition, faithfulness verification, visual description, and engagement assessment. Although we show robustness across backbones (GPT-4o, Gemini-3, Qwen-3), these models may still introduce biases, hallucinations, or sensitivity to prompt phrasing. While the multi-agent structure mitigates single-model failure modes, the framework does not eliminate dependence on the underlying model capabilities.
\item \textbf{Evaluation Rather Than Generation.}
\EffectivePresentationScorer{} is designed as an evaluation and diagnostic framework, not a generative model. While we discuss how its outputs could be used as training signals or reinforcement rewards, we do not implement a closed-loop generation system in this work. Demonstrating end-to-end gains in video generation quality through direct optimization against these diagnostics is left to future research.
\item \textbf{Human Annotation Scope.}
Although we collect carefully screened human judgments and analyze agreement in detail, our human evaluation is limited to a fixed number of papers and annotators within NLP and ML domains. Generalization to other scientific fields, educational levels, or broader audiences requires further validation.
\item \textbf{Benchmark Scale.} EffectivePresentation-EvalBench currently comprises 20 papers and 140 videos. This scale is a deliberate consequence of our controlled, within-paper A/B design, which requires seven matched variants and screened annotators per paper to support causal attribution of instructional failures. Because the evaluation framework is modular and dataset-agnostic, scaling to larger and more domain-diverse corpora is straightforward and is an important direction for follow-up work.
\item \textbf{Focus on Presentation-Style Videos.} Our study targets long-form, presentation-style scientific videos (e.g., narrated slides, conference talks), where structured explanation, prerequisite ordering, and claim-based reasoning are central to learning. The framework is not directly evaluated on other instructional formats such as interactive tutorials, code walkthroughs, or highly cinematic explanatory videos. We emphasize, however, that the core abstractions of EffectivePresentationScorer—decomposing understanding into claim-level reasoning, dependency structure, and delivery quality—are fundamentally format-agnostic. Adapting the framework to other genres primarily requires extending (i) the input representation, e.g., incorporating interaction traces for tutorials or execution traces for code walkthroughs, and (ii) the notion of delivery, e.g., interactivity or learner-driven pacing in place of fixed narration. Evaluating EffectivePresentationScorer on such formats is an important direction for future work.
\end{enumerate}

\paragraph{Use of AI assistants.} The authors used AI tools (OpenAI's ChatGPT, Google Gemini, Anthropic's Claude) for coding assistance during data analysis and visualization, and as a writing assistant limited to paraphrasing for conciseness. All substantive content, analysis, and conclusions are the authors' own work.

\bibliography{custom}

\begin{thebibliography}{35}
\providecommand{\natexlab}[1]{#1}

\bibitem[{Ackermans et~al.(2025)Ackermans, {de Koning}, and Jarodzka}]{ACKERMANS2025102137}
Kevin Ackermans, Björn~B. {de Koning}, and Halszka Jarodzka. 2025.
\newblock \href {https://doi.org/10.1016/j.learninstruc.2025.102137} {Instructional videos and deeper processing: Insights and applications}.
\newblock \emph{Learning and Instruction}, 98:102137.

\bibitem[{Adams(2015)}]{Adams2015-nv}
Nancy~E Adams. 2015.
\newblock Bloom's taxonomy of cognitive learning objectives.
\newblock \emph{J Med Libr Assoc}, 103(3):152--153.

\bibitem[{Ayres and Ackermans(2025)}]{AYRES2025102077}
Paul Ayres and Kevin Ackermans. 2025.
\newblock \href {https://doi.org/10.1016/j.learninstruc.2024.102077} {Some do's and don'ts of educational videos}.
\newblock \emph{Learning and Instruction}, 96:102077.

\bibitem[{Bansal et~al.(2024)Bansal, Lin, Xie, Zong, Yarom, Bitton, Jiang, Sun, Chang, and Grover}]{bansal2024videophyevaluatingphysicalcommonsense}
Hritik Bansal, Zongyu Lin, Tianyi Xie, Zeshun Zong, Michal Yarom, Yonatan Bitton, Chenfanfu Jiang, Yizhou Sun, Kai-Wei Chang, and Aditya Grover. 2024.
\newblock \href {https://arxiv.org/abs/2406.03520} {Videophy: Evaluating physical commonsense for video generation}.
\newblock \emph{Preprint}, arXiv:2406.03520.

\bibitem[{Cao et~al.(2026)Cao, Nguyen, Yavuz, Yu, Wang, Bharaj, and Francis}]{cao2026developingauthenticsimulatedlearners}
Jie Cao, Ha~Nguyen, Selim Yavuz, Boran Yu, Shuguang Wang, Pavneet~Kaur Bharaj, and Dionne~Cross Francis. 2026.
\newblock \href {https://arxiv.org/abs/2604.04361} {Developing authentic simulated learners for mathematics teacher learning: Insights from three approaches with large language models}.
\newblock \emph{Preprint}, arXiv:2604.04361.

\bibitem[{He et~al.(2024{\natexlab{a}})He, Jiang, Zhang, Ku, Soni, Siu, Chen, Chandra, Jiang, Arulraj, Wang, Do, Ni, Lyu, Narsupalli, Fan, Lyu, Lin, and Chen}]{he-etal-2024-videoscore}
Xuan He, Dongfu Jiang, Ge~Zhang, Max Ku, Achint Soni, Sherman Siu, Haonan Chen, Abhranil Chandra, Ziyan Jiang, Aaran Arulraj, Kai Wang, Quy~Duc Do, Yuansheng Ni, Bohan Lyu, Yaswanth Narsupalli, Rongqi Fan, Zhiheng Lyu, Bill~Yuchen Lin, and Wenhu Chen. 2024{\natexlab{a}}.
\newblock \href {https://doi.org/10.18653/v1/2024.emnlp-main.127} {{V}ideo{S}core: Building automatic metrics to simulate fine-grained human feedback for video generation}.
\newblock In \emph{Proceedings of the 2024 Conference on Empirical Methods in Natural Language Processing}, pages 2105--2123, Miami, Florida, USA. Association for Computational Linguistics.

\bibitem[{He et~al.(2024{\natexlab{b}})He, Jiang, Zhang, Ku, Soni, Siu, Chen, Chandra, Jiang, Arulraj, Wang, Do, Ni, Lyu, Narsupalli, Fan, Lyu, Lin, and Chen}]{he2024videoscore}
Xuan He, Dongfu Jiang, Ge~Zhang, Max Ku, Achint Soni, Sherman Siu, Haonan Chen, Abhranil Chandra, Ziyan Jiang, Aaran Arulraj, Kai Wang, Quy~Duc Do, Yuansheng Ni, Bohan Lyu, Yaswanth Narsupalli, Rongqi Fan, Zhiheng Lyu, Yuchen Lin, and Wenhu Chen. 2024{\natexlab{b}}.
\newblock \href {https://arxiv.org/abs/2406.15252} {Videoscore: Building automatic metrics to simulate fine-grained human feedback for video generation}.
\newblock \emph{ArXiv}, abs/2406.15252.

\bibitem[{Huang et~al.(2024{\natexlab{a}})Huang, He, Yu, Zhang, Si, Jiang, Zhang, Wu, Jin, Chanpaisit, Wang, Chen, Wang, Lin, Qiao, and Liu}]{huang2023vbench}
Ziqi Huang, Yinan He, Jiashuo Yu, Fan Zhang, Chenyang Si, Yuming Jiang, Yuanhan Zhang, Tianxing Wu, Qingyang Jin, Nattapol Chanpaisit, Yaohui Wang, Xinyuan Chen, Limin Wang, Dahua Lin, Yu~Qiao, and Ziwei Liu. 2024{\natexlab{a}}.
\newblock {VBench}: Comprehensive benchmark suite for video generative models.
\newblock In \emph{Proceedings of the IEEE/CVF Conference on Computer Vision and Pattern Recognition}.

\bibitem[{Huang et~al.(2024{\natexlab{b}})Huang, Zhang, Xu, He, Yu, Dong, Ma, Chanpaisit, Si, Jiang, Wang, Chen, Chen, Wang, Lin, Qiao, and Liu}]{huang2024vbenchcomprehensiveversatilebenchmark}
Ziqi Huang, Fan Zhang, Xiaojie Xu, Yinan He, Jiashuo Yu, Ziyue Dong, Qianli Ma, Nattapol Chanpaisit, Chenyang Si, Yuming Jiang, Yaohui Wang, Xinyuan Chen, Ying-Cong Chen, Limin Wang, Dahua Lin, Yu~Qiao, and Ziwei Liu. 2024{\natexlab{b}}.
\newblock \href {https://arxiv.org/abs/2411.13503} {Vbench++: Comprehensive and versatile benchmark suite for video generative models}.
\newblock \emph{Preprint}, arXiv:2411.13503.

\bibitem[{Ku et~al.(2024)Ku, Jiang, Wei, Yue, and Chen}]{ku-etal-2024-viescore}
Max Ku, Dongfu Jiang, Cong Wei, Xiang Yue, and Wenhu Chen. 2024.
\newblock \href {https://doi.org/10.18653/v1/2024.acl-long.663} {{VIES}core: Towards explainable metrics for conditional image synthesis evaluation}.
\newblock In \emph{Proceedings of the 62nd Annual Meeting of the Association for Computational Linguistics (Volume 1: Long Papers)}, pages 12268--12290, Bangkok, Thailand. Association for Computational Linguistics.

\bibitem[{Liao et~al.(2024)Liao, Lu, Zhang, Wan, Wang, Zhao, Zuo, Ye, and Wang}]{liao2024evaluation}
Mingxiang Liao, Hannan Lu, Xinyu Zhang, Fang Wan, Tianyu Wang, Yuzhong Zhao, Wangmeng Zuo, Qixiang Ye, and Jingdong Wang. 2024.
\newblock Evaluation of text-to-video generation models: A dynamics perspective.
\newblock \emph{arXiv preprint arXiv:2407.01094}.

\bibitem[{Liu et~al.(2025)Liu, Whitehouse, Yu, Mahon, Saxena, Zhao, Qiu, Lapata, and Demberg}]{liu-etal-2025-talk}
Dongqi Liu, Chenxi Whitehouse, Xi~Yu, Louis Mahon, Rohit Saxena, Zheng Zhao, Yifu Qiu, Mirella Lapata, and Vera Demberg. 2025.
\newblock \href {https://doi.org/10.18653/v1/2025.acl-long.310} {What is that talk about? a video-to-text summarization dataset for scientific presentations}.
\newblock In \emph{Proceedings of the 63rd Annual Meeting of the Association for Computational Linguistics (Volume 1: Long Papers)}, pages 6187--6210, Vienna, Austria. Association for Computational Linguistics.

\bibitem[{Liu et~al.(2024{\natexlab{a}})Liu, Cun, Liu, Wang, Zhang, Chen, Liu, Zeng, Chan, and Shan}]{liu2024evalcrafterbenchmarkingevaluatinglarge}
Yaofang Liu, Xiaodong Cun, Xuebo Liu, Xintao Wang, Yong Zhang, Haoxin Chen, Yang Liu, Tieyong Zeng, Raymond Chan, and Ying Shan. 2024{\natexlab{a}}.
\newblock \href {https://arxiv.org/abs/2310.11440} {Evalcrafter: Benchmarking and evaluating large video generation models}.
\newblock \emph{Preprint}, arXiv:2310.11440.

\bibitem[{Liu et~al.(2024{\natexlab{b}})Liu, Zhang, Li, Yan, Gao, Chen, Yuan, Huang, Sun, Gao, He, and Sun}]{liu2024sorareviewbackgroundtechnology}
Yixin Liu, Kai Zhang, Yuan Li, Zhiling Yan, Chujie Gao, Ruoxi Chen, Zhengqing Yuan, Yue Huang, Hanchi Sun, Jianfeng Gao, Lifang He, and Lichao Sun. 2024{\natexlab{b}}.
\newblock \href {https://arxiv.org/abs/2402.17177} {Sora: A review on background, technology, limitations, and opportunities of large vision models}.
\newblock \emph{Preprint}, arXiv:2402.17177.

\bibitem[{Mayer(2002)}]{Mayer2024ThePP}
Richard~E Mayer. 2002.
\newblock Multimedia learning.
\newblock In \emph{Psychology of learning and motivation}, volume~41, pages 85--139. Elsevier.

\bibitem[{McConnell et~al.(2017)McConnell, Chapman, Czajka, Jones, Ryker, and Wiggen}]{McConnell2017InstructionalUA}
David~A. McConnell, LeeAnna~Young Chapman, Charles~Douglas Czajka, Jason~P. Jones, Katherine Ryker, and Jennifer Wiggen. 2017.
\newblock \href {https://api.semanticscholar.org/CorpusID:85462730} {Instructional utility and learning efficacy of common active learning strategies}.
\newblock \emph{Journal of Geoscience Education}, 65:604 -- 625.

\bibitem[{Mondal et~al.(2024)Mondal, S, Natarajan, Garimella, Bandyopadhyay, and Boyd-Graber}]{mondal-etal-2024-presentations}
Ishani Mondal, Shwetha S, Anandhavelu Natarajan, Aparna Garimella, Sambaran Bandyopadhyay, and Jordan Boyd-Graber. 2024.
\newblock \href {https://doi.org/10.18653/v1/2024.eacl-long.163} {Presentations by the humans and for the humans: Harnessing {LLM}s for generating persona-aware slides from documents}.
\newblock In \emph{Proceedings of the 18th Conference of the European Chapter of the Association for Computational Linguistics (Volume 1: Long Papers)}, pages 2664--2684, St. Julian{'}s, Malta. Association for Computational Linguistics.

\bibitem[{Niekrenz and Spreckelsen(2024)}]{Niekrenz2024-hb}
Lukas Niekrenz and Cord Spreckelsen. 2024.
\newblock How to design effective educational videos for teaching evidence-based medicine to undergraduate learners - systematic review with complementing qualitative research to develop a practicable guide.
\newblock \emph{Med Educ Online}, 29(1):2339569.

\bibitem[{Radford et~al.(2021)Radford, Kim, Hallacy, Ramesh, Goh, Agarwal, Sastry, Askell, Mishkin, Clark, Krueger, and Sutskever}]{radford2021learningtransferablevisualmodels}
Alec Radford, Jong~Wook Kim, Chris Hallacy, Aditya Ramesh, Gabriel Goh, Sandhini Agarwal, Girish Sastry, Amanda Askell, Pamela Mishkin, Jack Clark, Gretchen Krueger, and Ilya Sutskever. 2021.
\newblock \href {https://arxiv.org/abs/2103.00020} {Learning transferable visual models from natural language supervision}.
\newblock \emph{Preprint}, arXiv:2103.00020.

\bibitem[{Ram{\'o}n-Arbu{\'e}s et~al.(2025)Ram{\'o}n-Arbu{\'e}s, Bl{\'a}zquez-Ornat, Sagarra-Romero, Benito-Ruiz, Ant{\'o}n-Solanas, and G{\'o}mez-Torres}]{Ramon-Arbues2025-ao}
Enrique Ram{\'o}n-Arbu{\'e}s, Isabel Bl{\'a}zquez-Ornat, Luc{\'\i}a Sagarra-Romero, Eva Benito-Ruiz, Isabel Ant{\'o}n-Solanas, and Piedad G{\'o}mez-Torres. 2025.
\newblock Students' perceptions of creating educational videos as a teaching and learning strategy.
\newblock \emph{Nurse Educ}, 50(4):E219--E224.

\bibitem[{Reimers and Gurevych(2019)}]{reimers-gurevych-2019-sentence}
Nils Reimers and Iryna Gurevych. 2019.
\newblock \href {https://doi.org/10.18653/v1/D19-1410} {Sentence-{BERT}: Sentence embeddings using {S}iamese {BERT}-networks}.
\newblock In \emph{Proceedings of the 2019 Conference on Empirical Methods in Natural Language Processing and the 9th International Joint Conference on Natural Language Processing (EMNLP-IJCNLP)}, pages 3982--3992, Hong Kong, China. Association for Computational Linguistics.

\bibitem[{Reimers and Gurevych(2020)}]{reimers-2020-multilingual-sentence-bert}
Nils Reimers and Iryna Gurevych. 2020.
\newblock \href {https://arxiv.org/abs/2004.09813} {Making monolingual sentence embeddings multilingual using knowledge distillation}.
\newblock In \emph{Proceedings of the 2020 Conference on Empirical Methods in Natural Language Processing}. Association for Computational Linguistics.

\bibitem[{Reiter(2025)}]{reiter2025evaluaterealworldimpact}
Ehud Reiter. 2025.
\newblock \href {https://doi.org/10.1162/coli.a.18} {We should evaluate real-world impact}.
\newblock \emph{Computational Linguistics}, 51(4):1419--1431.

\bibitem[{Sun et~al.(2021)Sun, Hou, Wang, Zhang, and Wang}]{sun-etal-2021-d2s}
Edward Sun, Yufang Hou, Dakuo Wang, Yunfeng Zhang, and Nancy X.~R. Wang. 2021.
\newblock \href {https://doi.org/10.18653/v1/2021.naacl-main.111} {{D}2{S}: Document-to-slide generation via query-based text summarization}.
\newblock In \emph{Proceedings of the 2021 Conference of the North American Chapter of the Association for Computational Linguistics: Human Language Technologies}, pages 1405--1418, Online. Association for Computational Linguistics.

\bibitem[{Sweller(1994)}]{SWELLER1994295}
John Sweller. 1994.
\newblock \href {https://doi.org/10.1016/0959-4752(94)90003-5} {Cognitive load theory, learning difficulty, and instructional design}.
\newblock \emph{Learning and Instruction}, 4(4):295--312.

\bibitem[{Sweller(2024)}]{SWELLER2024102423}
John Sweller. 2024.
\newblock \href {https://doi.org/10.1016/j.lindif.2024.102423} {Cognitive load theory and individual differences}.
\newblock \emph{Learning and Individual Differences}, 110:102423.

\bibitem[{Tsukagoshi et~al.(2021)Tsukagoshi, Sasano, and Takeda}]{tsukagoshi-etal-2021-defsent}
Hayato Tsukagoshi, Ryohei Sasano, and Koichi Takeda. 2021.
\newblock \href {https://doi.org/10.18653/v1/2021.acl-short.52} {{D}ef{S}ent: Sentence embeddings using definition sentences}.
\newblock In \emph{Proceedings of the 59th Annual Meeting of the Association for Computational Linguistics and the 11th International Joint Conference on Natural Language Processing (Volume 2: Short Papers)}, pages 411--418, Online. Association for Computational Linguistics.

\bibitem[{Unterthiner et~al.(2019)Unterthiner, van Steenkiste, Kurach, Marinier, Michalski, and Gelly}]{unterthiner2019accurategenerativemodelsvideo}
Thomas Unterthiner, Sjoerd van Steenkiste, Karol Kurach, Raphael Marinier, Marcin Michalski, and Sylvain Gelly. 2019.
\newblock \href {https://arxiv.org/abs/1812.01717} {Towards accurate generative models of video: A new metric \& challenges}.
\newblock \emph{Preprint}, arXiv:1812.01717.

\bibitem[{Watts et~al.(2025)Watts, Liu, Ober, Song, Jusino-Del~Valle, Zhai, Wang, and Liu}]{educsci15111507}
Field~M. Watts, Lei Liu, Teresa~M. Ober, Yi~Song, Euvelisse Jusino-Del~Valle, Xiaoming Zhai, Yun Wang, and Ninghao Liu. 2025.
\newblock \href {https://doi.org/10.3390/educsci15111507} {A framework for designing an ai chatbot to support scientific argumentation}.
\newblock \emph{Education Sciences}, 15(11).

\bibitem[{Wiedemer et~al.(2025)Wiedemer, Li, Vicol, Gu, Matarese, Swersky, Kim, Jaini, and Geirhos}]{wiedemer2025videomodelszeroshotlearners}
Thaddäus Wiedemer, Yuxuan Li, Paul Vicol, Shixiang~Shane Gu, Nick Matarese, Kevin Swersky, Been Kim, Priyank Jaini, and Robert Geirhos. 2025.
\newblock \href {https://arxiv.org/abs/2509.20328} {Video models are zero-shot learners and reasoners}.
\newblock \emph{Preprint}, arXiv:2509.20328.

\bibitem[{Wu et~al.(2021)Wu, Minervini, Stenetorp, and Riedel}]{wu-etal-2021-training}
Yuxiang Wu, Pasquale Minervini, Pontus Stenetorp, and Sebastian Riedel. 2021.
\newblock \href {https://doi.org/10.18653/v1/2021.acl-short.57} {Training adaptive computation for open-domain question answering with computational constraints}.
\newblock In \emph{Proceedings of the 59th Annual Meeting of the Association for Computational Linguistics and the 11th International Joint Conference on Natural Language Processing (Volume 2: Short Papers)}, pages 447--453, Online. Association for Computational Linguistics.

\bibitem[{Yang et~al.(2025)Yang, Fan, Sun, Li, Zeng, Han, Zhai, Liu, Cao, and Zha}]{yang2025videogen}
Yuhang Yang, Ke~Fan, Shangkun Sun, Hongxiang Li, Ailing Zeng, FeiLin Han, Wei Zhai, Wei Liu, Yang Cao, and Zheng-Jun Zha. 2025.
\newblock Videogen-eval: Agent-based system for video generation evaluation.
\newblock \emph{arXiv preprint arXiv:2503.23452}.

\bibitem[{Yin et~al.(2025)Yin, Zhang-Li, Yu, Li, Tu, Wang, Liu, Liu, Hou, Li, and Xu}]{yin2025lecevalautomatedmetricmultimodal}
Joy Lim~Jia Yin, Daniel Zhang-Li, Jifan Yu, Haoxuan Li, Shangqing Tu, Yuanchun Wang, Zhiyuan Liu, Huiqin Liu, Lei Hou, Juanzi Li, and Bin Xu. 2025.
\newblock \href {https://arxiv.org/abs/2505.02078} {Leceval: An automated metric for multimodal knowledge acquisition in multimedia learning}.
\newblock \emph{Preprint}, arXiv:2505.02078.

\bibitem[{Zheng et~al.(2025)Zheng, Guan, Kong, Zhang, Zheng, Zhou, Lin, Lu, Han, and Sun}]{zheng-etal-2025-pptagent}
Hao Zheng, Xinyan Guan, Hao Kong, Wenkai Zhang, Jia Zheng, Weixiang Zhou, Hongyu Lin, Yaojie Lu, Xianpei Han, and Le~Sun. 2025.
\newblock \href {https://doi.org/10.18653/v1/2025.emnlp-main.728} {{PPTA}gent: Generating and evaluating presentations beyond text-to-slides}.
\newblock In \emph{Proceedings of the 2025 Conference on Empirical Methods in Natural Language Processing}, pages 14413--14429, Suzhou, China. Association for Computational Linguistics.

\bibitem[{Zhu et~al.(2025)Zhu, Lin, and Shou}]{zhu2025paper2videoautomaticvideogeneration}
Zeyu Zhu, Kevin~Qinghong Lin, and Mike~Zheng Shou. 2025.
\newblock \href {https://arxiv.org/abs/2510.05096} {Paper2video: Automatic video generation from scientific papers}.
\newblock \emph{Preprint}, arXiv:2510.05096.

\end{thebibliography}
\section*{Appendix}

The appendix provides additional details supporting the methodology, dataset construction, evaluation protocol, and analysis presented in the main paper. It is organized as follows:

\begin{itemize}
    \item \textbf{Related Work (Section~\ref{sec:related_work_eval})} Extended discussion of prior work on video generation evaluation, including benchmark suites, dynamics-aware metrics, and industrial systems such as NotebookLM.

    \item \textbf{Dataset Construction (Section~\ref{sec:human_eval_details}).} Detailed description of \textbf{EffectivePresentation-EvalBench}, including background question generation, annotator recruitment and screening, utility-defining evaluation questions, and controlled video perturbations.

    \item \textbf{Dataset Statistics~\ref{sec:Datasetstats}} 

    \item \textbf{Implementation Details (Section~\ref{appendix:baselines}} End-to-end implementation details of \EffectivePresentationScorer{}, including PDF preprocessing, video preprocessing, multimodal representations, and agent-specific prompts.


\end{itemize}

\begin{table*}[t]
\tiny
\centering
\footnotesize
\setlength{\tabcolsep}{4pt}
\renewcommand{\arraystretch}{1.15}

\begin{tabular}{ccc|cccc}
\toprule
\multicolumn{3}{c|}{\textbf{Contiguous Frames}} 
& \textbf{Human} 
& \textbf{EPS} 
& \textbf{PQ} 
& \textbf{ViS} \\
\midrule

\includegraphics[width=0.18\linewidth]{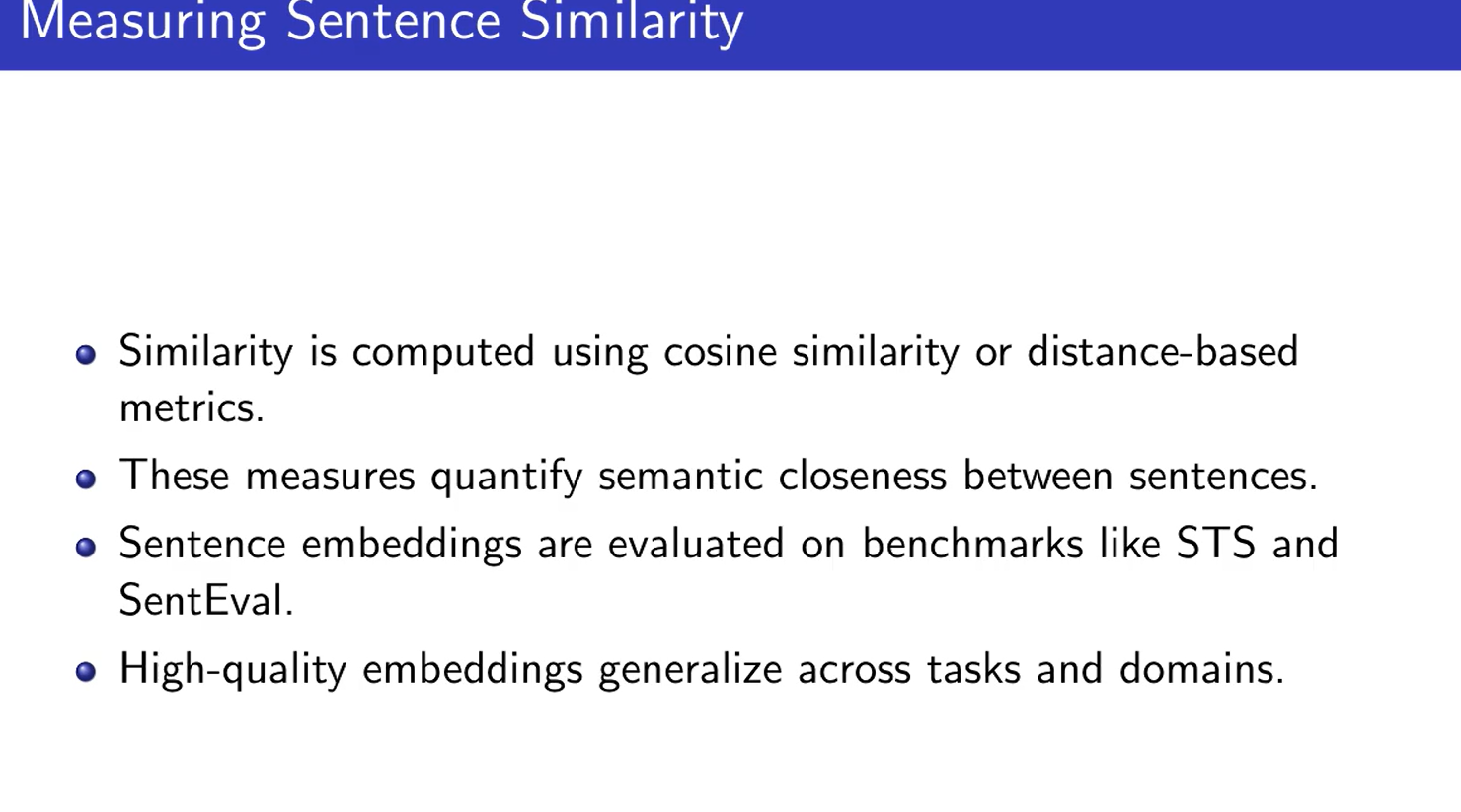} &
\includegraphics[width=0.18\linewidth]{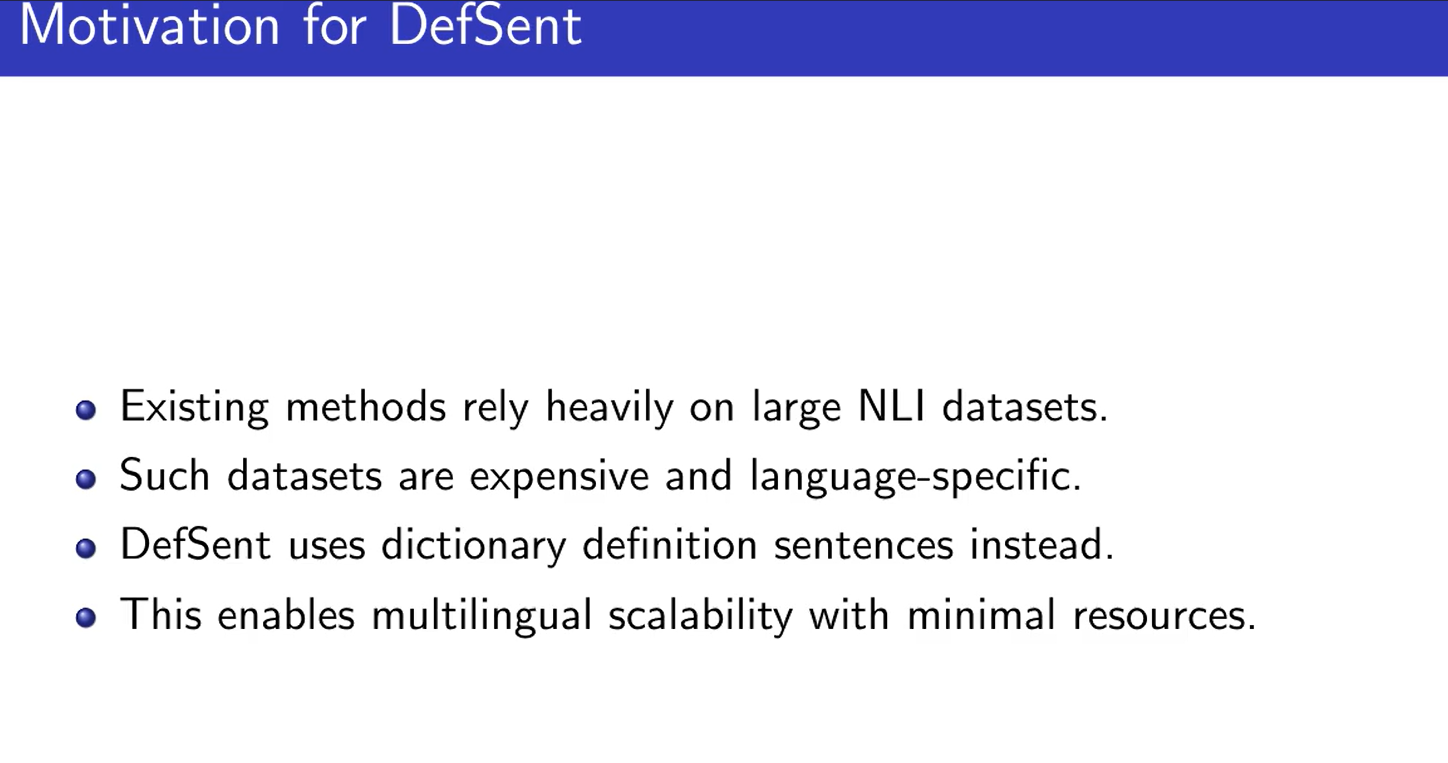} &
\includegraphics[width=0.18\linewidth]{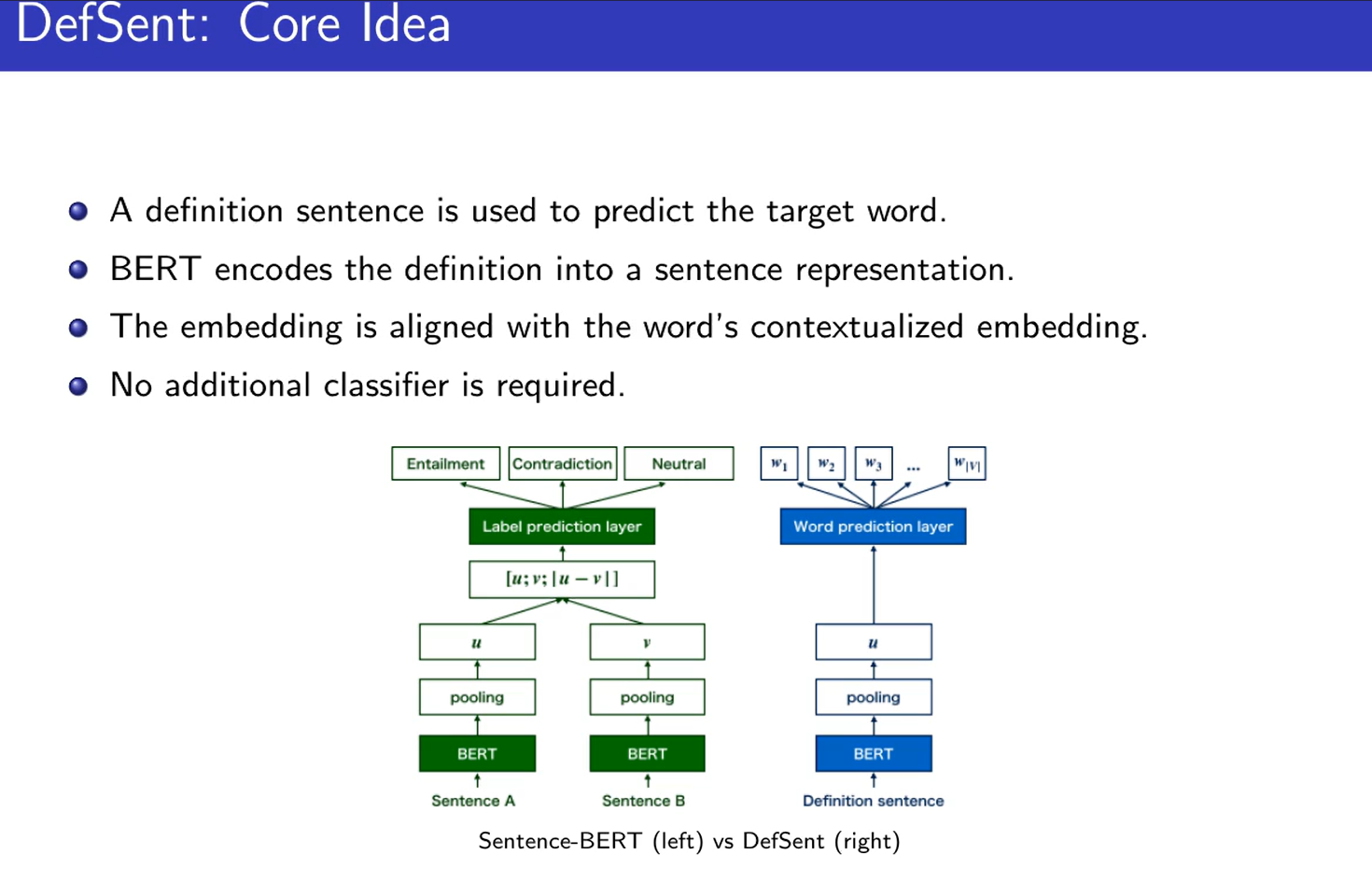} 
&
\textbf{L} 
&
\textbf{L} 
&
\textbf{H} 
&
\textbf{H} \\

\multicolumn{3}{l|}{\emph{Low coverage: key concepts introduced without explanation or grounding}} 
& & & & \\
\midrule

\includegraphics[width=0.18\linewidth]{figures/Screenshot_2025-12-30_183406.png} &
\includegraphics[width=0.18\linewidth]{figures/Screenshot_2025-12-30_183419.png} &
\includegraphics[width=0.18\linewidth]{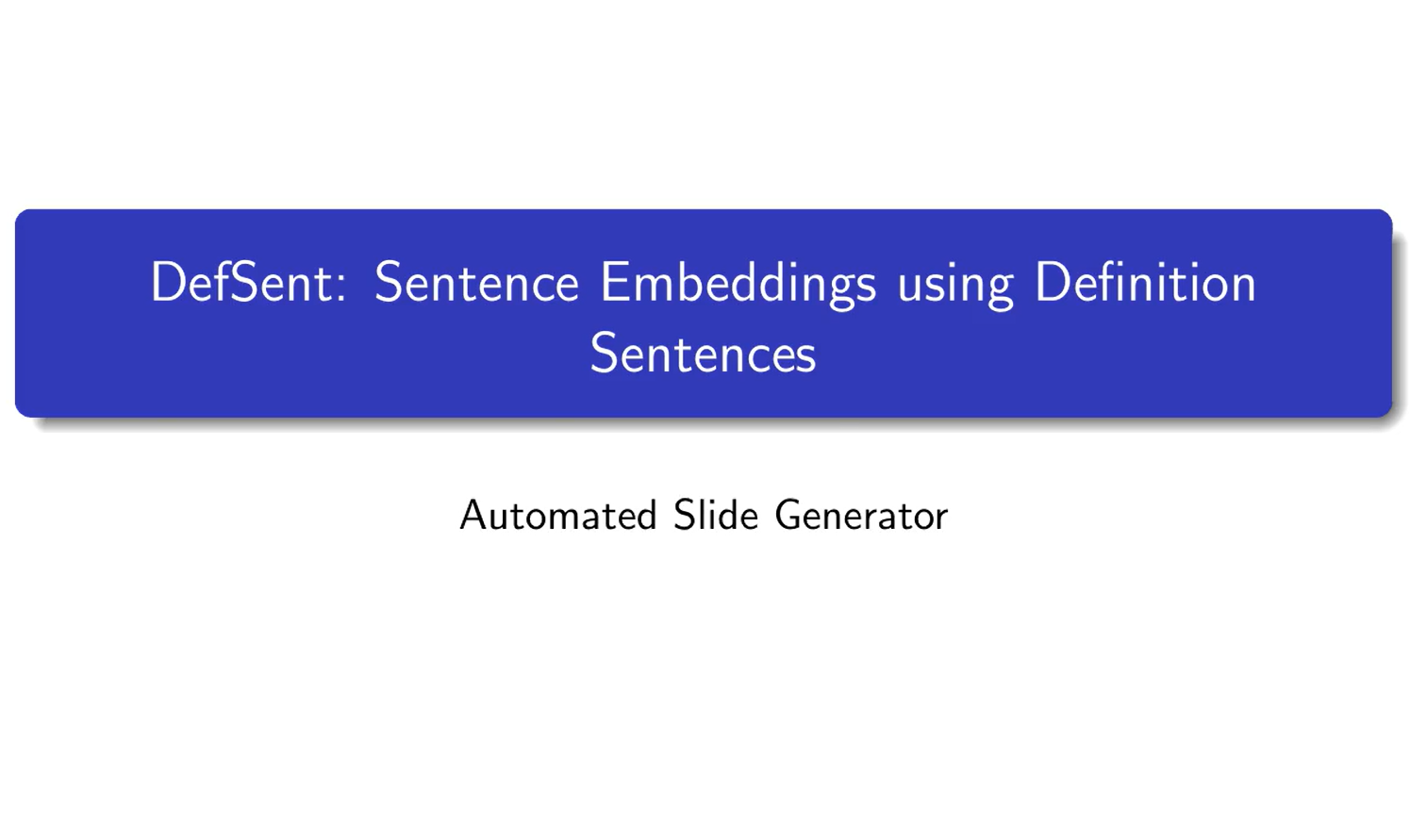} 
&
\textbf{L} 
&
\textbf{L} 
&
\textbf{H} 
&
\textbf{H} \\

\multicolumn{3}{l|}{\emph{Bad coherence: motivation, method, and core idea presented out of logical order}} 
& & & & \\
\midrule

\includegraphics[width=0.18\linewidth]{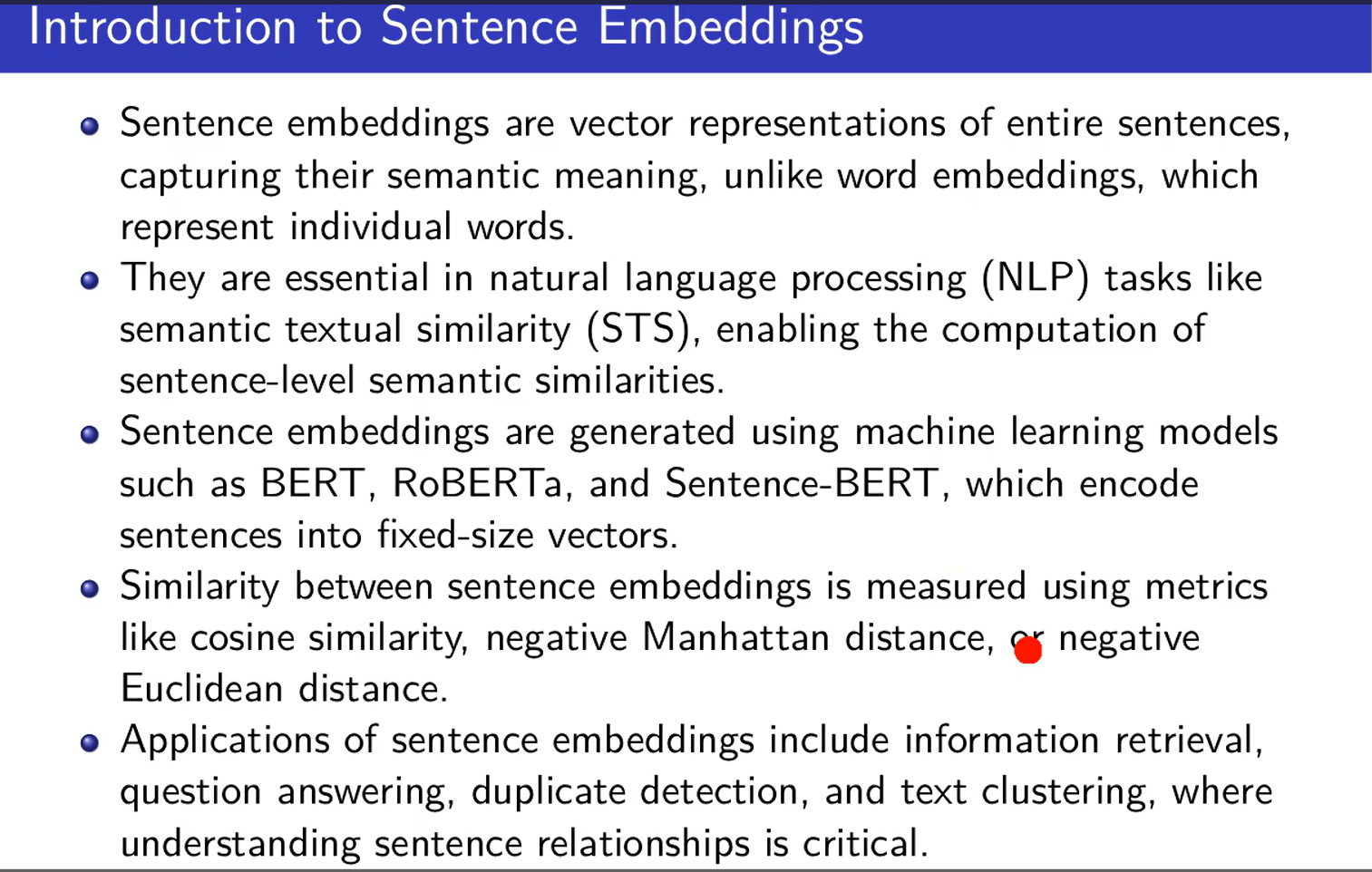} &
\includegraphics[width=0.18\linewidth]{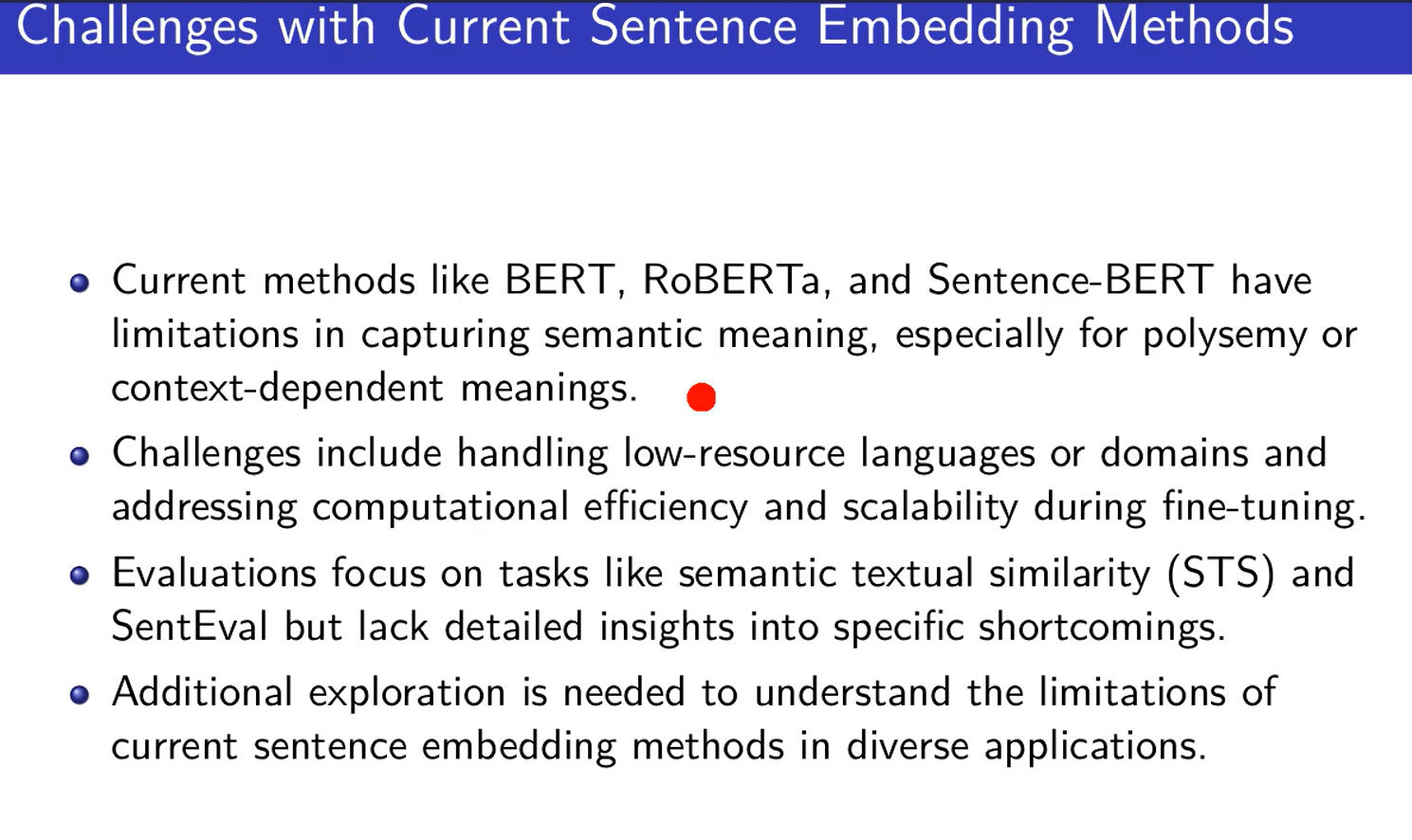} &
\includegraphics[width=0.18\linewidth]{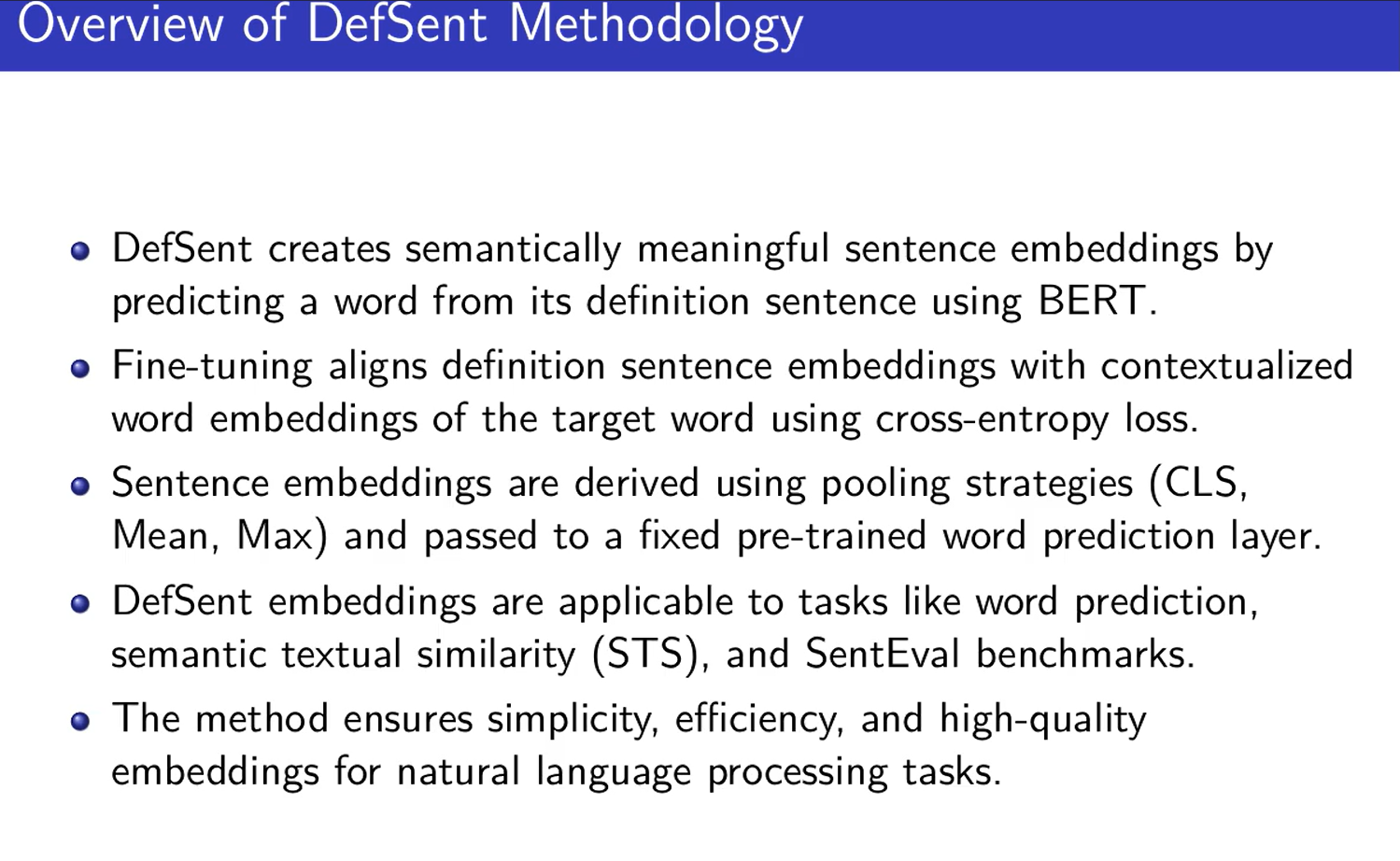} 
&
\textbf{M} 
&
\textbf{M} 
&
\textbf{H} 
&
\textbf{H} \\

\multicolumn{3}{l|}{\emph{Dense Slides with Cursor}} 
& & & & \\
\midrule

\includegraphics[width=0.18\linewidth]{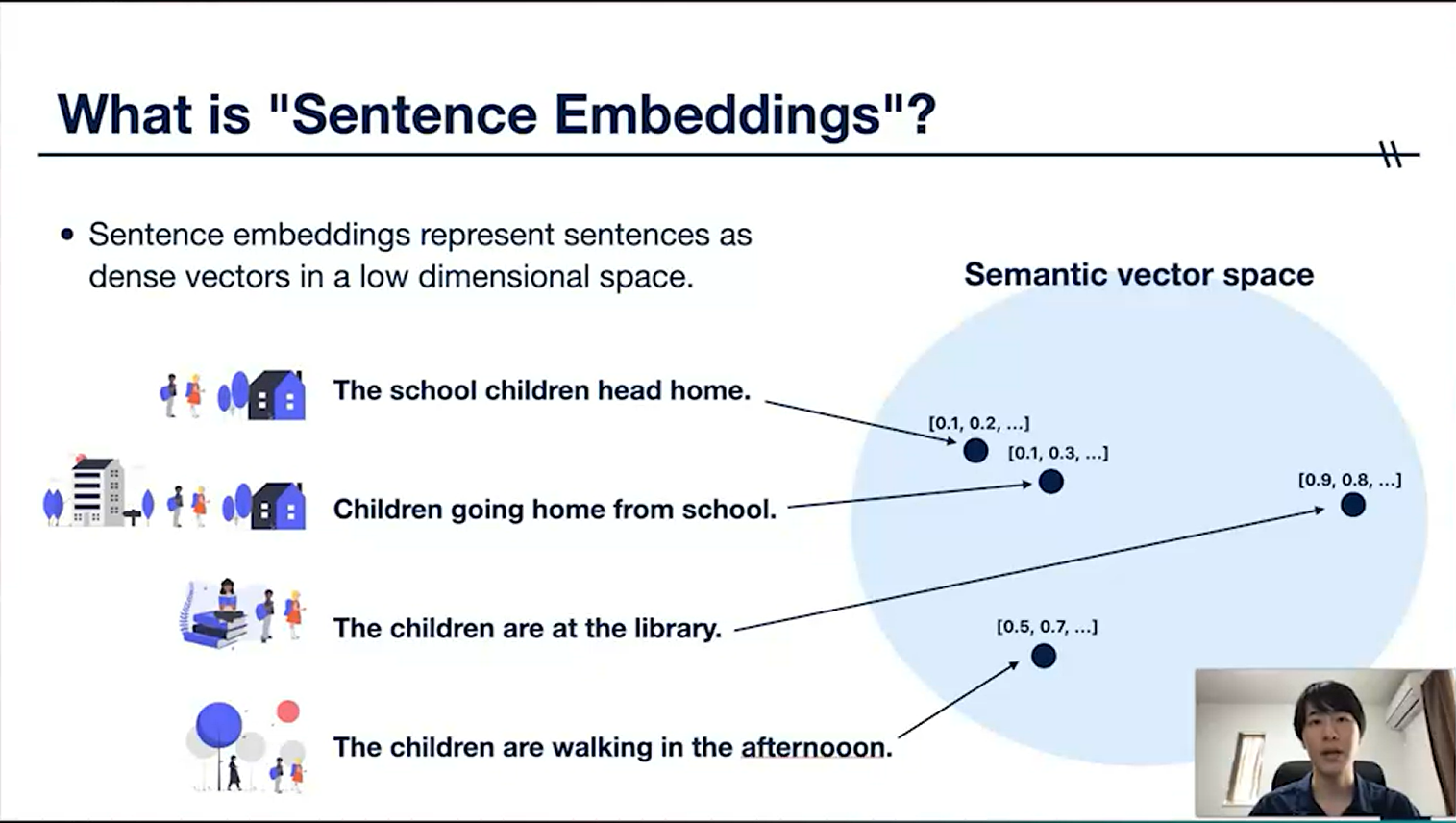} &
\includegraphics[width=0.18\linewidth]{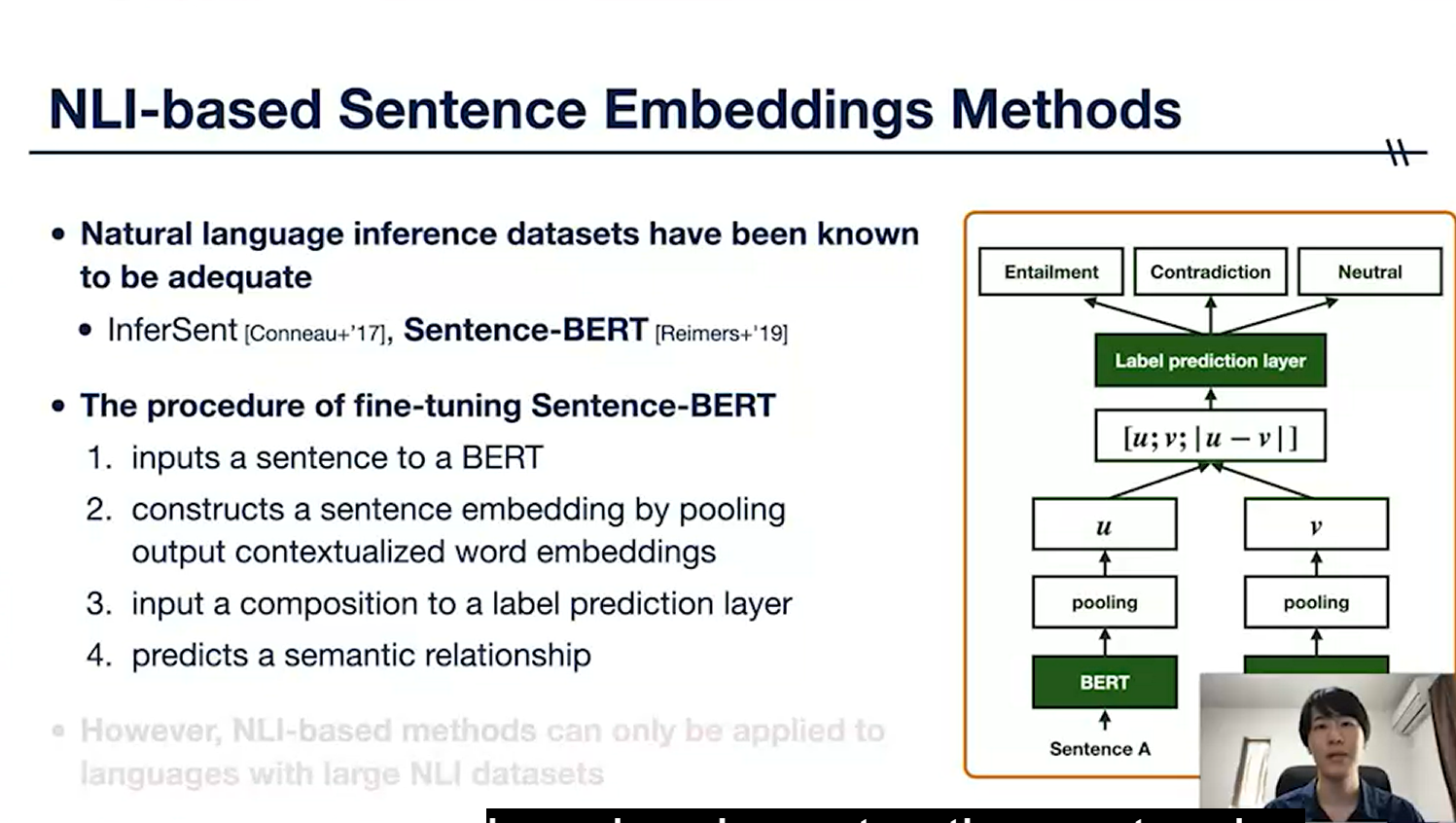} &
\includegraphics[width=0.18\linewidth]{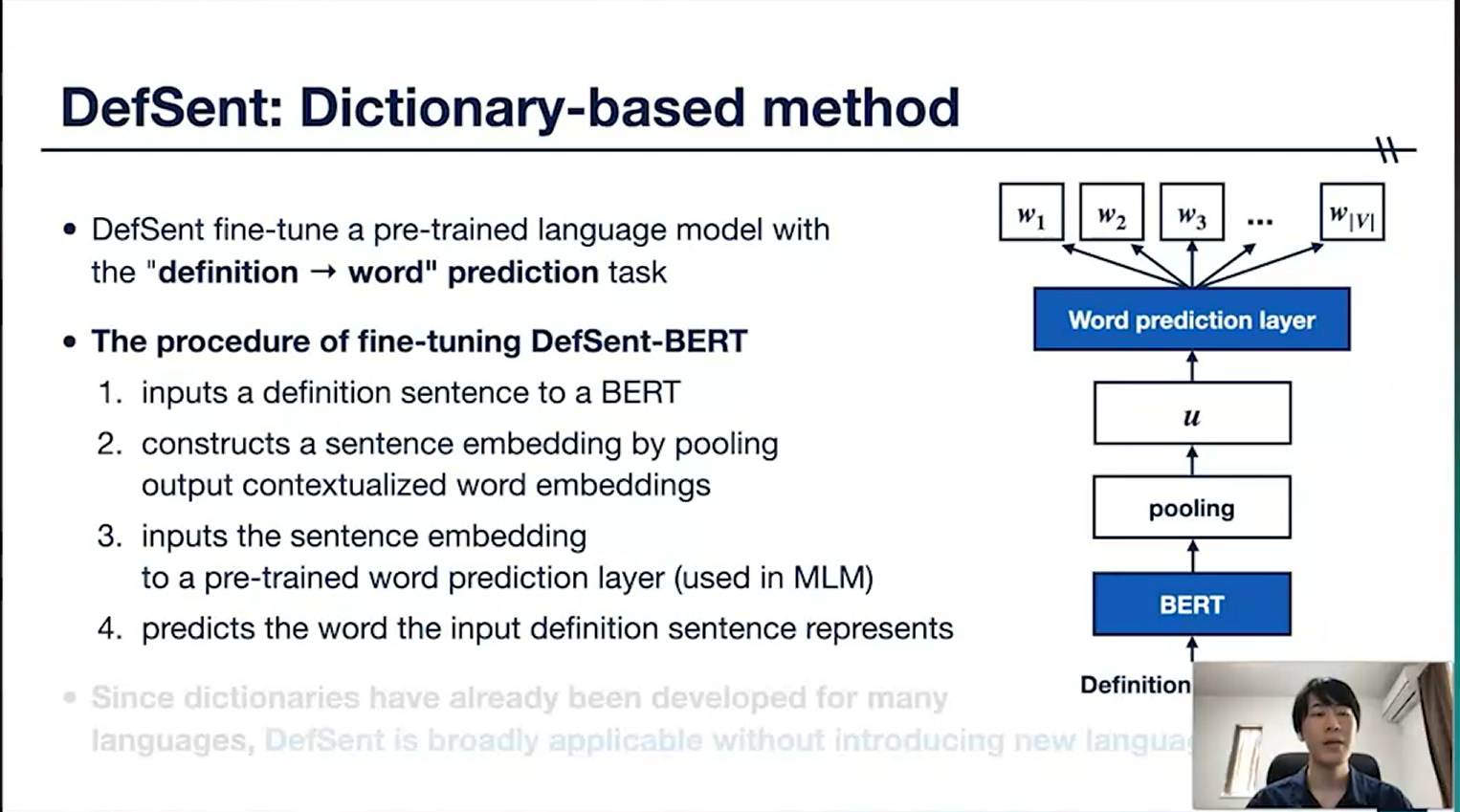} 
&
\textbf{H} 
&
\textbf{H} 
&
\textbf{H} 
&
\textbf{H} \\

\multicolumn{3}{l|}{\emph{Human-created: explicit background, step-by-step explanation, coherent narrative}} 
& & & & \\

\bottomrule
\end{tabular}

\caption{
Qualitative comparison using contiguous slide frames.
The first row illustrates a generated video with low explanatory coverage, where concepts are introduced without grounding.
The second row shows a coherence failure caused by improper ordering of motivation and mechanism.
The third row presents a human-created slide deck with explicit background, coherent structure, and guided explanation.
Human judgments and \EffectivePresentationScorer{} (EPS) consistently penalize coverage and coherence failures, while PresentQuiz (PQ) and VideoScore (ViS) are less sensitive to these instructional breakdowns and often overestimate utility based on surface-level content presence or visual quality.
}
\label{tab:qualitative_case_study_three_regimes}
\end{table*}

\section{Robustness Related to Methodological HyperParameter Sensitivity}
\label{app:hypsensitivity}

\paragraph{Calibration of Aggregation Weights ($\lambda$) and Claim Importance $I(c)$.}
The aggregation weights $\lambda$ used in the Meta-Evaluator (Eq.~12) were calibrated on a held-out validation subset containing 25\% of the paper–video pairs that were not used in the final evaluation. 
Weights were selected to maximize rank correlation with human-measured utility while preserving interpretability of the components. 
The final weights were set to:
$\lambda_{\text{cov}}=0.30$, $\lambda_{\text{faith}}=0.25$, $\lambda_{\text{coh}}=0.20$, $\lambda_{\text{del}}=0.15$, and $\lambda_{\text{eng}}=0.10$,
reflecting the intuition that missing or incorrect explanations should incur stronger penalties than presentation quality.

To assess robustness, we varied each $\lambda$ by $\pm20\%$ while keeping the remaining weights normalized. Across these perturbations, Kendall’s $\tau$ with human rankings varied only from $0.55$–$0.59$ (baseline $0.58$), and pairwise ranking accuracy varied from $0.77$–$0.81$ (baseline $0.80$), indicating that the evaluation is not highly sensitive to precise weight values.
Claim importance scores $I(c)$ are assigned by the Importance Agent on a continuous scale $[0,1]$ based on the claim’s explanatory role in the paper’s reasoning chain. 
Across the dataset, the mean importance for terminal causal claims was $0.82$, compared to $0.46$ for supporting context claims.

\paragraph{Inter-Annotator Agreement for Human Utility Scores.}
Each $(\text{paper}, \text{video})$ pair in \textbf{EffectivePresentation-EvalBench} is evaluated by three independent annotators. 
Agreement was measured using both Cohen’s $\kappa$ (pairwise) and Krippendorff’s $\alpha$ (multi-annotator). 
We observe substantial agreement with $\kappa=0.71$ and $\alpha=0.68$ across all videos. 
When restricting to reasoning-oriented questions (non-recall), agreement remains stable at $\kappa=0.66$, indicating that annotators consistently identify instructional quality differences even for higher-level conceptual explanations.

\paragraph{Robustness of Claim Detection and Temporal Alignment.}
To evaluate the reliability of claim presence and timestamp detection, we manually audited a randomly sampled subset of 120 claim–video pairs (about 10\% of the dataset). 
Human verification confirmed claim presence labels with 92.4\% agreement and first-introduction timestamp alignment within a tolerance of ±3 seconds in 88.7\% of cases. 
Most discrepancies occurred when explanations were implicitly implied rather than explicitly stated.

\paragraph{Normalization of Difficulty and Effort Signals.}
Difficulty and effort are reported by annotators on 5-point Likert scales. 
To mitigate annotator-specific bias, scores are normalized per annotator before aggregation. 
We evaluated three aggregation strategies: raw averaging, per-annotator z-scoring, and median normalization. 
Across these schemes, paper–video ranking correlation varied by less than 0.02 in Kendall’s $\tau$, indicating that the final utility estimates are robust to the aggregation choice.

\subsection{Cross-Backbone Agreement and Disagreement Analysis.}
\label{sec:llm-judge-robustness}
To assess whether EffectivePresentationScorer's per-claim decisions are robust to the underlying VLM, we evaluate the same paper–video pairs using three independent backbones (GPT-4o, Gemini-3, Qwen-VL) and compute pairwise agreement on the three primary categorical decisions: claim presence, faithfulness, and coherence. We observe 87.2\% agreement on presence, 84.5\% on faithfulness, and 81.3\% on coherence, indicating that the multi-agent pipeline yields stable categorical judgments across model families rather than backbone-specific artifacts.
To characterize residual disagreements, we manually audit a stratified sample of 120 claim–video pairs in which at least two backbones disagree. Disagreements are not uniformly distributed: approximately 62\% arise from implicit reasoning steps where the video conveys a claim through inference rather than direct statement, 23\% involve underspecified causal claims for which the paper itself does not pin down a single mechanism, and 15\% reflect ambiguous phrasing in the narration or slide text. These categories coincide with cases of higher human disagreement, suggesting that residual errors reflect intrinsic task difficulty rather than systematic LLM bias. Because each agent operates over an explicit, paper-grounded claim set with a constrained decision space, reliance on model priors is bounded, and disagreement is concentrated precisely where principled disagreement is expected.
\section{Related Work}
\label{sec:related_work_eval}

Early evaluation of text-to-video (T2V) models relied heavily on distributional and perceptual similarity metrics (e.g., IS/FID/FVD) and vision--language alignment (e.g., CLIP-based similarity), which primarily reflect visual quality and prompt relevance but provide limited diagnostic insight about \emph{why} a video is good or bad for a user’s goal. VBench \cite{huang2023vbench} formalized the need for more fine-grained evaluation by decomposing ``video generation quality'' into 16 hierarchical, disentangled dimensions with tailored prompt suites, evaluation pipelines, and per-dimension human preference validation, aiming to better align automated scores with human perception and to reveal model-specific failures.

VBench++ \cite{huang2024vbenchcomprehensiveversatilebenchmark} expands this line of work beyond T2V to also include image-to-video evaluation, and further introduces a \emph{trustworthiness} component (e.g., cultural fairness, bias, safety) aligned with human perception through annotations and experiments. It additionally releases resources intended to better support long-video evaluation (e.g., VBench-Long) while continuing to emphasize a disaggregated, multi-dimensional view of quality and condition consistency.

EvalCrafter \cite{liu2024evalcrafterbenchmarkingevaluatinglarge} proposes an ``exhaustive'' evaluation pipeline with a large prompt set (700 prompts) and a suite of objective metrics spanning visual quality, motion quality, temporal consistency, and text--video alignment. Beyond reporting many metrics, it fits coefficients to combine them into a final leaderboard score that better correlates with user opinions than naive averaging.

DEVIL \cite{liao2024evaluation} argues that common evaluation protocols largely ignore \emph{dynamics}, an essential property of video content, and shows that models can ``cheat'' existing metrics by generating low-motion videos that score well. It introduces a benchmark stratified by dynamics grades, defines dynamics scores across temporal granularities, and incorporates dynamics into metric refinement, reporting strong consistency with human ratings.

VideoGen-Eval \cite{yang2025videogen} emphasizes that evaluation has lagged behind rapid advances in video generation and attributes failures of prior benchmarks to unstructured prompts, fixed dimension definitions, and static/inflexible operators that break under content and distribution shift. It proposes a benchmark with structured prompts (explicitly encoding components such as camera/background/subject/style/lighting) and an agent-based evaluation pipeline combining an LLM prompt-structurer, an MLLM-based content judge, and tools for temporally dense checks, with the goal of improving alignment to human preferences.

In short, the state of the art has made major progress on \emph{how good a video looks} and \emph{how well it matches a textual prompt}, and has begun to incorporate additional axes such as dynamics and trustworthiness.
However, current evaluation paradigms still lack a robust, source-grounded notion of \emph{communicative success} for scientific and educational videos: whether the content is faithful to the underlying document, whether explanations are complete and prerequisite-aware, and whether the presentation order supports learning rather than merely producing visually fluent clips.

\paragraph{Industrial Prior Work.}
Google's NotebookLM  \footnote{\url{https://notebooklm.google/}} represents an important step toward document-grounded video generation, yet its design highlights a key limitation for our problem setting. Specifically, while NotebookLM can automatically generate videos from user-provided materials, it does not provide any meaningful mechanism for customizing the generation process with a user-specific goal in mind. The system emphasizes faithful rendering of input content, but it lacks ways for users to steer the narrative, adapt the style, or align the output with a particular communicative objective (e.g., education, persuasion, or engagement). As a result, NotebookLM remains largely a content-to-video pipeline rather than a framework that will generate videos on which goal-driven video utility can be measured. NotebookLM generated videos can measure the success through measuring fidelity to source material and not by which the generated video achieves its intended purpose for the end user.

\section{Comparison with LecEval}
\label{app:leceval}
We compared \EffectivePresentationScorer{} with LecEval \cite{yin2025lecevalautomatedmetricmultimodal} on a subset of 30 lecture-style videos. 
Across this subset, both methods show similar trends for clarity and coherence; however, EPS better differentiates videos with missing causal explanations. 
Specifically, EPS achieves Kendall’s $\tau=0.54$ with human rankings on the subset, compared to $\tau=0.47$ for LecEval. 
Qualitative inspection shows that LecEval tends to reward fluent presentations even when key reasoning steps are omitted, whereas EPS penalizes missing claims through its claim-coverage and dependency modeling.

\begin{table*}[t]
\centering
\small
\setlength{\tabcolsep}{6pt}
\begin{tabular}{p{3.2cm} p{3.2cm} p{8.6cm}}
\toprule
\rowcolor{gray!15}
\textbf{Question Source} & \textbf{Cognitive Level} & \textbf{Example Questions (DefSent)} \\
\midrule

\rowcolor{blue!8}
\multirow{3}{*}{\textbf{Citation-Grounded}} 
& \textbf{Big-Picture} 
& What is the main takeaway message of the paper, and why does it matter for sentence embedding research? \\

\rowcolor{blue!8}
& \textbf{Contribution} 
& What does the paper contribute compared to prior sentence embedding methods, and why is this contribution significant? \\

\rowcolor{blue!8}
& \textbf{Limitation} 
& What limitation of NLI-based sentence embedding methods does DefSent overcome, as highlighted by subsequent work? \\

\midrule

\rowcolor{green!8}
\multirow{3}{*}{\textbf{Bloom’s Taxonomy}} 
& \textbf{Recall} 
& What datasets were used to evaluate DefSent? \\

\rowcolor{green!8}
& \textbf{Understand} 
& What are the key components of the DefSent method, and how are dictionary definition sentences used during training? \\

\rowcolor{orange!10}
& \textbf{Analyze / Apply} 
& Why does training on dictionary definition sentences improve sentence embeddings, and how does this affect downstream semantic similarity tasks? \\

\rowcolor{orange!10}
\multirow{2}{*}{\textbf{Bloom’s Taxonomy}} 
& \textbf{Analyze} 
& How does DefSent compare to Sentence-BERT in terms of performance and supervision cost? \\

\rowcolor{orange!10}
& \textbf{Apply / Reason} 
& How could DefSent be extended to multilingual semantic similarity tasks, and what assumptions does this rely on? \\

\bottomrule
\end{tabular}
\caption{
Example evaluation questions generated for the DefSent paper \cite{tsukagoshi-etal-2021-defsent}.
\textcolor{blue!70!black}{Blue} rows indicate citation-grounded, community-level takeaway questions.
\textcolor{green!70!black}{Green} rows correspond to recall and comprehension questions.
\textcolor{orange!80!black}{Orange} rows denote higher-order reasoning and application questions.
This stratification enables fine-grained analysis of recall versus reasoning utility.
}
\label{tab:example_questions_defsent_colored}
\end{table*}

\section{EffectivePresentation-EvalBench}
\label{sec:human_eval_details}
This section describes a controlled human evaluation protocol to define utility and annotate videos from the papers with utility estimates by the humans to construct a utility-scored dataset of paper-video pairs, EffectivePresentation-EvalBench.
This is designed to isolate and measure the \emph{instructional utility} of scientific presentation videos.
The protocol proceeds in six stages, each targeting a distinct source of variability in human judgment.

First, in \S\ref{sec:background_questions}, we construct paper-specific \emph{background screening questions} that test prerequisite knowledge required to understand each paper, explicitly excluding the paper’s novel contributions.
These questions ensure that annotator performance reflects instructional quality rather than gaps in prior expertise.
We then validate the relevance and stability of these screening questions through independent expert review and inter-annotator agreement analysis in \S\ref{sec:background_validation}.

Next, \S\ref{sec:annotator_screening} describes annotator recruitment, eligibility screening, and answer verification.
Annotators are filtered based on background question performance, with correctness assessed via paper-independent gold answers and automated semantic grading, ensuring that only qualified annotators proceed to the main study.

In \S\ref{sec:utility_questions}, we define what it means for a video to be instructionally useful by constructing a fixed set of \emph{paper-grounded evaluation questions}.
These questions are derived from community-recognized contributions identified via citation analysis and are expanded across multiple cognitive levels using Bloom’s taxonomy, yielding a principled operationalization of instructional utility.

To enable controlled comparisons, \S\ref{appendix:controlled_perturbations} details how we generate multiple video variants per paper through systematic content- and delivery-level perturbations.
All variants explain the same source document and are evaluated against the same question set, allowing differences in performance to be causally attributed to specific instructional failures.

We then describe annotator assignment and experimental design in \S\ref{subsec:annotatorassignment}, where each video variant is evaluated by multiple independent annotators under a balanced, A/B-style protocol that controls for fatigue and learning effects.
Finally, \S\ref{sec:human_utility} formalizes how human responses are aggregated into a scalar \emph{utility score}.
This aggregation jointly accounts for answer correctness, perceived difficulty, and reported cognitive effort, producing a learner-centered measure of instructional quality that reflects not only whether viewers can answer questions, but how accessible the explanation is.

Together, these components form a rigorous human evaluation framework that disentangles prior knowledge, question design, presentation quality, and cognitive burden, enabling fine-grained analysis of instructional success and failure in generated scientific videos.

\subsection{Background Question Generation for Annotator Screening}
\label{sec:background_questions}

To ensure that annotator judgments reflect the instructional quality of generated videos rather than gaps in prior knowledge, we construct a set of background screening questions for each paper. These questions are designed to test prerequisite knowledge required to understand the paper and its associated explanatory videos, while explicitly excluding the paper’s novel contributions. We generate these questions through a multi-step, prompt-controlled process that progressively extracts, filters, calibrates, and validates prerequisite concepts.

\paragraph{Step 1: Prerequisite Concept Extraction.}
Given a paper $p$, we first identify the set of prerequisite background concepts that a reader must already understand in order to follow the paper’s explanation. Using the full paper text as input, we prompt a large language model with the following instruction:
\begin{quote}
\ttfamily
You are given a scientific paper.

List the prerequisite background concepts a reader must already understand
to follow this paper.

Guidelines:
Exclude the paper’s novel methods, results, or contributions.
Focus on foundational ideas assumed by the paper.
Target undergraduate or early graduate-level knowledge.
Return 8--15 concise concepts.
\end{quote}
This step yields an initial set of background concepts $\mathcal{B}(p) = \{b_1, b_2, \dots, b_K\}$, which may include general modeling paradigms, commonly used objectives, or standard evaluation notions referenced but not introduced by the paper.

\paragraph{Step 2: Screening-Relevant Concept Filtering.}
Not all extracted background concepts are suitable for annotator screening. We therefore perform a second filtering step in which the language model selects only those concepts that are strictly necessary to understand an explanatory video of the paper and that can be assessed independently of the paper itself. Concepts that are too vague, overly specialized, or difficult to test with a short objective question are removed. The model is prompted as follows:
\begin{quote}
\ttfamily
Given the following background concepts extracted from a paper:

[CONCEPT LIST]

Select the concepts that:
Are necessary to understand an explanatory video of the paper,
Can be assessed without reading the paper,
Can be tested with a short, objective question.

For each selected concept, provide a one-line justification.
\end{quote}
This produces a refined prerequisite set $\mathcal{B}^\ast(p) \subset \mathcal{B}(p)$, which forms the basis for question generation.

\paragraph{Step 3: Background Question Generation.}
For each prerequisite concept $b \in \mathcal{B}^\ast(p)$, we generate a single background screening question intended to test prior knowledge. The question-generation prompt enforces that the question is answerable without reading the paper, has a clearly correct answer, and targets conceptual understanding rather than factual trivia or advanced derivations. The exact prompt used is:
\begin{quote}
\ttfamily
Generate a background screening question for the following concept:

Concept: [b]

Requirements:
The question must be answerable without reading the paper.
It must have a single, clearly correct answer.
It should test conceptual understanding, not trivia or advanced derivations.

Return the result in the following format:
Question:
Correct Answer:
Brief Explanation:
\end{quote}
Collectively, this results in a set of background questions $\mathcal{Q}_{bg}(p) = \{(q_i, a_i)\}_{i=1}^{N}$, where $N$ is fixed to ten questions per paper for screening consistency.

\paragraph{Step 4: Difficulty and Scope Calibration.}
To ensure that screening rigor is consistent across papers, we perform an additional calibration step that adjusts question difficulty and scope. Each generated question is reviewed by the language model under a prompt that enforces mid-level conceptual difficulty, ensuring that the question is neither trivially answerable nor excessively technical. The calibration prompt is:
\begin{quote}
\ttfamily
Review the following background question.

Determine whether it:
Tests basic conceptual understanding,
Is not answerable by surface-level familiarity,
Does not require advanced mathematics or specialized edge cases.

If the question is too easy or too hard, rewrite it.
Otherwise, respond with: No change.
\end{quote}
Questions identified as too trivial or too specialized are rewritten, while those that already satisfy the criteria are retained unchanged.

\paragraph{Step 5: Automatic Question Validation.}
Finally, we apply an automated validation step to each background question to verify its suitability for screening. The language model is prompted to assess whether the question can be answered without reading the paper, whether the correct answer is unambiguous, and whether the question cleanly maps to a single prerequisite concept. The validation prompt is:
\begin{quote}
\ttfamily
Given the following question and answer:

[QUESTION]
[ANSWER]

Answer the following:
Can this question be answered without reading the paper?
Is the correct answer unambiguous?
Which prerequisite concept does this question test?

If any criterion is not satisfied, suggest a revised question.
\end{quote}
If a question fails any of these checks, it is revised and re-validated. Only questions that pass all validation criteria are included in the final screening set used for annotator eligibility assessment.

\subsection{Validation of Background Questions}
\label{sec:background_validation}

To validate the quality of the generated background screening questions, two co-authors independently review all background questions for two representative papers. Each question is rated on a 5-point relevance scale, where 1 denotes not relevant and 5 denotes highly relevant. Relevance is defined as whether the question appropriately targets prerequisite knowledge that a viewer must possess in order to understand the paper and its explanatory videos, without requiring access to the paper itself.

Across all reviewed background questions, both annotators assign consistently high relevance scores. No question receives a rating below 3 from either annotator, indicating that all questions are judged to meaningfully assess prerequisite understanding rather than paper-specific content. Table~\ref{tab:bg_question_relevance} reports the mean and standard deviation of relevance scores, demonstrating low variance across questions.

To assess consistency between annotators, we compute inter-annotator agreement using both Pearson correlation and Cohen’s $\kappa$. As shown in Table~\ref{tab:bg_question_agreement}, agreement scores indicate substantial consistency in relevance judgments, suggesting that the notion of prerequisite relevance is well-defined and reliably assessed. Together, these results confirm that the background screening questions are both appropriate and stable for filtering annotators based on prior knowledge.

\begin{table}[t]
\centering
\small
\begin{tabular}{lcc}
\toprule
\textbf{Question Type} & \textbf{Mean-Relevance} & \textbf{Std. Dev.} \\
\midrule
Background & 4.12 & 0.29 \\
\bottomrule
\end{tabular}
\caption{Mean and standard deviation of relevance scores (1--5) for background screening questions, as rated by two co-authors.}
\label{tab:bg_question_relevance}
\end{table}

\begin{table}[t]
\centering
\small
\begin{tabular}{lcc}
\toprule
\textbf{Question Type} & \textbf{Pearson $r$} & \textbf{Cohen’s $\kappa$} \\
\midrule
Background Questions & 0.64 & 0.68 \\
\bottomrule
\end{tabular}
\caption{Inter-annotator agreement on relevance ratings for background screening questions.}
\label{tab:bg_question_agreement}
\end{table}

\subsection{Annotator Recruitment, Screening, and Answer Verification}
\label{sec:annotator_screening}

We recruit annotators through Upwork and restrict participation to individuals with foundational familiarity in the relevant research area. To ensure that subsequent utility judgments reflect instructional quality rather than deficiencies in prior knowledge, all annotators undergo a background knowledge screening before participating in the main study.

For each background screening question, we first construct a canonical gold answer using LLM-based knowledge extraction. Specifically, given a background question, we prompt a large language model operating in a closed-book setting to produce a concise, self-contained answer grounded in general domain knowledge rather than the paper itself. To ensure answer reliability, two co-authors independently review all generated gold answers for three representative papers. This review assesses both factual correctness and alignment with the intended prerequisite concept. Any discrepancies between the LLM-generated answers and the co-authors’ judgments are resolved through discussion, resulting in a verified set of gold answers used for screening.

\paragraph{Annotator Demographics.}  In our study, each of the 140 videos is evaluated by 3 independent annotators (420 total evaluation instances) . Annotators are recruited from graduate and senior undergraduate pools with prior exposure to machine learning and NLP, ensuring baseline domain familiarity. Specifically, our annotator pool consists of approximately 38\% PhD students, 44\% MS students, and 18\% senior undergraduates, all of whom have completed at least one course or project in ML/NLP. To ensure task-relevant expertise, we employ a paper-specific screening protocol, where annotators must correctly answer 7/10 prerequisite questions before evaluating videos . This ensures that all annotators possess sufficient background knowledge of the paper’s domain (e.g., core ML/NLP concepts) and are capable of judging instructional clarity. In addition, we collect self-reported familiarity levels (on a 5-point scale) with the topic of each paper; across the dataset, annotators report an average familiarity of 3.7/5, indicating moderate-to-strong prior exposure. We emphasize that our evaluation protocol is designed to be robust to annotator variability, as eligibility is determined via objective screening rather than fixed demographics. This allows the benchmark to be replicated across different populations (e.g., novices vs. experts) by adjusting the screening criteria, supporting broader generalization.
During screening, annotators are explicitly instructed to rely solely on their own knowledge when answering background questions. The screening interface prominently displays the following instruction:
\begin{quote}
\ttfamily
Please answer the following questions using only your own knowledge.
Do NOT use any large language model, search engine, external website,
book, notes, or other reference material.
These questions are intended to assess your existing background knowledge.
\end{quote}
Annotators must acknowledge this instruction before proceeding.

\paragraph{Background Knowledge Screening.}
Each annotator completes a screening consisting of ten background questions per paper, targeting prerequisite concepts required to understand the paper and its explanatory videos. Screening is conducted across a total of twenty papers. Annotators who correctly answer at least seven out of ten questions for a paper are deemed eligible to evaluate videos for that paper. This threshold is chosen to balance screening strictness with annotator retention while ensuring sufficient prerequisite knowledge.

To assess the correctness of annotator responses at scale, we compare each human-provided answer against the verified gold answer using GPT-4o as an automated judge. The model is prompted to evaluate semantic correctness rather than surface-level string match. The exact prompt used for answer evaluation is:
\begin{quote}
\ttfamily
You are given a background screening question, a gold answer,
and a human-provided answer.

Question:
[QUESTION]

Gold Answer:
[GOLD ANSWER]

Human Answer:
[HUMAN ANSWER]

Determine whether the human answer is correct.
The answer should be marked correct if it is semantically equivalent
to the gold answer, even if phrased differently.

Return only one of the following labels:
Correct
Incorrect
\end{quote}
Per-question correctness labels are aggregated to compute each annotator’s screening score.

Out of 35 initial participants, 24 annotators meet the eligibility criterion and are retained for the study. This screening procedure ensures that downstream evaluations are conducted by annotators with adequate prerequisite knowledge, thereby isolating instructional quality as the primary factor influencing utility judgments.

\subsection{Utility-Defining Evaluation Questions}
\label{sec:utility_questions}

To evaluate instructional utility, we construct a fixed set of paper-grounded evaluation questions that define what it means for a video to be \emph{useful} for understanding a scientific paper. The resulting question set $\mathcal{Q}$ operationalizes instructional utility by jointly reflecting (i) the paper’s core contributions as perceived by the research community and (ii) graded cognitive demands that distinguish surface-level recall from deeper conceptual understanding and reasoning.

\paragraph{Community-Grounded Contribution Discovery.}
Rather than generating evaluation questions solely from the target paper text, we ground question construction in how the broader research community contextualizes the work. Let $\mathcal{P}$ denote a target paper and $\mathrm{Cite}(\mathcal{P})$ the set of papers that cite it. For each $\mathcal{P}$, we select the $m=10$ most recent citing papers and extract citation snippets in which $\mathcal{P}$ is explicitly referenced. Each citation snippet is normalized into a structured acknowledgement record capturing the attributed claim, the aspect of the work being cited (e.g., method, objective, evaluation), and the citation stance.

Citation normalization is performed using the following prompt:
\begin{quote}
\ttfamily
You are given a citation snippet that references a target paper.

Extract a structured acknowledgement with:
(1) the main claim attributed to the paper,
(2) the aspect being cited (e.g., method, objective, evaluation, finding),
(3) the citation stance (e.g., use, extension, comparison).

Return a concise, self-contained record.
\end{quote}

Semantically equivalent or paraphrastic acknowledgement records are canonicalized and clustered into contribution groups. To avoid idiosyncratic or one-off interpretations, we retain only consensus contributions that appear across multiple citing papers. These clusters define the paper’s community-recognized contributions and form the semantic backbone of the utility evaluation.

Table~\ref{tab:citation_ack_example} illustrates this process for an example paper, showing how multiple citation snippets are normalized and aggregated into a single contribution group.

\begin{table}[t]
\centering
\small
\setlength{\tabcolsep}{6pt}
\begin{tabular}{p{0.43\linewidth} p{0.22\linewidth} p{0.10\linewidth}}
\toprule
\textbf{Citation Snippet} & \textbf{Attributed Claim} & \textbf{Aspect} \\
\midrule
``We adopt DefSent to improve semantic clustering of sentences.'' 
& Sentence-level objectives improve clustering 
& Method \\
``DefSent shows better generalization due to sentence-based supervision.'' 
& Sentence-level supervision improves generalization 
& Finding \\
``Unlike word-level embeddings, DefSent captures sentence semantics.'' 
& Sentence-level representations capture semantics 
& Method \\
\bottomrule
\end{tabular}
\caption{Example citation snippets normalized into structured acknowledgement records and clustered into a consensus contribution.}
\label{tab:citation_ack_example}
\end{table}

\paragraph{Question Generation from Community Contributions.}
From each consensus contribution cluster, we generate evaluation questions that probe the paper’s main idea, novelty, and significance relative to prior work. Question generation is driven by the following prompt:
\begin{quote}
\ttfamily
You are given a community-recognized contribution of a paper.

Generate evaluation questions that assess whether a video explains:
(1) what the contribution is,
(2) why it is novel or important,
(3) how it differs from prior work.

Do not assume the viewer has read the paper.
\end{quote}
This ensures that evaluation targets what the community itself identifies as the paper’s core message, rather than peripheral details.

\paragraph{Cognitive Coverage via Bloom’s Taxonomy.}
To capture varying depths of understanding, we further expand the question set $\mathcal{Q}$ using Bloom’s taxonomy. For each citation-grounded contribution, we generate questions at multiple cognitive levels. Recall questions assess whether a video conveys essential definitions or factual statements, such as the name of a method or the objective it optimizes. Higher-level questions assess understanding and reasoning, including explanations of how components interact, why a design choice matters, or how the method improves upon prior approaches. In some cases, questions require analysis or application, such as reasoning about mechanisms or consequences under changed conditions. The prompt used to control cognitive level is:
\begin{quote}
\ttfamily
Given a paper contribution, generate questions at different Bloom levels:
Recall: factual definition or statement.
Understand: explanation of components or intuition.
Analyze / Apply: reasoning about mechanisms or effects.

Ensure each question targets a distinct level of understanding.
\end{quote}

Concrete examples of recall- and reasoning-level questions derived from the same contribution are shown in Table~\ref{tab:example_questions_defsent_colored}.

\paragraph{Paper-Grounded Gold Answer Construction.}
For each generated evaluation question, we extract a paper-grounded gold answer using a two-stage pipeline. First, a retriever identifies the most relevant sections of the paper corresponding to the question. Second, an answer extractor based on GPT-4o produces a concise answer constrained to be textually grounded in the retrieved content. All automatically generated questions and answers are manually reviewed by the first author to correct factual errors, clarify ambiguous phrasing, and remove cases that admit multiple interpretations.

\subsection{Utility-Defining Evaluation Questions}
\label{sec:question-validation}
To assess the pedagogical validity of automatically generated evaluation questions, we manually audit an expanded subset of 5 papers comprising approximately 60 questions. Across this subset, 88–91\% of questions are judged to be well-formed, unambiguous, and pedagogically aligned with the cited contribution, confirming that the generated questions largely reflect meaningful instructional targets. The distribution of question types follows the intended stratification across Bloom's taxonomy: Recall (about 30\%), Comprehension (about 25\%), Application (about 20\%), and Analysis/Reasoning (about 25\%), grounded in citation-derived contributions and paired with verified gold answers.
We further analyze the 9–12\% of questions flagged as flawed during manual review. These failures fall into three consistent categories: (i) overly compositional reasoning questions that require implicit intermediate steps not explicitly scaffolded by the paper, (ii) underspecified questions where key constraints (e.g., the experimental setting being asked about) are missing, and (iii) minor phrasing ambiguities that admit multiple reasonable interpretations. These failure modes are consistent with well-documented challenges in educational assessment design and are not arbitrary LLM errors. Importantly, the EffectivePresentationScorer framework is agnostic to the source of questions: expert-authored, curriculum-aligned, or learner-generated question sets can be substituted directly, and our use of LLM-generated questions is a scalable instantiation of the pipeline rather than a constraint of the method.

\subsection{Controlled Perturbations in Video Generation}
\label{appendix:controlled_perturbations}

To analyze instructional utility under controlled conditions, we generate multiple presentation-style video variants for each paper using a structured, prompt-driven pipeline. All variants are derived from the same source document and share identical generation parameters, except for explicitly injected perturbations. This design allows us to attribute observed differences in learning outcomes to specific content and delivery modifications.

For each paper $\mathcal{D}$, we first construct a canonical presentation plan that serves as the unperturbed reference. 
The plan is derived from the paper’s structure and consists of:
(i) a sequence of slides generated using the few-shot pipeline expert-persona aware slide generation of \cite{mondal-etal-2024-presentations}
(ii) aligned narration text, and
(iv) approximate timing constraints.
This reference plan reflects a faithful, pedagogically reasonable explanation of the paper and is used as the starting point for all controlled variants.

\paragraph{Perturbation Planning and Execution}
Controlled perturbations are introduced through a two-stage process: a planning phase that determines \emph{what} to perturb and \emph{how}, followed by an execution phase that applies the perturbation to the presentation plan.

\paragraph{Perturbation Planning.}
In the planning phase, the model identifies critical instructional elements whose modification is likely to affect understanding, such as prerequisite explanations, concept ordering, or time allocation.

\begin{quote}
\ttfamily
Given a presentation plan, identify important instructional elements
(e.g., prerequisites, ordering, emphasis, timing) that could be perturbed.
Select one or more elements and specify how they should be modified
to reduce instructional quality.
\end{quote}

The output of this phase is a perturbation plan specifying the target elements and the intended modification.

\paragraph{Perturbation Execution.}
In the execution phase, the perturbation plan is applied to the canonical presentation while keeping all other aspects fixed.

\begin{quote}
\ttfamily
Apply the following perturbation plan to the presentation.
Modify only the specified elements and keep all other content unchanged.
Return the updated slide content, narration, and timing.
\end{quote}

\begin{table*}[t]
\centering
\small
\setlength{\tabcolsep}{6pt}
\rowcolors{2}{gray!6}{white}
\begin{tabular}{p{0.15\linewidth} p{0.14\linewidth} p{0.30\linewidth} p{0.25\linewidth}}
\toprule
\textbf{Perturbation Level} & \textbf{Perturbation Type} & \textbf{Description} & \textbf{Intended Instructional Failure} \\
\midrule
Content 
& Prerequisite Omission 
& Removes prerequisite concepts such as task definitions, notation, or prior methods based on dependency graphs derived from the paper structure. 
& Main contribution becomes hard to follow due to missing background assumptions. \\

Content 
& Concept Reordering 
& Permutes the order of key concepts or steps (e.g., results before methods, usage before definitions) while preserving slide granularity. 
& Violates pedagogical dependencies and disrupts logical flow. \\

\midrule
Delivery 
& Temporal Misallocation 
& Redistributes narration time unevenly across sections, overemphasizing background or rushing through complex mechanisms. 
& Insufficient time for understanding key concepts. \\

Delivery 
& Audio--Visual Misalignment 
& Desynchronizes narration from slide visuals (e.g., references figures before or after they appear). 
& Breaks cross-modal grounding between speech and visuals. \\

Delivery 
& Visual Density Manipulation 
& Adds or removes figures, equations, or text blocks to create overly dense or overly sparse slides. 
& Increases cognitive load or underutilizes visual support. \\

\midrule
Joint 
& Content--Delivery Interaction 
& Combines multiple perturbations (e.g., prerequisite omission with accelerated pacing or concept reordering with time misallocation). 
& Produces compounded instructional failures resembling realistic generation errors. \\

\bottomrule
\end{tabular}
\caption{Summary of controlled perturbations applied to generate instructional video variants. Each perturbation targets a specific content or delivery dimension while holding the source paper and evaluation questions fixed.}
\label{tab:controlled_perturbations}
\end{table*}

The various kinds of perturbations are illustrated in Table~\ref{tab:controlled_perturbations} and are described as follows:
\paragraph{1) Content-Level Perturbations}
Content perturbations modify \emph{what information is included or emphasized}, without changing the underlying paper. We introduce these perturbations by editing the presentation plan prior to slide and narration generation.

\paragraph{1.1. Prerequisite Omission.}
To simulate incomplete background explanations, we selectively remove prerequisite concepts (e.g., task definitions, notation, or prior methods) that are required to understand the main contribution. The removed content is chosen based on dependency graphs induced from the paper structure, ensuring that omissions target genuine conceptual prerequisites rather than arbitrary details.

\paragraph{1.2. Concept Reordering.}
To test sensitivity to pedagogical ordering, we permute the order in which key concepts or steps are introduced. For example, experimental results may be presented before methodological details, or definitions may be delayed until after usage. Reorderings respect slide granularity but violate logical dependencies identified in the reference plan.

\paragraph{2) Delivery-Level Perturbations}
Delivery perturbations alter \emph{how information is presented} while keeping slide content largely fixed. These perturbations are applied during narration synthesis and temporal alignment.

\paragraph{2.1. Temporal Misallocation.}
We vary the amount of narration time assigned to different sections. Some variants devote disproportionate time to background or setup, leaving insufficient time for key results, while others rush through explanations of complex mechanisms.

\paragraph{2.2 Audio--Visual Misalignment.}
In these variants, narration is intentionally desynchronized from slide visuals. For example, references to figures or equations occur before or after the corresponding visual appears, disrupting cross-modal grounding.

\paragraph{2.3. Visual Density Manipulation.}
We alter the number of visual elements per slide by adding or removing figures, equations, or text blocks. High-density variants overload slides with information, while low-density variants underutilize visual space, relying heavily on narration.

\paragraph{3. Joint Content--Delivery Perturbations}
In practice, instructional failures often arise from interactions between content and delivery. We therefore generate variants that combine multiple perturbations, such as omitting prerequisites while simultaneously accelerating pacing, or reordering concepts while misallocating time. These joint perturbations produce qualitatively different failure modes than isolated edits and better reflect realistic generation errors. 

For each paper, we generate six automatically produced variants:
two focusing primarily on content-level perturbations, three on delivery-level perturbations, and one combining both (As shown in Table~\ref{tab:controlled_perturbations}).
Along with the human-authored conference presentation, this yields a controlled set of seven videos per paper that explain the same source document but differ systematically in instructional quality.

Because all variants share the same paper and question set, differences in human performance and model-based utility estimates can be directly attributed to specific content and delivery perturbations, enabling targeted analysis of instructional success and failure.

\subsection{Annotator Assignment and Paper-Video Utility Estimation}
\label{subsec:annotatorassignment}
For the main study, each paper is associated with seven video variants, enabling an explicit \emph{A/B-style} comparison in which all variants explain the same source document but differ in controlled content or delivery properties. 
Because the paper and evaluation question set are held fixed, differences in annotator judgments can be attributed to the injected perturbations rather than topic, difficulty, or personal preference.
Each video variant is independently annotated by three distinct annotators, which reduces the impact of individual subjectivity and yields repeated measurements per variant that are robust to annotator-specific noise. 
To control for fatigue and learning effects, each annotator is assigned at most 20 videos spanning multiple papers and variants, preventing a single worker from repeatedly seeing highly similar content and becoming systematically more familiar with a paper over time. 
This design maintains a manageable annotation load while ensuring balanced coverage across variants, thereby supporting reliable within-paper comparisons akin to controlled A/B tests.

\subsection{Human Utility Aggregation from Answers, Difficulty, and Effort}
\label{sec:human_utility}

We aggregate human judgments into a learner-centered utility estimate that reflects not only whether a video enables correct answers, but also how cognitively demanding the viewing experience is. The central intuition is that a video is less instructionally useful if correct answers require excessive mental effort or are perceived as difficult to extract from the explanation.

\paragraph{Annotator Elicitation.}
For each video variant $v$, paper $\mathcal{P}$, and evaluation question $q \in \mathcal{Q}_{\mathcal{P}}$, each annotator provides three signals. First, they produce a free-text answer to the question based solely on the video. Second, they report perceived difficulty on a 5-point Likert scale, where 1 denotes very easy and 5 denotes very difficult. Difficulty reflects how hard it was to answer the question from the video, not the inherent difficulty of the paper. Third, annotators self-report perceived viewing effort on a 5-point Likert scale, capturing cognitive strain such as the need to concentrate intensely, rewatch segments, or mentally infer missing steps.

Annotators are instructed as follows:
\begin{quote}
\ttfamily
Answer the question using only what you learned from the video.
Do not use external resources.
Then rate how difficult it was to answer this question based on the video alone
(1 = very easy, 5 = very difficult).
Finally, rate how much mental effort the video required for this question
(1 = very low effort, 5 = very high effort).
\end{quote}

\paragraph{Answer Correctness.}
Free-text answers are graded against paper-grounded gold answers using GPT-4o as an automated judge. Each answer is assigned a discrete correctness score $s_a(v,q) \in \{0, 0.5, 1\}$ for annotator $a$, corresponding to incorrect, partially correct, and fully correct answers. This discrete scale captures meaningful distinctions between incomplete understanding and complete correctness while remaining robust to paraphrasing.

\paragraph{Difficulty and Effort Normalization.}
Self-reported difficulty $d_a(v,q)$ and effort $e_a(v,q)$ are normalized to the unit interval to make them comparable and interpretable as penalties:
\[
\widetilde{d}_a(v,q) = \frac{d_a(v,q) - 1}{4} \]
\[
\widetilde{e}_a(v,q) = \frac{e_a(v,q) - 1}{4}.
\]
This normalization maps the lowest perceived burden to zero penalty and the highest to maximal penalty.

\paragraph{Per-Annotator Utility.}
For each annotator $a$, we define question-level utility as
\begin{equation}
u_a(v,q) = s_a(v,q)\,\bigl(1 - ,\widetilde{d}_a(v,q)\bigr)\,\bigl(1 - ,\widetilde{e}_a(v,q)\bigr),
\end{equation}
Intuitively, correctness provides the base signal of learning, while difficulty and effort act as multiplicative penalties that downweight answers obtained under high cognitive strain. This formulation reflects the pedagogical intuition that explanations should be not only correct but also accessible.

\paragraph{Aggregation Across Annotators.}
Each video variant is annotated by three independent annotators. To reduce individual subjectivity while preserving sensitivity to variation, we average per-annotator utilities:
\begin{equation}
\bar{u}(v,q) = \frac{1}{3}\sum_{a=1}^{3} u_a(v,q).
\end{equation}
This aggregation treats annotators as repeated measurements of the same instructional experience, yielding a stable estimate of how a typical viewer would perform on the question.

\paragraph{Paper- and Variant-Level Utility.}
Question-level utilities are averaged to obtain a paper-level utility score for each video:
\begin{equation}
U(v,\mathcal{P}) = \frac{1}{|\mathcal{Q}_{\mathcal{P}}|} \sum_{q \in \mathcal{Q}_{\mathcal{P}}} \bar{u}(v,q).
\end{equation}
Finally, we compute a variant-level utility score by averaging across all evaluated papers:
\begin{equation}
U(v) = \frac{1}{|\mathcal{P}|} \sum_{\mathcal{P}} U(v,\mathcal{P}).
\end{equation}

\paragraph{Interpretation.}
This hierarchical aggregation yields a robust estimate of instructional utility that balances learning outcomes with perceived cognitive burden. Videos that enable correct answers with low reported difficulty and effort receive high utility, while videos that require viewers to struggle—despite eventual correctness—are appropriately penalized. Averaging across annotators, questions, and papers ensures that utility reflects systematic instructional quality rather than idiosyncratic viewer behavior.

\section{Description of EffectivePresentation-EvalBench}
\label{sec:Datasetstats}
\textbf{EffectivePresentation-EvalBench} comprises 20 research papers, each paired with a controlled set of seven presentation-style videos, yielding 140 videos in total (Table~\ref{tab:dataset_stats}). 
Videos are long-form and instructional in nature, averaging 28.6 slides and 7.8 minutes in duration, with substantial textual and visual content including figures and equations.
For each paper, six automatically generated variants are produced via controlled content- and delivery-level perturbations, alongside one human-authored reference video, enabling fine-grained analysis of instructional failures under matched conditions.

The benchmark spans a diverse range of core NLP and ML research areas (Table~\ref{tab:topic_distribution}), including representation learning, information extraction, evaluation and benchmarking, multimodal and vision--language reasoning, NLP applications, and learning and optimization.
This diversity ensures that evaluation results are not tied to a single subfield, while maintaining sufficient conceptual complexity to stress-test instructional quality across different types of scientific content.

\begin{table*}[t]
\centering
\small
\setlength{\tabcolsep}{6pt}
\begin{tabular}{c l|cc|cc}
\toprule
\textbf{VLM} 
& \textbf{Evaluation Method} 
& \multicolumn{2}{c|}{\textbf{Recall Questions}} 
& \multicolumn{2}{c}{\textbf{Non-Recall Questions}} \\
\cmidrule(lr){3-4} \cmidrule(lr){5-6}
& 
& \textbf{Kendall $\tau$ $\uparrow$} 
& \textbf{Pair-Acc $\uparrow$}
& \textbf{Kendall $\tau$ $\uparrow$} 
& \textbf{Pair-Acc $\uparrow$} \\
\midrule

\multirow{4}{*}{\rotatebox{90}{\textbf{GPT-4o}}}
& \textbf{EffectivePresentationScorer} 
& \textbf{0.58} & \textbf{0.80} 
& \textbf{0.53} & \textbf{0.76} \\

& Single-Agent QA (F + T) 
& 0.49 & 0.72 
& 0.28 & 0.56 \\

& Single-Agent QA (T-only) 
& 0.47 & 0.70 
& 0.25 & 0.54 \\

& Holistic LLM Utility Rating 
& 0.45 & 0.69 
& 0.26 & 0.55 \\

\midrule

\multirow{4}{*}{\rotatebox{90}{\textbf{Gemini-3}}}
& \textbf{EffectivePresentationScorer} 
& \textbf{0.55} & \textbf{0.67} 
& \textbf{0.56} & \textbf{0.61} \\

& Single-Agent QA (F + T) 
& 0.43 & 0.60 
& 0.32 & 0.46 \\

& Single-Agent QA (T-only) 
& 0.41 & 0.57 
& 0.27 & 0.55 \\

& Holistic Utility Rating 
& 0.40 & 0.55 
& 0.33 & 0.58 \\

\midrule

\multirow{4}{*}{\rotatebox{90}{\textbf{Qwen-3}}}
& \textbf{EffectivePresentationScorer} 
& \textbf{0.57} & \textbf{0.64} 
& \textbf{0.56} & \textbf{0.61} \\

& Single-Agent QA (F + T) 
& 0.43 & 0.57 
& 0.32 & 0.46 \\

& Single-Agent QA (T-only) 
& 0.40 & 0.54 
& 0.27 & 0.51 \\

& Holistic Utility Rating 
& 0.45 & 0.56 
& 0.29 & 0.45 \\

\midrule

& VideoScore~\cite{he-etal-2024-videoscore}
&  0.44 & 0.68 
& 0.20 & 0.52 \\

& EvalCrafter~\cite{liu2024evalcrafterbenchmarkingevaluatinglarge} 
& 0.46 & 0.69 
& 0.22 & 0.53 \\

& PPTEval~\cite{zheng-etal-2025-pptagent} 
& 0.46 & 0.70
& 0.41 & 0.59 \\

\bottomrule
\end{tabular}

\caption{
Paper-level Kendall’s $\tau$ and pairwise ranking accuracy with respect to
human-measured utility, separated into recall-only and non-recall questions.
Results are grouped by VLM backbone for compactness.
Rows highlighted in green correspond to \EffectivePresentationScorer{}.
}

\label{tab:recall_vs_reasoning_vertical_vlm}
\end{table*}

\section{Impact of VLM/LLM Backbones on Utility Evaluation
}
Table~\ref{tab:recall_vs_reasoning_vertical_vlm} answers the question:
does \EffectivePresentationScorer{} work because it uses a strong LLM, or because the evaluation framework itself is better? To isolate this, the same evaluation methods are instantiated with different VLM/LLM backbones—GPT-4o, Gemini-3, and Qwen-3—and their outputs are compared against human judgments of video usefulness.

For each backbone, the table compares four ways of evaluating a paper-to-video system. The first is \EffectivePresentationScorer{}. The other three baselines rely on a single model making a holistic judgment: answering questions using frames and transcript of the video variants, answering using transcript alone, or directly assigning a usefulness score without decomposition.

Performance is measured by how closely each method’s ranking of video variants matches human rankings. Two complementary metrics are used. 
Kendall’s tau captures how similar the overall ranking order is, while pairwise accuracy measures how often the evaluator chooses the better video when humans are shown a pair.
Importantly, results are split into recall questions, which mainly test whether facts are mentioned, and non-recall questions, which test whether ideas are explained coherently and with correct reasoning.

Across all three backbones, \EffectivePresentationScorer{} consistently shows the strongest alignment with human judgments. This holds even when the underlying model changes, which indicates that the improvements are not simply due to using a stronger LLM. Instead, they come from the structure of the evaluation itself. Stronger backbones such as GPT-4o do increase absolute scores slightly, but the relative advantage of \EffectivePresentationScorer{} remains stable across models. In contrast, single-agent and holistic baselines benefit much less from stronger backbones, suggesting that unstructured prompting fails to fully leverage model capacity.

The most important pattern appears in non-recall questions. These questions require multi-step reasoning, correct ordering of concepts, and faithful explanation of why claims hold, rather than surface-level coverage. Single-agent QA and holistic scoring perform poorly here, because they tend to reward the presence of keywords or fluent narration even when explanations are incomplete, out of order, or weakly grounded in the paper. \EffectivePresentationScorer{} performs substantially better because it explicitly checks whether prerequisite ideas are introduced, whether reasoning chains are intact, and whether explanations are consistent with the source paper.

The bottom rows compare against existing video evaluation benchmarks such as VideoScore, EvalCrafter, and PresentQuiz. These methods show reasonable performance on recall questions but struggle on non-recall questions, reinforcing the idea that current benchmarks primarily capture answerability or surface quality rather than explanatory usefulness. Overall, the table demonstrates that structured, paper-grounded, multi-agent evaluation is more important than the choice of backbone, especially when the goal is to measure educational utility rather than visual quality or factual recall.

\section{Implementation Details of EffectivePresentationScorer}

In this section, we will ilustrate how we preprocess the scientific papers, the associated videos (human-made or automatically generated), 
then also discuss the nitty-gritty details and prompts used in \EffectivePresentationScorer{} in Section~\ref{sec:methodology}.

\subsection{PDF Preprocessing and Structured Document Extraction.}
To enable paper-grounded evaluation and claim-level reasoning, we first preprocess scientific PDFs into a structured, machine-readable representation that preserves section boundaries, figures, and references. Given the variability and complexity of scientific PDF layouts,
We use the SciPDF Parser \footnote{\url{https://github.com/titipata/scipdf_parser}}, a Python wrapper built on top of GROBID, to convert each scientific PDF into a structured dictionary representation. 
GROBID \footnote{\url{https://github.com/kermitt2/grobid}} performs machine-learning–based segmentation of scholarly documents, enabling reliable extraction of titles, abstracts, section headings, section text, references, and metadata. 
Prior to parsing, we run a GROBID service locally, following the recommended Docker-based deployment, which exposes a REST API (default port 8070) for PDF processing. 
We use the latest stable GROBID version, as newer releases substantially improve section segmentation and citation parsing.

Given a PDF file or a direct URL to a PDF, SciPDF Parser returns a normalized dictionary containing the paper title, abstract, a list of sections with headings and text, references with bibliographic fields, detected figures with captions and identifiers, and document-level metadata such as the DOI. We configure the parser to return section text either as a single concatenated string or as a list of paragraphs, depending on downstream needs. This structured representation serves as the foundation for paper context retrieval, claim decomposition, and importance estimation, as all subsequent agents operate over section-level or paragraph-level text rather than raw PDFs. For analyses requiring full fidelity to the original markup, we additionally retain the XML output produced by GROBID, which allows inspection of fine-grained structural elements.

While GROBID provides high-quality textual segmentation, reliable extraction of figures and their associated captions requires specialized tooling. We therefore complement SciPDF parsing with pdffigures2 \footnote{\url{https://github.com/allenai/pdffigures2}}, a dedicated system for identifying figures, tables, captions, and their bounding boxes from scientific PDFs. 
pdffigures2 processes PDFs at the page level and outputs structured JSON metadata along with extracted figure images, including figure identifiers, caption text, and positional information.

In our pipeline, we apply pdffigures2 to folders containing only PDF files, producing a parallel figure index for each paper. This allows us to align figure images and captions with the corresponding sections extracted by SciPDF, enabling downstream agents to reason jointly over textual explanations and visual evidence. 
The extracted figures are later used when generating video frames, constructing visual descriptions, and evaluating visual reasoning during delivery and faithfulness assessment.

The outputs of SciPDF Parser and pdffigures2 are merged into a unified document representation that exposes section-level text, paragraph boundaries, reference metadata, and figure-caption pairs in a consistent schema. This preprocessing step ensures that all subsequent components of EffectivePresentationScorer operate on clean, structured, and paper-grounded inputs, rather than noisy PDF artifacts. By decoupling document parsing from evaluation logic, the pipeline remains modular and robust to improvements in underlying PDF parsing tools.

\subsection{Video Preprocessing and Multimodal Frame Representation}

To enable fine-grained, claim-level evaluation of instructional delivery, we preprocess each presentation video into a temporally aligned multimodal representation that explicitly exposes visual content, narration, and timestamps. This representation serves as the common substrate for presence detection, faithfulness verification, coherence analysis, and delivery-quality assessment.

Given a video $v$, we first decode it into a sequence of frames using a fixed frame sampling strategy. Frames are extracted at a uniform temporal interval of 1 second, chosen to balance visual coverage with computational efficiency, ensuring that all slide transitions and major visual changes are captured. 
For slide-based presentation videos, this frame-level sampling is sufficient to recover distinct slide content while avoiding redundant near-identical frames. 
Each extracted frame is associated with its corresponding timestamp interval in the original video.

In parallel, we obtain the video narration by extracting the audio track and applying automatic speech recognition (gtts) \footnote{\url{https://github.com/pndurette/gTTS}} to produce a timestamped transcript. 
The transcript is segmented into utterances aligned with time intervals, yielding narration spans that can be directly associated with extracted frames. 
This alignment ensures that each frame or short sequence of frames can be paired with the narration spoken during the same temporal window, enabling joint reasoning over what is shown and what is said.
To enable language-based evaluators to reason over visual evidence, we convert
each extracted video frame into a structured textual description using a
vision--language model. For every frame, the model is prompted using Figure~\ref{fig:prompt-instructional-visual-description} to produce an instructional-level description that captures semantically meaningful visual elements, such as diagrams, plots, tables, equations, highlighted text, and overall slide structure.  This textual representation allows downstream agents to reason uniformly over narration and visual content using natural language, enabling consistent claim detection, faithfulness verification, coherence analysis, and delivery-quality assessment.

\begin{figure}[t]
\centering
\fcolorbox{promptborder}{promptgray}{
\begin{minipage}{0.95\columnwidth}
\textcolor{promptblue}{\textbf{Task: Instructional Visual Description}}

You are given a single video frame extracted from a scientific presentation.
Your task is to describe the visual content in a way that supports instructional
understanding and reasoning.

Focus on:
- diagrams, plots, tables, or equations,
- highlighted or emphasized text,
- relationships between visual elements,
- how the visual content contributes to explaining a concept or idea.

Do not describe low-level visual details such as colors, fonts, or decorative
layout unless they convey meaning.
Do not infer information that is not visually present.

\textbf{Video Frame:} \{Frame image $s_i$\}

Return a concise textual description of the frame that captures its explanatory
content.
\end{minipage}}
\caption{Prompt used to generate instructional-level textual descriptions of individual video frames for multimodal video representation.}
\label{fig:prompt-instructional-visual-description}
\end{figure}


The result of this preprocessing step is a structured multimodal video representation
\[
\mathcal{V} = \{(s_i, t_i, a_i, d_i)\}_{i=1}^{M},
\]
where $s_i$ denotes the $i$-th frame image, $t_i$ its timestamp interval, $a_i$ the aligned narration text, and $d_i$ the corresponding textual description of the visual content. Consecutive frames with identical or near-identical visual descriptions are optionally merged into longer segments to reduce redundancy and stabilize downstream reasoning.

This temporally aligned multimodal representation enables precise localization of where claims are introduced, explained, or omitted within a video. Presence agents operate by scanning $(a_i, d_i)$ pairs to detect whether a claim is explicitly conveyed, faithfulness agents compare these descriptions against paper excerpts, coherence agents reason over the order in which claims first appear via their timestamps $t_i$, and delivery agents estimate explanation depth by aggregating narration length and descriptive richness across contiguous segments. By grounding all evaluation components in this unified frame-level representation, the pipeline ensures consistent, interpretable, and reproducible analysis of instructional video content.

\subsection{Agents involved in \EffectivePresentationScorer{}}
\subsubsection{Question Decomposition into Claims.}
\label{appendix:decomposition}
To ensure that question decomposition is grounded in the source paper rather than
prior model knowledge, we explicitly retrieve and condition on relevant paper
text when constructing the decomposition prompt. Given a question $q$ and paper
$p$, we first preprocess the paper into a structured form using the SciPDF
parser, which provides section-level text for the abstract, introduction,
methods, and results. We restrict retrieval to these sections, as they are most
likely to contain explanatory and causal content relevant to answering $q$.

We then split the selected sections into sentence-level units and embed each sentence using a pretrained sentence embedding model \cite{reimers-2020-multilingual-sentence-bert} (e.g., a Sentence-BERT–style encoder using SBERT \footnote{\url{https://github.com/huggingface/sentence-transformers}}). The question $q$ is embedded using the same model, and cosine similarity is used to retrieve the top-$k$ sentences most semantically similar to the question. To preserve local coherence and avoid fragmentary evidence, we expand each retrieved sentence with a fixed window of neighboring sentences from the same section. The resulting spans are concatenated and truncated to a fixed context budget, forming a compact paper context window.
This retrieved context is passed verbatim to the question decomposition prompt in Figure \ref{fig:prompt-question-decomposition}, and GPT4-o is explicitly instructed to derive claims only from the
provided paper text. By grounding decomposition in embedding-based retrieval over the paper itself, we prevent reliance on background knowledge and ensure that each generated claim corresponds to an explicit statement or mechanism described in the source document. This grounding is critical for enabling downstream agents to reliably assess claim presence, faithfulness, ordering, and delivery quality
with respect to both the paper and the video.

\subsubsection{Claim Importance Estimation.}
\label{appendix:claim-importance-estimation}
In addition to determining whether claims are present, faithful, and coherently ordered, our evaluation framework must distinguish which claims matter most for understanding a paper’s core contribution. We therefore assign each claim an importance score that reflects its conceptual necessity for answering a given question, grounded entirely in the source paper. Claim importance is designed to capture \emph{what must be explained well} for a learner to reason correctly, rather than what is merely mentioned frequently or appears visually salient.

For a paper $p$ and question $q$, we begin with the paper-grounded claim set $C(q)$ obtained during question decomposition, along with the dependency structure over claims. Each claim corresponds to a statement or reasoning step supported by the paper text. To estimate its importance, we retrieve the paper spans most relevant to the claim, including sentences from the abstract, introduction, method, and results sections that explicitly or implicitly support it. Using only this retrieved paper context, we assess the rhetorical role of the claim with respect to the question. In particular, we determine whether the claim functions primarily as background context, a methodological component, an intermediate explanatory step, or a core causal or outcome-defining statement that directly answers the question.

This assessment is performed using a constrained LLM prompt that classifies the claim’s role conditioned on the paper text, without access to the video. Claims that articulate the paper’s main causal mechanism, justification, or takeaway for the question are assigned higher importance, while claims that serve as prerequisites or supporting context receive lower importance. The resulting scores are normalized to the range $[0,1]$ within each question, ensuring that importance is interpreted relatively rather than absolutely. As a consequence, downstream explanatory claims that are critical for reasoning receive higher weight during aggregation than background claims, and videos that omit or weakly explain such claims are penalized more strongly even if surface-level coverage is high.

We operationalize claim importance estimation using the following prompt.

\begin{center}
\fcolorbox{promptborder}{promptgray}{
\begin{minipage}{0.95\columnwidth}
\textcolor{promptblue}{\textbf{Task: Claim Importance Assessment}}

You are given a scientific paper excerpt and a single claim derived from a
question about the paper. Your task is to assess how important this claim is
for explaining the paper’s core contribution with respect to the question.

Base your judgment \emph{only} on the paper text provided.
Do not consider how often the claim appears or how it is presented in the video.
Do not assess correctness or explanation quality.

A claim is more important if it:
directly expresses the paper’s main contribution,
describes a key causal or explanatory mechanism,
or connects the method to the primary outcome relevant to the question.

A claim is less important if it:
serves primarily as background,
defines terminology,
or provides supporting context that enables but does not complete the explanation.

\textbf{Question:} \{Question $q$\}

\textbf{Claim:} \{Claim $c$\}

\textbf{Paper Context:} \{Retrieved paper spans supporting $c$\}

Return a scalar importance score $I(c)\in[0,1]$ and a brief justification grounded
in the paper text.
\end{minipage}}
\end{center}

These importance scores are subsequently used to weight coverage, faithfulness,
coherence, and delivery diagnostics during aggregation, ensuring that evaluation
prioritizes the claims that are most critical for reasoning and understanding.

\begin{figure}[t]
\centering
\promptbox{%
\begin{minipage}{0.97\linewidth}
\raggedright\small
\textcolor{promptblue}{\textbf{Task: Paper-Grounded Question Decomposition}}\\[-2pt]
\hrulefill

\vspace{4pt}
You are given a scientific paper and a question about that paper.
Your task is to decompose the question into a sequence of minimal,
paper-grounded claims that must be explained to correctly answer it.

\vspace{4pt}
\textbf{Guidelines:}\\
\begin{tabular}{@{}p{0.97\linewidth}@{}}
Derive claims \emph{only} from the content of the paper.\\
Do not introduce claims based on prior knowledge or intuition.\\
Each claim should correspond to a specific statement, mechanism,
or finding described or implied by the paper.\\
If one claim depends on understanding another, explicitly mark this dependency.
\end{tabular}

\vspace{6pt}
\textbf{Paper Context (excerpt):}\\
\texttt{DefSent trains sentence embeddings using entailment and contradiction
signals. Unlike word-level objectives, sentence-level supervision encourages
representations that capture global semantic meaning, leading to improved
generalization across downstream tasks.}

\vspace{6pt}
\textbf{Question:}\\
Why does DefSent improve generalization?

\vspace{6pt}
\textbf{Illustrative Decomposition (example):}\\
\begin{tabular}{@{}p{0.97\linewidth}@{}}
Claim 1: DefSent is trained using sentence-level supervision rather than word-level objectives.\\
Claim 2: Sentence-level supervision captures global semantic meaning in embeddings.\\
Claim 3: Capturing global semantic meaning improves performance on downstream tasks.\\
Claim 4: Improved downstream performance reflects better generalization.\\
\\
Dependency structure:\\
Claim~2 depends on Claim~1.\\
Claim~3 depends on Claim~2.\\
Claim~4 depends on Claim~3.
\end{tabular}

\vspace{6pt}
\textbf{Output format:}\\
\texttt{Return the claims and their dependency relations in structured JSON.}\\
\texttt{Do not answer the question or summarize the paper.}
\end{minipage}%
}
\caption{Prompt used for paper-grounded question decomposition into dependency-structured claims.}
\label{fig:prompt-question-decomposition}
\end{figure}

\subsubsection{Claim Presence Detection in Videos.}
\label{appendix:presence}
After decomposing a question into a set of paper-grounded claims, we determine
whether each claim is explicitly conveyed in the video. Claim presence detection
operates purely at the level of the video and is intentionally separated from
faithfulness or correctness assessment. The goal of this step is to identify
\emph{where} and \emph{whether} a claim is expressed in the visual or narrated
content, enabling subsequent reasoning about ordering, explanation quality, and
faithfulness.

Given a claim $c \in C(q)$ and the multimodal video representation
$\mathcal{V}=\{(s_i,t_i,a_i,d_i)\}_{i=1}^{M}$, we scan the video timeline in
temporal order and evaluate each segment independently. Each segment consists of
a textual description of the frame or slide content and the aligned narration
spoken during the same timestamp interval. We consider a claim to be present if
it is directly stated in the narration or if it is clearly conveyed by the
combination of visual content and narration. Importantly, the model is
explicitly instructed not to infer unstated information, extrapolate from partial
evidence, or assess whether the claim is faithful to the paper. This strict
definition ensures that presence detection captures only explicit communicative
signals in the video.
We implement claim presence detection using a dedicated prompt that conditions on
a single claim and the full, time-ordered video timeline. The prompt asks the
model to return a binary presence decision, along with the earliest timestamp at
which the claim appears and a short justification grounded in the frame
description or narration. This timestamp localization is later used by the
coherence and delivery agents to reason about prerequisite ordering and temporal
allocation. The exact prompt used for claim presence detection is shown in
Figure~\ref{fig:prompt-claim-presence-detection}.

Claims that are not detected as present contribute zero coverage and delivery value in downstream aggregation. By isolating presence detection from faithfulness and reasoning, this step provides a clean and interpretable signal about what information the video actually communicates, forming the basis for subsequent coherence, delivery-quality, and reasoning evaluation.

\begin{figure}[t]
\centering
\promptbox{%
\begin{minipage}{0.97\linewidth}
\raggedright\small
\textcolor{promptblue}{\textbf{Task: Claim Presence Detection from Video}}\\[-2pt]
\hrulefill

\vspace{4pt}
You are given a single claim derived from a scientific paper and a
time-ordered video timeline. Each video segment contains:

\vspace{2pt}
\begin{tabular}{@{}p{0.97\linewidth}@{}}
\emph{(1)} A textual description of the visual frame(s);\\
\emph{(2)} The narration spoken during the same timestamp.
\end{tabular}

\vspace{6pt}
Your task is to determine whether the claim is \emph{explicitly present}
anywhere in the video.

\vspace{4pt}
\textbf{Important constraints:}\\
\begin{tabular}{@{}p{0.97\linewidth}@{}}
Do \emph{not} assess correctness, faithfulness, or completeness relative to the paper.\\
Only check whether the claim is stated or clearly conveyed in the video.\\
Do not infer unstated information.
\end{tabular}

\vspace{6pt}
A claim is considered \emph{present} if it is:
\begin{tabular}{@{}p{0.97\linewidth}@{}}
Directly stated in the narration;\\
Or clearly expressed by the visual content together with narration.
\end{tabular}

\vspace{6pt}
\textbf{Claim:}\\
\texttt{\{Claim $c$\}}

\vspace{6pt}
\textbf{Video Timeline:}\\
\texttt{\{Timestamped frame descriptions and narration from $\mathcal{V}$\}}

\vspace{6pt}
\textbf{Return:}\\
\texttt{Presence: Yes / No}\\
\texttt{Earliest Timestamp: <time if present>}\\
\texttt{Justification: <cite frame description or narration>}
\end{minipage}%
}
\caption{Prompt used to detect whether a paper-grounded claim is explicitly
present in the video timeline based on frame-level visual descriptions and
aligned narration.}
\label{fig:prompt-claim-presence-detection}
\end{figure}









\subsubsection{Claim Faithfulness Verification via Paper--Video Comparison.}
\label{appendix:faithfulness}
For claims that are detected as present in the video, we next assess whether the video conveys those claims \emph{faithfully} with respect to the source paper. Faithfulness verification evaluates whether the explanation provided by the video preserves the paper’s stated mechanism, conditions, and scope, and does not contradict, exaggerate, or oversimplify the original claim. This step is explicitly separated from presence detection and delivery quality, and focuses solely on semantic alignment between the paper and the video.

Given a claim $c$ that satisfies $P(c,v)=1$, we first retrieve the portions of the paper most relevant to the claim. Using the structured paper representation obtained during PDF preprocessing, we restrict retrieval to explanatory sections such as the abstract, introduction, methods, and results. We split these sections into sentence-level units and embed them using the same sentence embedding model employed during question decomposition. The claim text is embedded using the same encoder, and cosine similarity is used to retrieve the top-$k$ paper sentences most relevant to $c$. To preserve contextual completeness, each retrieved sentence is expanded with a fixed window of surrounding sentences from the same section. The resulting paper spans form a compact evidence set that reflects how the paper defines, qualifies, and justifies the claim.

In parallel, we collect the video evidence associated with the claim by selecting all video segments in which the claim was detected as present. Each segment includes a textual description of the visual content and the aligned narration, along with the corresponding timestamp. These paper and video evidence sets are then jointly provided to a dedicated faithfulness verification prompt, which asks the model to compare how the claim is expressed in the video against how it is stated in the paper.

The faithfulness verifier is instructed to reason explicitly about whether the
video preserves the paper’s causal explanation, constraints, and stated
assumptions, and to identify cases where the video omits key conditions,
generalizes beyond the paper’s claims, or introduces unsupported implications.
Importantly, the prompt does not assess usefulness, explanation depth, or
presentation quality; it only evaluates semantic alignment. The output consists
of a categorical faithfulness judgment and a short diagnostic justification,
which is later used by the meta-evaluator to penalize hallucinated or distorted
explanations. The exact prompt used for claim faithfulness verification is shown
in Figure~\ref{fig:prompt-claim-faithfulness-verification}.

\begin{figure}[!t]
\centering
\promptbox{%
\begin{minipage}{0.97\linewidth}
\raggedright\small
\textcolor{promptblue}{\textbf{Task: Claim Faithfulness Verification via Paper--Video Comparison}}\\[-2pt]
\hrulefill

\vspace{4pt}
You are given a claim that has already been detected as present in the video.
Your task is to determine whether the way this claim is expressed in the video
is \emph{faithful} to the scientific paper.

\vspace{4pt}
\textbf{Definition of faithfulness:}\\
\begin{tabular}{@{}p{0.97\linewidth}@{}}
The video does not contradict the paper;\\
The video does not oversimplify or exaggerate the paper’s claim;\\
The video preserves the paper’s stated mechanism, conditions, and scope.
\end{tabular}

\vspace{6pt}
You are given the following evidence:

\vspace{2pt}
\begin{tabular}{@{}p{0.97\linewidth}@{}}
\emph{(A)} Relevant excerpts retrieved from the paper;\\
\emph{(B)} Video frame descriptions and narration where the claim appears.
\end{tabular}

\vspace{6pt}
You may internally consider questions such as:
\begin{tabular}{@{}p{0.97\linewidth}@{}}
What exactly does the paper claim, and under what conditions?\\
What causal mechanism or evidence does the paper provide?\\
Does the video preserve this mechanism, or omit or alter key details?\\
Does the video introduce implications not supported by the paper?
\end{tabular}

\vspace{6pt}
\textbf{Important constraints:}\\
\begin{tabular}{@{}p{0.97\linewidth}@{}}
Do not judge usefulness, explanation quality, or ordering.\\
Only assess semantic faithfulness to the paper.
\end{tabular}

\vspace{6pt}
\textbf{Claim:}\\
\texttt{\{Claim $c$\}}

\vspace{6pt}
\textbf{Paper Evidence:}\\
\texttt{\{Retrieved paper spans supporting or qualifying $c$\}}

\vspace{6pt}
\textbf{Video Evidence:}\\
\texttt{\{Timestamped frame descriptions and narration where $c$ appears\}}

\vspace{6pt}
\textbf{Return:}\\
\texttt{Faithfulness Label: Faithful / Partially Faithful / Unfaithful}\\
\texttt{Justification: <comparison of paper statements and video expression>}\\
\texttt{If Unfaithful: <what is missing, exaggerated, or contradicted>}
\end{minipage}%
}
\caption{Prompt used to verify whether a claim expressed in the video is faithful
to the corresponding paper-grounded evidence, based on retrieved paper excerpts
and frame-level video descriptions.}
\label{fig:prompt-claim-faithfulness-verification}
\end{figure}

Claims marked as partially faithful or unfaithful are penalized during utility
aggregation, even if they are present and well-delivered, reflecting the fact
that fluent explanations that distort or overgeneralize the paper’s claims do not
support correct understanding or reasoning.

\subsubsection{Cross-Claim Coherence and Prerequisite Ordering}
\label{appendix:coherence}
Beyond verifying whether individual claims are present and faithful, an
instructional video must also present claims in an order that supports
incremental understanding and reasoning. We therefore evaluate
\emph{cross-claim coherence}, which measures whether the video introduces claims
in a pedagogically valid sequence consistent with the dependency structure
derived from the source paper. This step focuses exclusively on temporal and
conceptual ordering and is intentionally decoupled from correctness,
faithfulness, or explanation quality.

Given a question $q$, we operate over the paper-grounded claim set $C(q)$ and its
directed dependency graph, where edges encode prerequisite relations between
claims inferred during question decomposition. Each claim $c \in C(q)$ is
associated with an earliest appearance timestamp obtained during claim presence
detection. This timestamp corresponds to the first video segment in which the
claim is explicitly conveyed, based on aligned narration and frame-level visual
descriptions.

To assess coherence, we compare the temporal ordering of claims in the video against prerequisite constraints derived from a paper-based dependency graph.
This dependency graph is constructed by analyzing how explanatory claims are introduced and justified in the paper: claims that rely on definitions, mechanisms, or assumptions introduced earlier in the paper are treated as dependents, while the concepts they rely on are treated as prerequisites.
Formally, the graph is a directed acyclic structure where an edge $c_i \rightarrow c_j$ indicates that understanding claim $c_j$ presupposes claim $c_i$, as inferred from causal language, definitional references, and explanatory structure in the paper text.

A coherence violation occurs when a dependent claim appears in the video before one or more of its prerequisite claims have been introduced.
Such violations indicate that the video assumes conceptual knowledge that has not yet been established, which can hinder comprehension even if all required claims are eventually covered.

We implement coherence evaluation using a dedicated prompt that conditions on the question, the set of claims, the paper-derived dependency relations, and the video claim timeline.
The prompt instructs the model to identify prerequisite violations and assess pedagogical coherence without judging claim correctness or faithfulness.
The output consists of a categorical coherence judgment and a list of violated claim pairs, each grounded in claim dependencies and video timestamps.
This structured output is later consumed by the meta-evaluator to penalize videos that cover the correct content but present it in a confusing or ill-ordered manner.
The exact prompt used for cross-claim coherence evaluation is shown in Figure~\ref{fig:prompt-cross-claim-coherence}.

\begin{figure}[!t]
\centering
\promptbox{%
\begin{minipage}{0.97\linewidth}
\raggedright\small
\textcolor{promptblue}{\textbf{Task: Cross-Claim Coherence}}\\[-2pt]
\hrulefill

\vspace{4pt}
You are given a set of claims that together answer a single question,
along with a dependency structure derived from the source paper.
You are also given a time-ordered video timeline indicating when
each claim first appears in the video.

\vspace{6pt}
Your task is to evaluate whether the video presents these claims
in a pedagogically coherent order.

\vspace{4pt}
\textbf{Definition of coherence:}\\
\begin{tabular}{@{}p{0.97\linewidth}@{}}
Prerequisite claims are introduced before dependent claims;\\
The ordering of claims follows the conceptual structure implied by the paper.
\end{tabular}

\vspace{6pt}
\textbf{Important constraints:}\\
\begin{tabular}{@{}p{0.97\linewidth}@{}}
Do not judge correctness or faithfulness of claims.\\
Do not assess explanation quality or delivery.\\
Only assess ordering and prerequisite consistency across claims.
\end{tabular}

\vspace{6pt}
\textbf{Question:}\\
\texttt{\{Question $q$\}}

\vspace{6pt}
\textbf{Claims:}\\
\texttt{\{Claim $c_1$, Claim $c_2$, \ldots, Claim $c_n$\}}

\vspace{6pt}
\textbf{Claim Dependencies (from paper):}\\
\texttt{\{Directed edges indicating prerequisite relations between claims\}}

\vspace{6pt}
\textbf{Video Claim Timeline:}\\
\texttt{\{For each claim $c_i$, the earliest timestamp interval where it appears in the video,}\\
\texttt{\quad along with the associated narration or frame description\}}

\vspace{6pt}
Analyze whether the video violates any prerequisite ordering constraints.
Identify all claim pairs where a dependent claim appears before its prerequisite.

\vspace{6pt}
\textbf{Return:}\\
\texttt{Coherence Judgment: Coherent / Partially Coherent / Incoherent}\\
\texttt{Prerequisite Violations: <ordered claim pairs>}\\
\texttt{Explanation: <grounded in dependencies and video timestamps>}
\end{minipage}%
}
\caption{Prompt used to evaluate cross-claim coherence by comparing the temporal
ordering of claims in the video against the paper-derived prerequisite
dependency structure.}
\label{fig:prompt-cross-claim-coherence}
\end{figure}

Videos that exhibit prerequisite violations receive lower coherence scores even
when all claims are present and faithful, reflecting the fact that incorrect
ordering can impede reasoning and learning. By isolating coherence as a separate
diagnostic dimension, our framework distinguishes between videos that merely
cover the right content and those that present it in a conceptually sound and
instructionally effective manner.





\subsubsection{Delivery Quality  Evaluation}
\label{appendix:delivery}
Beyond verifying that required claims are present and faithful to the source paper, an instructional video must also \emph{explain} those claims in a way that supports reasoning.
We therefore evaluate \emph{delivery quality} as a measure of how clearly, thoroughly, and causally a video explains the claims necessary to answer a given question.
Delivery quality is explicitly conditioned on claim presence and focuses on explanatory adequacy rather than surface fluency, narration style, or visual polish alone.

For a paper $p$, question $q$, and video $v$, we operate over the paper-grounded claim set $\mathcal{C}(q)$ obtained during question decomposition, along with each claim’s importance score $I(c)\in[0,1]$.
Claim importance reflects the role a claim plays in the paper’s explanatory structure: claims that directly connect the method to the paper’s core contribution receive higher importance, while background or supporting claims receive lower weights.
The video is represented as a temporally ordered multimodal sequence $\mathcal{V}=\{(s_i,t_i,a_i,d_i)\}_{i=1}^{M}$, where each segment consists of a slide image $s_i$, its timestamp interval $t_i$, aligned narration $a_i$, and a vision--language description $d_i$.

Delivery quality is computed at the claim level and aggregated at the question level.
For each claim $c$ that is present in the video, we estimate how well it is explained by combining four complementary signals.
First, $\pi(c,v)\in\{0,1\}$ indicates whether the claim is explicitly stated in the video.
Second, $\hat{T}(c,v)$ measures the normalized amount of narration time devoted to explaining $c$, computed from the contiguous video segments where the narration or visuals focus on that claim.
This captures whether the video allocates sufficient temporal emphasis to the claim.
Third, $Q(c,v)\in[0,1]$ captures the qualitative richness of the explanation, as judged by a dedicated LLM prompt that evaluates whether the explanation provides causal or mechanistic reasoning, introduces intermediate concepts where needed, and goes beyond merely stating the claim.
Importantly, this assessment does not judge correctness relative to the paper—which is handled by the Faithfulness Agent—but instead evaluates whether the explanation itself supports understanding.
Finally, $A(c,v)\in[0,1]$ measures audio--visual alignment, assessing whether the narration explaining $c$ is temporally synchronized with relevant visual content such as figures, equations, or slide text.

We combine these signals into a delivery score defined as
$D_{\text{del}}(q,v)=\sum_{c\in\mathcal{C}(q)} I(c)\,\pi(c,v)\,\hat{T}(c,v)\,Q(c,v)\,A(c,v)$.
This formulation ensures that explanations of high-importance claims contribute more strongly to delivery quality, while claims that are briefly mentioned, weakly explained, poorly grounded in visuals, or entirely absent receive low or zero contribution.
As a result, weak or rushed explanations of central causal claims are penalized more heavily than equally weak explanations of supporting context.

Delivery quality interacts naturally with coherence.
While the Coherence Agent ensures that claims are introduced in a prerequisite-consistent order, the Delivery Agent determines whether each step in the reasoning chain is explained with sufficient depth and grounding to enable the viewer to infer downstream conclusions.
Consequently, videos that are fluent but shallow, or polished yet rushed, receive low delivery scores despite high engagement or coverage, whereas videos that progressively motivate concepts and clearly explain causal links score higher on reasoning-oriented questions.

For example, although $v_A$ mentions all required claims, it allocates little time and weak visual grounding to explaining why clustering leads to generalization, resulting in a low delivery contribution for the most important claim $c_3$.
Video $v_B$ provides richer and better-aligned explanations overall but suffers from ordering violations, while $v_C$ delivers strong explanations for $c_1$ and $c_2$ yet contributes nothing for $c_3$ due to its absence.
Details of the prompts used to operationalize $Q(c,v)$ and $A(c,v)$ are provided in Appendix~\ref{appendix:delivery}.

\begin{center}
\fcolorbox{promptborder}{promptgray}{
\begin{minipage}{0.95\columnwidth}
\textcolor{promptblue}{\textbf{Task: Explanation Quality Assessment}}

You are given a single claim and the video segments where this claim is explained.
Each segment includes a textual description of the visual content and the aligned narration.

Your task is to assess how well the video explains the claim.
Focus on whether the explanation provides reasoning support rather than merely stating the claim.

Consider whether the explanation:
clarifies why the claim holds,
introduces intermediate concepts or mechanisms if needed,
and connects the claim to surrounding ideas in a way that supports understanding.

Do not judge factual correctness relative to the paper.
Do not assess ordering or prerequisite satisfaction.
Only evaluate the explanatory depth and clarity of the explanation.

\textbf{Claim:} \{Claim $c$\}

\textbf{Video Segments:} \{Timestamped frame descriptions and narration where $c$ appears\}

Return a scalar explanation quality score $Q(c,v)\in[0,1]$ and a brief justification grounded in the video content.
\end{minipage}}
\end{center}

By separating delivery quality from coverage, faithfulness, and coherence, this design allows the evaluator to diagnose why a video may fail to support reasoning even when all required claims are present. As demonstrated in our experiments, delivery quality plays a critical role in distinguishing videos that merely enumerate correct facts from those that genuinely enable learners to understand and reason about a paper’s core contributions.

\subsubsection{Engagement Score Calculation}
\label{appendix:engagement}
In addition to content coverage, faithfulness, coherence, and delivery quality, we estimate viewer engagement as a complementary signal that captures how dynamically the video narration is delivered. Engagement is not intended to measure educational correctness or reasoning support, but rather the expressive qualities of speech that are commonly associated with attentiveness and listener interest. Our engagement score is therefore computed solely from the audio track of the video and is decoupled from visual or textual content.

Given a video $v$, we first extract its audio stream using a standard video processing library and resample it to a fixed sampling rate. The resulting audio signal is then analyzed using a lightweight acoustic feature extraction pipeline. Specifically, we compute pitch contours using a pitch-tracking algorithm and retain only non-zero pitch values corresponding to voiced speech. The standard deviation of these pitch values is used to quantify pitch variation, which serves as a proxy for vocal expressiveness. In parallel, we compute short-time root mean square (RMS) energy over the audio signal and estimate both the mean energy and its standard deviation. The ratio of energy variation to mean energy captures loudness dynamics, distinguishing monotonous narration from expressive delivery.

To approximate speaking rate, we apply beat tracking over the audio signal and use the estimated tempo as a coarse proxy for speech rhythm. Although originally designed for music, beat tracking provides a stable estimate of temporal pacing that correlates with perceived speech speed in presentation-style videos. Each acoustic feature is normalized to lie in the range $[0,1]$ using empirically chosen upper bounds to prevent extreme values from dominating the score.

The final engagement factor is computed as a weighted combination of normalized pitch variation, energy variation, and tempo, with higher weight assigned to pitch and energy dynamics and lower weight to tempo. This aggregation reflects the intuition that expressive intonation and loudness modulation contribute more strongly to perceived engagement than raw speaking speed. The resulting scalar engagement score provides a coarse but interpretable estimate of narration dynamism.

We compute engagement scores independently for each video variant and store the resulting pitch, energy, tempo, and aggregate engagement values in a structured JSON format for downstream analysis. Engagement is reported as an auxiliary signal alongside utility scores and is not used to override content-based diagnostics. 

\subsection{Meta-Evaluator Utility Aggregation}
\label{appendix:meta-evaluation}
After computing claim-level and question-level diagnostics, we aggregate all
signals into a unified diagnostic record where each term
captures a distinct aspect of instructional utility. Coverage $\pi(q,v)$ reflects
whether the claims required to answer the question are explicitly present in the
video, faithfulness $F(q,v,p)$ measures semantic alignment between the video and
the source paper, coherence $C(q,v)$ evaluates prerequisite-consistent ordering
of claims, delivery quality $D_{\text{del}}(q,v)$ captures explanatory depth and
temporal emphasis, and engagement $E(q,v)$ estimates narration dynamism. Each
signal is normalized to lie in $[0,1]$ to enable meaningful aggregation.

The Meta-Evaluator computes question-level utility as a weighted combination of
these signals and reflect the
relative importance of content correctness and reasoning support over surface
qualities. In particular, coverage, faithfulness, and coherence receive higher
weight than engagement, ensuring that fluent but misleading or incomplete videos
are penalized appropriately.

Importantly, the Meta-Evaluator does not merely compute a scalar score. In
addition to aggregation, it produces a structured rationale that attributes
utility differences to specific diagnostic failures. For example, a video may
receive moderate utility due to shallow delivery despite full coverage, another
may be penalized primarily for prerequisite violations or faithfulness errors,
and a third may receive low utility even with strong delivery and engagement if
it omits a core causal claim. This explicit attribution allows the evaluator to
distinguish between missing content, misordered explanations, distorted claims,
and weak instructional depth, rather than conflating these factors into a single
opaque judgment.

We implement this aggregation and explanation step using a dedicated
Meta-Evaluator prompt that conditions on the full diagnostic record and instructs
the model to justify the final utility score in terms of the underlying signals.
The prompt explicitly discourages surface-level judgments based on fluency or
aesthetic quality alone and requires the rationale to reference coverage,
faithfulness, coherence, and delivery diagnostics. The exact prompt is shown in
Figure~\ref{fig:prompt-meta-evaluator}.

\begin{figure}[!t]
\centering
\promptbox{%
\begin{minipage}{0.97\linewidth}
\raggedright\small
\textcolor{promptblue}{\textbf{Task: Meta-Evaluator Utility Aggregation}}\\[-2pt]
\hrulefill

\vspace{4pt}
You are given a diagnostic record summarizing how a video performs when answering
a specific question about a scientific paper.

\vspace{4pt}
\textbf{Diagnostic signals include:}\\
\begin{tabular}{@{}p{0.97\linewidth}@{}}
Coverage of required claims;\\
Faithfulness to the source paper;\\
Coherence of claim ordering;\\
Delivery quality of explanations;\\
Engagement of narration.
\end{tabular}

\vspace{6pt}
Your task is to aggregate these signals into a final instructional utility judgment.

\vspace{4pt}
\textbf{You must:}\\
\begin{tabular}{@{}p{0.97\linewidth}@{}}
\emph{(1)} Compute an overall utility score using the provided diagnostics;\\
\emph{(2)} Explain why the score is high or low by explicitly referencing the diagnostics;\\
\emph{(3)} Attribute differences in utility to missing, distorted, misordered, or weakly
explained concepts, rather than surface presentation quality alone.
\end{tabular}

\vspace{6pt}
\textbf{Constraints:}\\
\begin{tabular}{@{}p{0.97\linewidth}@{}}
Do not introduce new evidence.\\
Do not rely on general impressions or fluency.
\end{tabular}

\vspace{6pt}
\textbf{Question:}\\
\texttt{\{Question $q$\}}

\vspace{6pt}
\textbf{Diagnostic Record:}\\
\texttt{\{P(q,v), F(q,v,p), C(q,v), D\textsubscript{del}(q,v), E(q,v)\}}

\vspace{6pt}
\textbf{Return:}\\
\texttt{Utility Score: $U(q,v,p)$}\\
\texttt{Rationale: <2--4 sentences grounded in the diagnostics>}
\end{minipage}%
}
\caption{Prompt used by the Meta-Evaluator to aggregate diagnostic signals into
a question-level utility score with an explicit, interpretable rationale.}
\label{fig:prompt-meta-evaluator}
\end{figure}

\subsection{Paper--Video Utility Aggregation}
\label{appendix:paper-video-utility}
While question-level utility captures whether a video successfully explains a
specific aspect of a paper, instructional videos are typically evaluated with
respect to multiple questions that probe different parts of the paper’s
contribution. Given a paper $p$ with a set of evaluation questions
$\mathcal{Q}(p)$, we therefore compute paper--video utility by averaging
question-level utilities:
\[
U(p,v)=\frac{1}{|\mathcal{Q}(p)|}\sum_{q\in\mathcal{Q}(p)}U(q,v,p).
\]

Aggregating across questions enables the Meta-Evaluator to identify systematic
patterns in a video’s strengths and weaknesses. For instance, a video may
consistently omit causal explanations across questions, repeatedly violate
prerequisite ordering, or frequently provide fluent but shallow explanations.
Because each question-level score is accompanied by a structured rationale,
these patterns remain interpretable at the paper level rather than collapsing
into an opaque average.

This paper--video utility score supports reliable comparison between video
variants generated from the same paper, while preserving diagnostic insight into
\emph{why} one variant is more useful than another. As a result, the evaluation
framework enables both quantitative ranking and qualitative error analysis,
bridging the gap between scalar metrics and human-interpretable judgments of
educational utility.





















\section{Experimental Details}
\subsection{Baseline Implementation Details}
\label{appendix:baselines}
For all single-agent and holistic baselines, we standardize the video input by selecting a small set of representative frames together with their aligned narration.
This design ensures that baseline methods receive a consistent, bounded view of the video, while remaining comparable to common evaluation practice in prior work. 
Representative frames are determined from the frame-level multimodal representation $\mathcal{V}=\{(s_i,t_i,a_i,d_i)\}_{i=1}^M$ constructed during video preprocessing. 
We first group consecutive frames with near-identical visual descriptions into segments corresponding to stable slide content.
For each segment, we select a single representative frame, typically the first frame after a slide transition, and associate it with the narration spoken during the same timestamp interval.
This procedure yields a concise set of frames that capture all major visual states of the video without redundancy. The corresponding narration spans are concatenated or truncated to fit the input budget of the baseline models.

Using this standardized input, we implement four baseline evaluation strategies: single-agent question answering using transcript only (Prompt~\ref{fig:prompt-qa-transcript-only}), single-agent question answering using both frames and transcript(Prompt~\ref{fig:prompt-qa-frames-transcript}), and a holistic utility rating that bypasses question answering entirely (Prompt~\ref{fig:prompt-holistic-utility}).
The exact prompts used for each baseline are shown below.
\begin{figure}[t]
\centering
\promptbox{%
\begin{minipage}{0.97\linewidth}
\raggedright\small
\textcolor{promptblue}{\textbf{Task: Single-Agent QA (Transcript-only)}}\\[-2pt]
\hrulefill

\vspace{4pt}
You are given a question about a scientific paper and the transcript of a video generated from that paper.

\vspace{4pt}
\textbf{Instructions:}
\begin{tabular}{@{}p{0.97\linewidth}@{}}
\emph{(i)} Answer using \emph{only} the transcript text provided.\\
\emph{(ii)} Do not use outside knowledge.\\
\emph{(iii)} If the transcript is insufficient, state that explicitly.
\end{tabular}

\vspace{6pt}
\textbf{Question:}\\
Why does DefSent improve generalization?

\vspace{6pt}
\textbf{Transcript (excerpt):}\\
\texttt{DefSent learns sentence representations using entailment-based supervision.
We train on sentence-level objectives rather than individual words.}

\vspace{6pt}
\textbf{Output:} Provide a concise answer grounded only in the transcript.
\end{minipage}%
}
\caption{Prompt used for single-agent question answering using only the video transcript, without access to visual information.}
\label{fig:prompt-qa-transcript-only}
\end{figure}

\begin{figure}[t]
\centering
\promptbox{%
\begin{minipage}{0.97\linewidth}
\raggedright\small
\textcolor{promptblue}{\textbf{Task: Single-Agent QA (Frames + Transcript)}}\\[-2pt]
\hrulefill

\vspace{4pt}
You are given a question about a scientific paper, representative video frames, and narration aligned to those frames.

\vspace{4pt}
\textbf{Instructions:}
\begin{tabular}{@{}p{0.97\linewidth}@{}}
\emph{(i)} Use both transcript and visual evidence.\\
\emph{(ii)} Do not assume information not supported by the inputs.
\end{tabular}

\vspace{6pt}
\textbf{Question:}\\
Why does DefSent improve generalization?

\vspace{6pt}
\textbf{Transcript (excerpt):}\\
\texttt{Sentence-level supervision encourages representations that capture global semantic meaning.}

\vspace{6pt}
\textbf{Representative Frame (description):}\\
\texttt{A diagram showing sentence embeddings forming tight semantic clusters, compared to dispersed word-level embeddings.}

\vspace{6pt}
\textbf{Output:} Provide a concise answer grounded in the transcript and the visual evidence.
\end{minipage}%
}
\caption{Prompt used for single-agent question answering with access to both representative video frames and aligned narration.}
\label{fig:prompt-qa-frames-transcript}
\end{figure}

\begin{figure}[t]
\centering
\promptbox{%
\begin{minipage}{0.97\linewidth}
\raggedright\small
\textcolor{promptblue}{\textbf{Task: LLM-as-Judge Answer Evaluation}}\\[-2pt]
\hrulefill

\vspace{4pt}
You are given a gold answer derived from the source paper and a model answer generated after watching the video.

\vspace{4pt}
\textbf{Scoring rubric (choose one):}\\
\textbf{1} = Correct and complete \quad
\textbf{0.5} = Partially correct or incomplete \quad
\textbf{0} = Incorrect or unsupported

\vspace{4pt}
Provide a brief justification.

\vspace{6pt}
\textbf{Gold Answer:}\\
\texttt{DefSent improves generalization by using sentence-level supervision that captures global semantic structure beyond individual words.}

\vspace{6pt}
\textbf{Model Answer:}\\
\texttt{DefSent generalizes better because it uses more training data.}

\vspace{6pt}
\textbf{Output format:}\\
\texttt{Score: <1|0.5|0>}\\
\texttt{Justification: <one or two sentences>}
\end{minipage}%
}
\caption{Prompt used to score model-generated answers against paper-derived gold answers in the single-agent QA baselines.}
\label{fig:prompt-llm-judge}
\end{figure}

\begin{figure}[t]
\centering
\promptbox{%
\begin{minipage}{0.97\linewidth}
\raggedright\small
\textcolor{promptblue}{\textbf{Task: Holistic Utility Rating}}\\[-2pt]
\hrulefill

\vspace{4pt}
You are given a summary of a video generated from a scientific paper.
Rate how useful the video is for answering the question.

\vspace{4pt}
\textbf{Guidelines:}
\begin{tabular}{@{}p{0.97\linewidth}@{}}
Consider correctness, clarity, and completeness.\\
Do \emph{not} decompose into sub-questions.\\
Do \emph{not} rely on stylistic quality alone.
\end{tabular}

\vspace{6pt}
\textbf{Question:}\\
Why does DefSent improve generalization?

\vspace{6pt}
\textbf{Video Summary:}\\
\texttt{The video introduces DefSent and describes sentence-level training, but does not explain how this leads to improved generalization or provide causal reasoning.}

\vspace{6pt}
\textbf{Output:} Assign a usefulness score from \textbf{1} (not useful) to \textbf{5} (very useful).
\end{minipage}%
}
\caption{Prompt used for holistic utility rating, where the model assigns a single usefulness score without explicit claim-level diagnostics.}
\label{fig:prompt-holistic-utility}
\end{figure}

\subsection{Implementation of Existing Metrics}
\label{appendix:existing-metrics}
We use the regularized version of VideoScore\footnote{\url{https://huggingface.co/TIGER-Lab/VideoScore}} ~\cite{he2024videoscore} as an existing automatic evaluation baseline. VideoScore is instantiated with the Mantis-8B-Idefics2 backbone\footnote{\url{https://huggingface.co/TIGER-Lab/Mantis-8B-Idefics2}} and produces per-video scores along multiple dimensions, including visual quality, temporal consistency, dynamic degree, and factual consistency, each rated on a 1--5 scale. For each paper, we apply VideoScore independently to all generated video variants and normalize the resulting scores to the range $[0,1]$ for comparability with other evaluation signals.

Next, we use the official PresentQuiz implementation released by
\citet{zhu2025paper2videoautomaticvideogeneration}
(\url{https://github.com/showlab/Paper2Video}) to automatically generate question–answer pairs from each paper. These paper-derived questions and gold answers are then used to evaluate the answerability of each corresponding video variant, following the protocol proposed in prior work. We use the accuracy as the final success criteria for each video variant, normalize in the range of 0-1, and use that to compare and rank the video variants per paper.

We additionally evaluate visual and temporal consistency using VBench metrics~\cite{huang2023vbench}, including Background Consistency, Subject Consistency, Temporal Flickering, Motion Smoothness, Dynamic Degree, Aesthetic Quality, and Imaging Quality. These metrics are computed over extracted video frames using the official VBench implementation\footnote{\url{https://github.com/Vchitect/VBench}}. For each video variant, we average the scores across all VBench dimensions and normalize the resulting value to the range $[0,1]$, which is used to rank video variants for the same paper.

Separately, we use PPTEval from the PPTAgent framework~\cite{zheng-etal-2025-pptagent} to assess presentation-level qualities, including design, content organization, and visual style. PPTEval assigns scores on a 1--5 scale for each dimension; we normalize these scores to $[0,1]$ and average them to obtain a single presentation-quality score per variant, which is used for comparative analysis across video variants for a single paper.

\begin{table*}[t]
\centering
\small
\setlength{\tabcolsep}{3pt}
\renewcommand{\arraystretch}{1.2}
\begin{tabular}{c c p{1.5cm} p{10.2cm}}
\toprule
\textbf{Frame} & \textbf{Visual} & \textbf{Time} & \textbf{Audio Narration } \\
\midrule

F1 &
\includegraphics[width=0.16\linewidth]{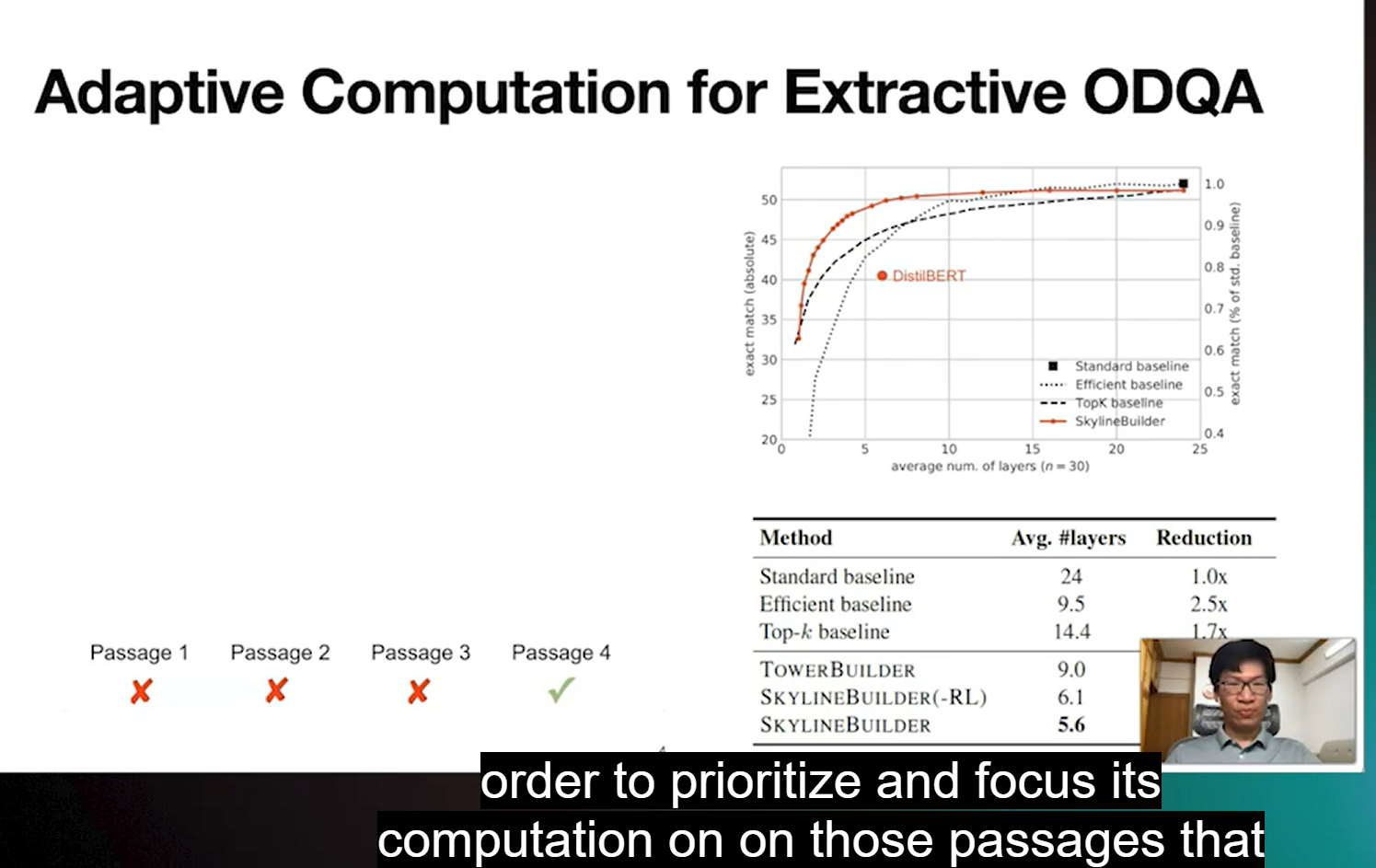} &
00:00--00:30 &
In open-domain question answering, the system retrieves many passages for each question, but only a few actually contain information needed to answer it.
Traditional models still process every retrieved passage using the full encoder depth, even though most passages are irrelevant.
This creates a large amount of wasted computation, because deep encoding is applied uniformly instead of selectively. \\
\midrule

F2 &
\includegraphics[width=0.16\linewidth]{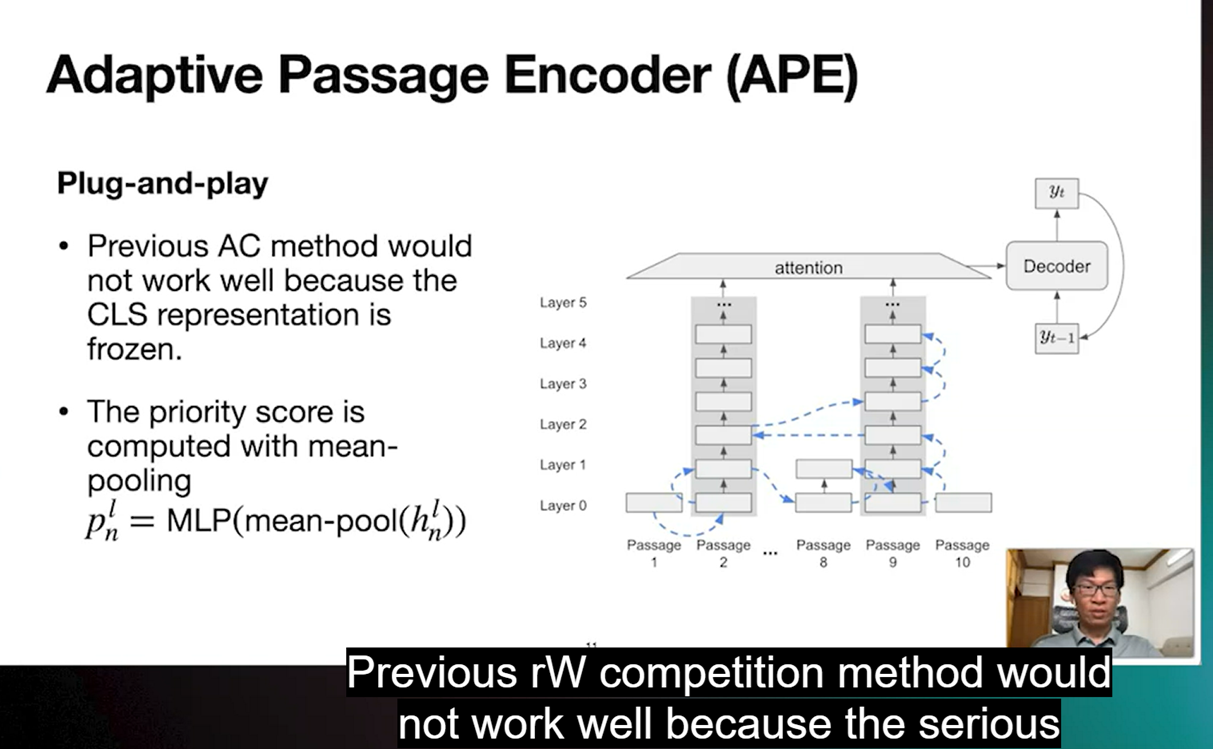} &
00:30--01:05 &
ADaptive Passage Encoder addresses this inefficiency under a key constraint: the base generative QA model remains frozen.
Instead of changing the model’s parameters, APE changes how computation is allocated.
It acts as a plug-and-play replacement for the encoder, allowing different passages to receive different amounts of computation.\\

\midrule

F3 &
\includegraphics[width=0.16\linewidth]{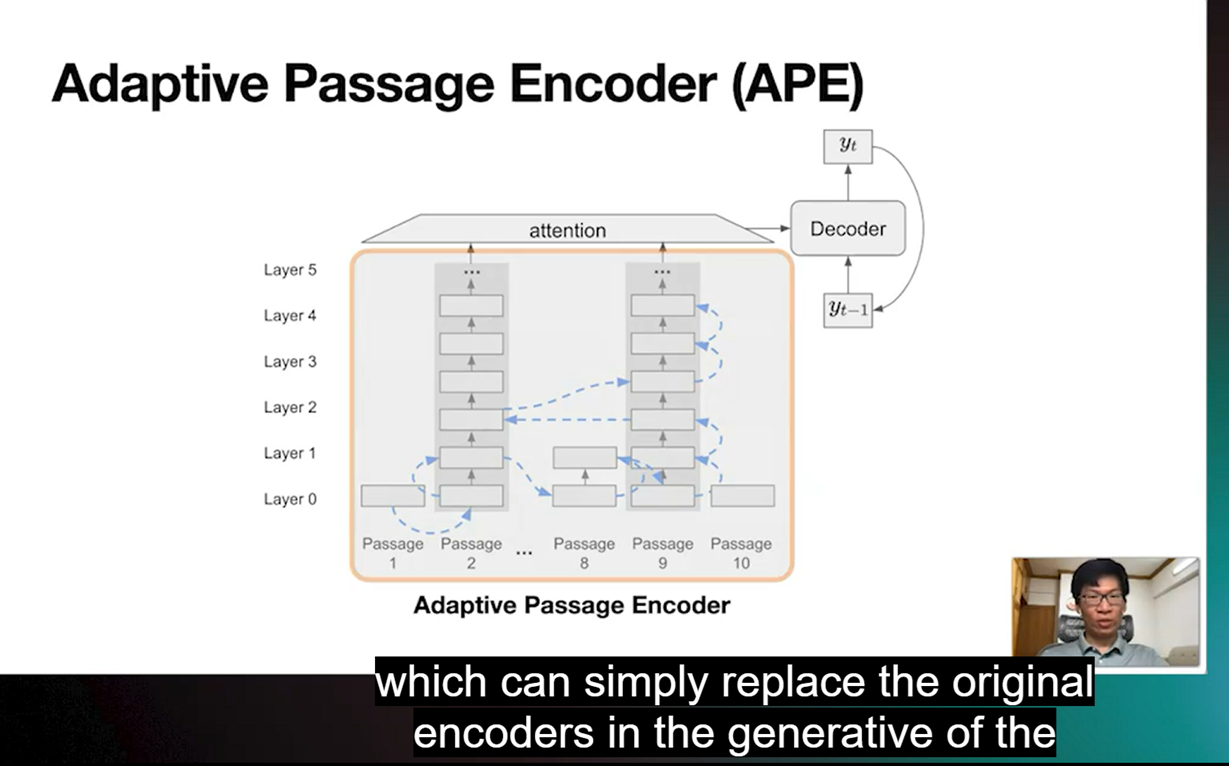} &
01:05--01:40 &
To decide which passages deserve more computation, APE introduces a lightweight component called HasAnswer.
After only a few shallow encoder layers, HasAnswer examines intermediate representations and predicts how likely each passage is to contain answer-relevant information.
This provides an early, low-cost signal about passage usefulness before committing to deeper computation. \\

\midrule

F4 &
\includegraphics[width=0.16\linewidth]{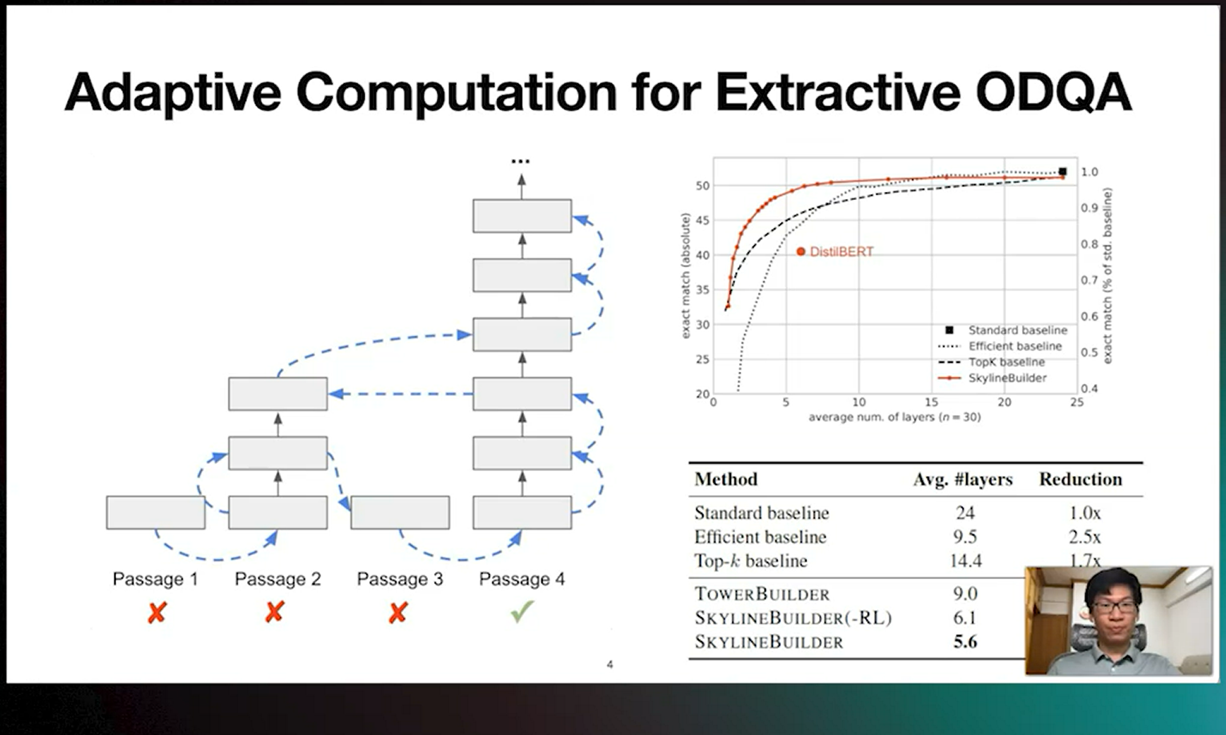} &
01:40--02:05 & Using the HasAnswer scores, a scheduler dynamically allocates encoder layers across passages.
Passages predicted to be irrelevant are exited early, saving computation.
Passages predicted to be promising are processed through more layers, producing richer representations.
In this way, computational resources are concentrated on passages most likely to support correct answering. \\

\midrule
F5 &
\includegraphics[width=0.16\linewidth]{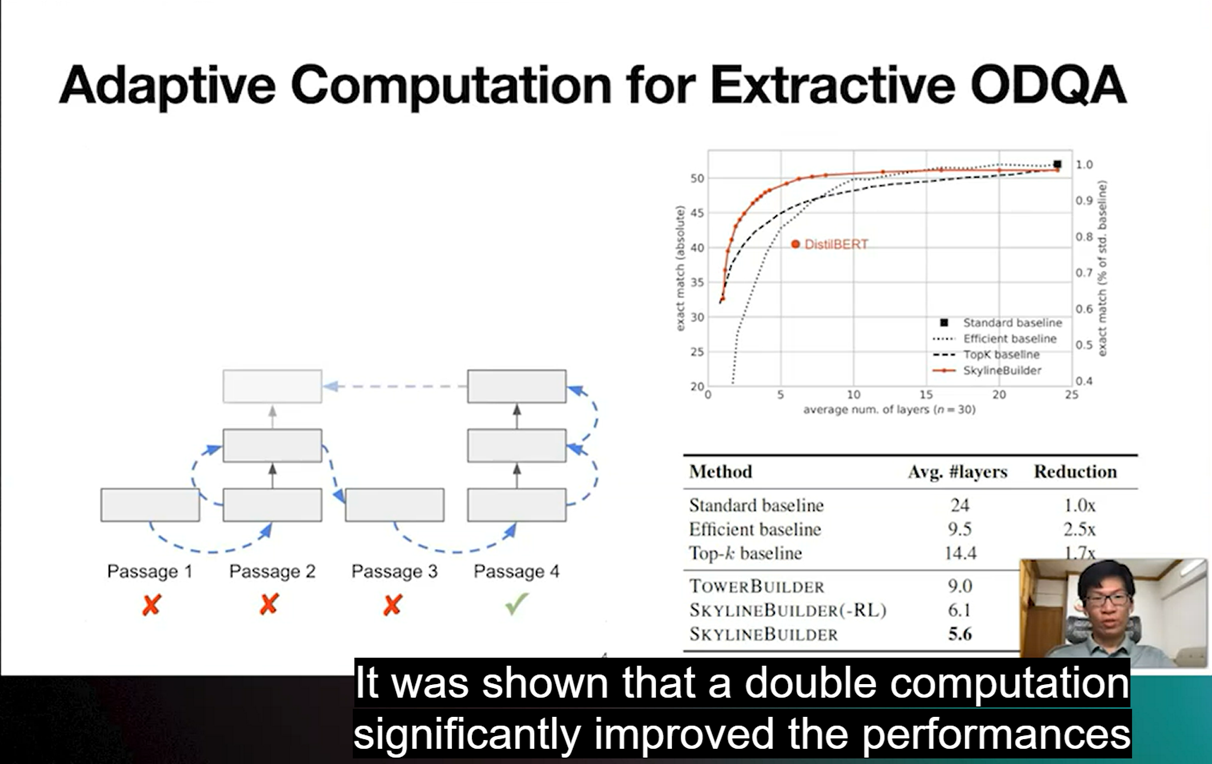} &
02:05--02:30 &
Empirical results show that APE computes far fewer encoder layers on average while maintaining comparable exact match performance.
This happens because the decoder primarily relies on deeply encoded, high-quality passages, rather than shallow representations of many irrelevant ones.
As a result, reducing computation on irrelevant passages does not hurt answer accuracy, since critical evidence is still processed thoroughly. \\

\bottomrule
\end{tabular}
\caption{Frame-level grounding with teaching-style narration: each row explains the concept step-by-step so a learner can follow the motivation, mechanism, and outcome of Adaptive Passage Encoder (APE) \cite{wu-etal-2021-training}.}
\label{tab:ape_frame_narration_teaching}
\end{table*}

\begin{table*}[t]
\centering
\small
\setlength{\tabcolsep}{6pt}
\renewcommand{\arraystretch}{1.25}
\begin{tabularx}{\textwidth}{c c c >{\raggedright\arraybackslash}m{0.45\textwidth}}
\toprule
\textbf{Frame} & \textbf{Time (s)} & \textbf{Visual Frame} & \textbf{Explanatory Narration (Rewritten)} \\
\midrule

\textbf{F2} & 0:12--0:16 &
\includegraphics[width=0.18\textwidth]{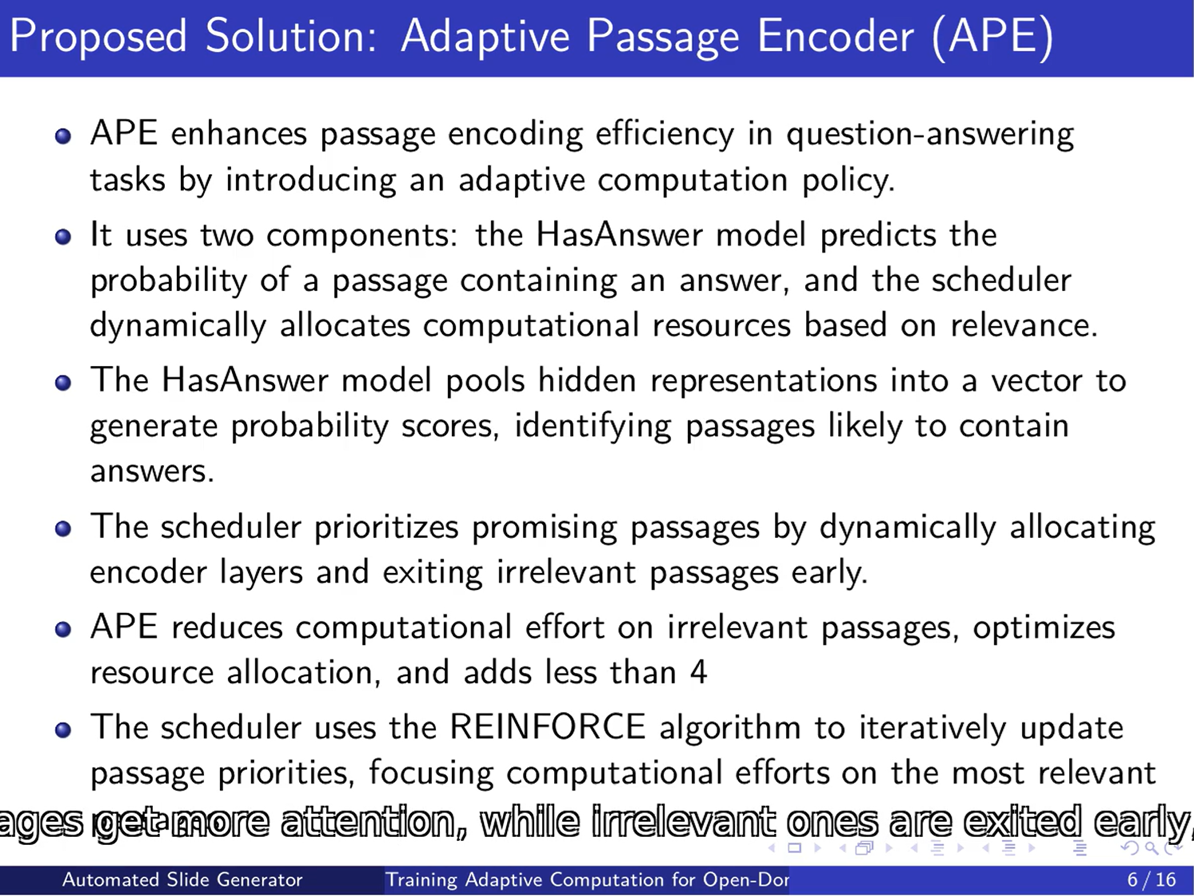} &
This slide introduces \emph{Adaptive Passage Encoder (APE)} as a shift from uniform computation to \emph{selective computation}. 
Instead of processing every retrieved passage with the same number of encoder layers, APE treats computation as a limited resource.
The key idea is that answer accuracy depends on deeply encoding only a few relevant passages, not all retrieved ones. \\

\midrule

\textbf{F3} & 0:16--0:20 &
\includegraphics[width=0.18\textwidth]{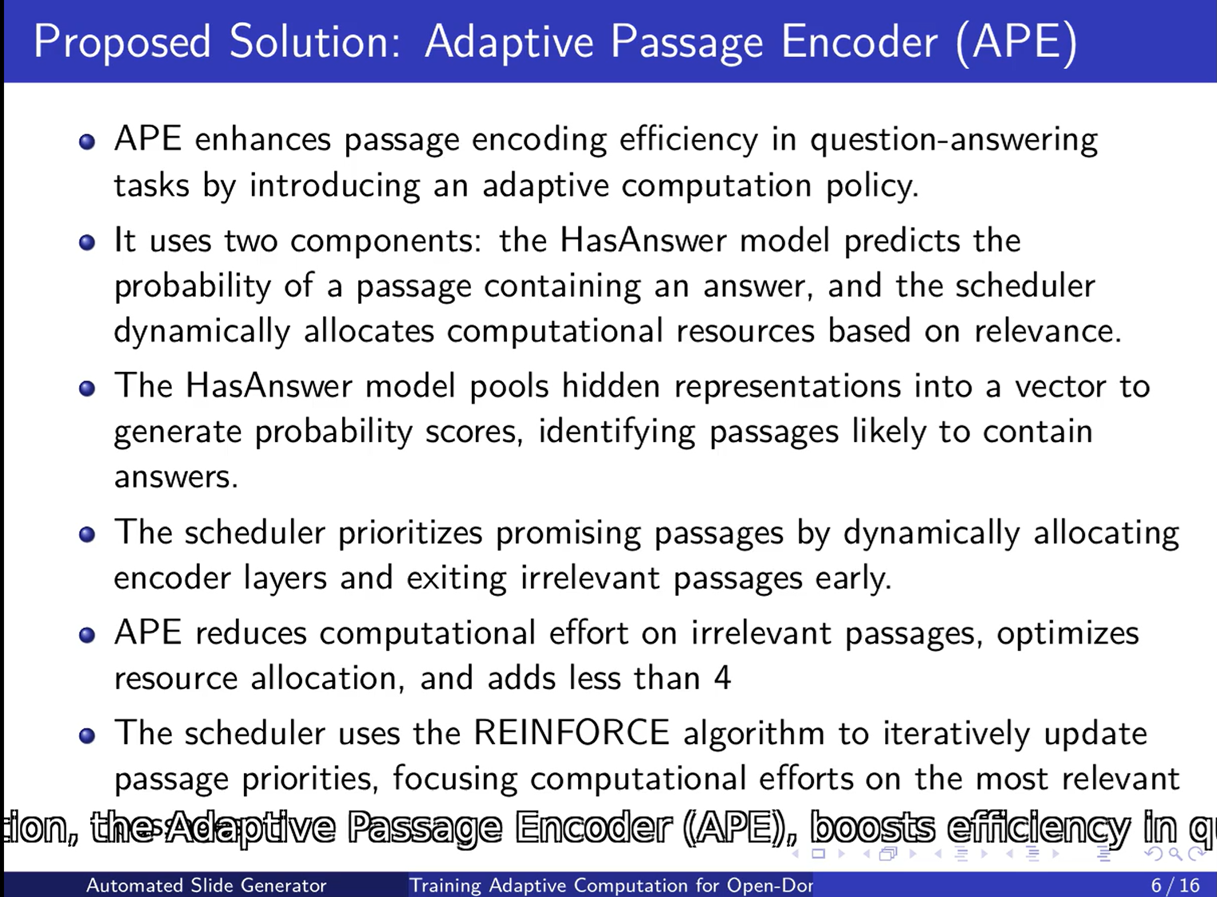} &
This slide explains the \emph{HasAnswer} component.
After only a few shallow encoder layers, HasAnswer pools hidden representations to estimate the probability that a passage contains answer evidence.
This early relevance signal is intentionally computed cheaply, allowing the system to decide \emph{before expensive processing} which passages are worth further computation. \\

\midrule

\textbf{F4} & 0:20--0:24 &
\includegraphics[width=0.18\textwidth]{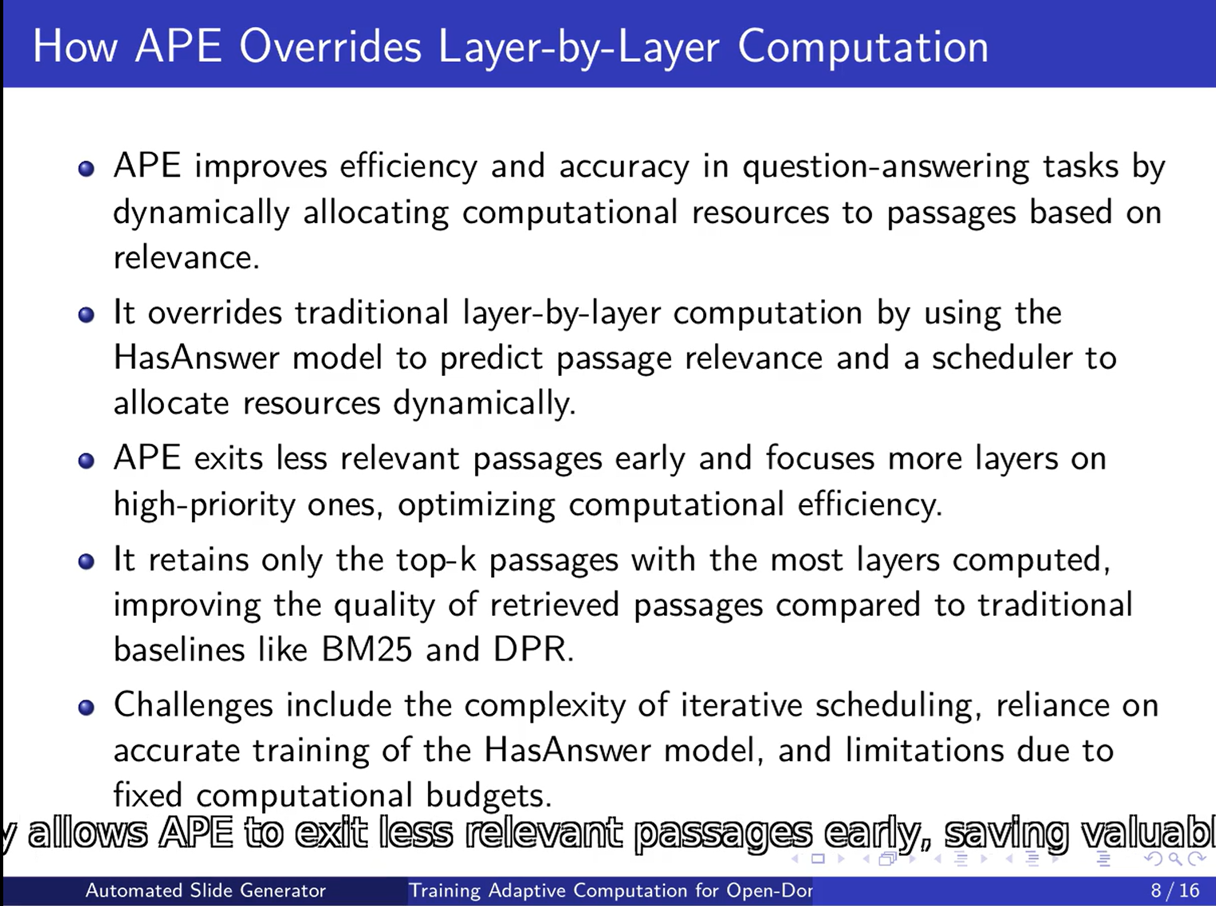} &
This slide describes the scheduler, which uses HasAnswer scores to dynamically allocate encoder depth.
Passages predicted to be irrelevant are exited early, while promising passages receive more layers.
Crucially, computation is not discarded but \emph{reallocated}, ensuring that answer-bearing passages are encoded more richly than in uniform baselines. \\

\midrule

\textbf{F5} & 0:24--0:28 &
\includegraphics[width=0.18\textwidth]{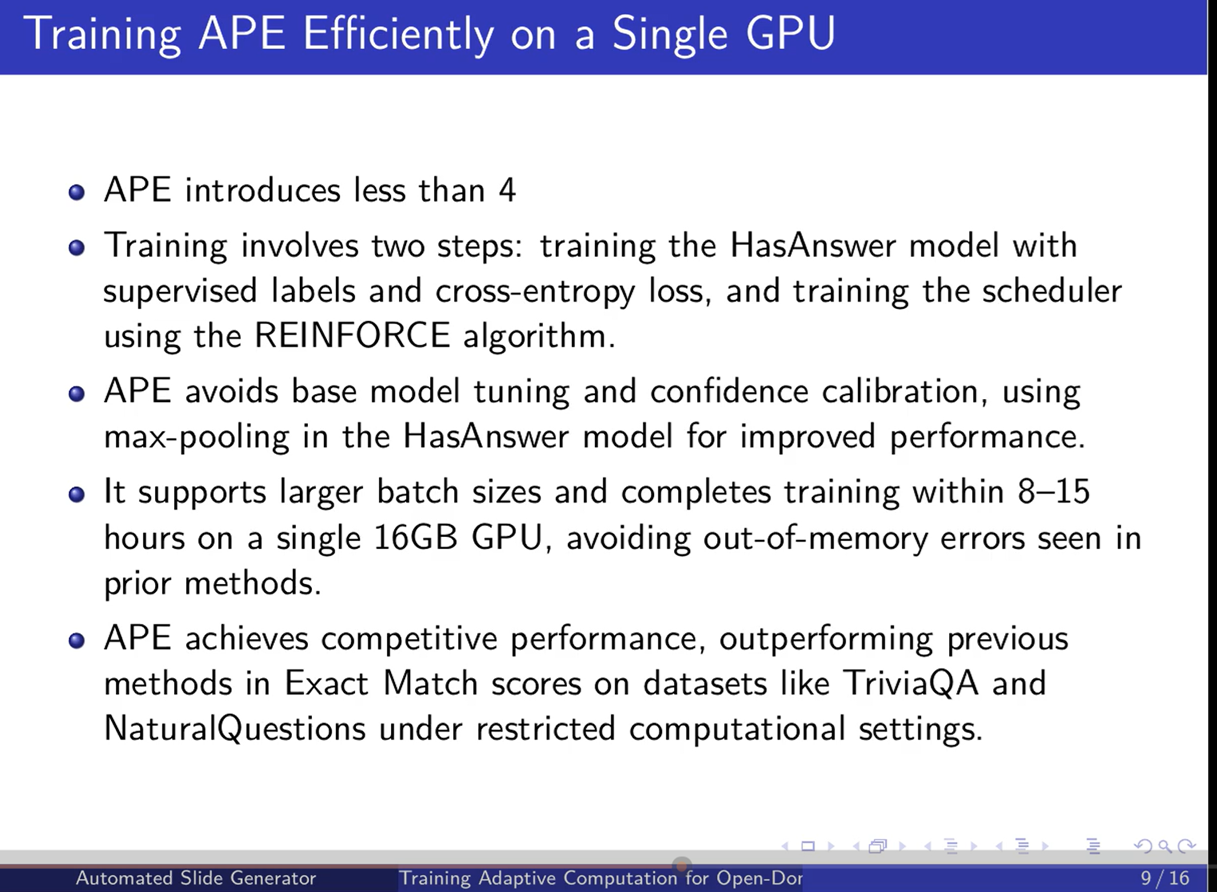} &
This slide explains why accuracy is preserved despite reduced computation.
Because the decoder primarily attends to the few deeply encoded passages, shallow representations of irrelevant passages do not hurt answer prediction.
As a result, APE achieves similar exact match scores using fewer average encoder layers, demonstrating an efficiency--accuracy trade-off enabled by adaptive scheduling. \\

\bottomrule
\end{tabularx}
\caption{Frame-level explanatory narration for Adaptive Passage Encoder (APE) \cite{wu-etal-2021-training}. The rewritten narration transforms read-through slides into a causal explanation that enables reasoning about why APE reduces computation without hurting answer accuracy.}
\label{tab:ape_frame_narration}
\end{table*}

\begin{table*}[t]
\centering
\small
\setlength{\tabcolsep}{6pt}
\rowcolors{2}{gray!8}{white}
\begin{tabular}{lccc}
\toprule
\textbf{Variant Control}
& \textbf{Human Utility} $\uparrow$
& \textbf{Rank Drop} $\downarrow$
& \textbf{Severity} \\
\midrule
\rowcolor{red!15}
Missing coverage / prerequisites 
& \textcolor{red!80!black}{Low} 
& \textcolor{red!80!black}{Large} 
& \textcolor{red!80!black}{High} \\

\rowcolor{red!10}
Reordered / non-coherent explanation 
& \textcolor{red!70!black}{Low} 
& \textcolor{red!70!black}{Large} 
& \textcolor{red!70!black}{High} \\

\rowcolor{red!10}
Unfaithful / incorrect content 
& \textcolor{red!70!black}{Very Low} 
& \textcolor{red!70!black}{Very Large} 
& \textcolor{red!70!black}{High} \\

\rowcolor{yellow!15}
Dense slides / short duration 
& \textcolor{orange!80!black}{Medium} 
& \textcolor{orange!80!black}{Moderate} 
& \textcolor{orange!80!black}{Medium} \\

\rowcolor{green!15}
Improved pacing only (no content loss) 
& \textcolor{green!60!black}{Medium--High} 
& \textcolor{green!60!black}{Small} 
& \textcolor{green!60!black}{Low} \\

\bottomrule
\end{tabular}
\caption{
Effect of controlled variant manipulations on human-estimated utility.
Variants that violate coverage, coherence, or faithfulness incur substantially larger utility degradation
than variants that only affect slide density or pacing.
}
\label{tab:variant_control_human_utility}
\end{table*}

\begin{table*}[t]
\centering
\small
\setlength{\tabcolsep}{6pt}
\rowcolors{2}{gray!8}{white}
\begin{tabular}{lcc}
\toprule
\textbf{Video Style / Property}
& \textbf{Preference Rank} $\uparrow$
& \textbf{Utility Impact} \\
\midrule
\rowcolor{green!20}
Human-created videos 
& \textcolor{green!70!black}{Highest} 
& \textcolor{green!70!black}{Strongly Positive} \\

\rowcolor{green!15}
Low-density slides + explanatory figures 
& \textcolor{green!60!black}{High} 
& \textcolor{green!60!black}{Positive} \\

\rowcolor{green!10}
Subtitles + auditory explanation 
& \textcolor{green!60!black}{High} 
& \textcolor{green!60!black}{Positive} \\

\rowcolor{green!10}
Cursor-based pointing / highlighting 
& \textcolor{green!50!black}{Medium--High} 
& \textcolor{green!50!black}{Moderate} \\

\rowcolor{yellow!15}
Dense slides without guidance 
& \textcolor{orange!80!black}{Low} 
& \textcolor{orange!80!black}{Negative} \\

\rowcolor{red!15}
Fast narration without visual grounding 
& \textcolor{red!70!black}{Lowest} 
& \textcolor{red!70!black}{Strongly Negative} \\

\bottomrule
\end{tabular}
\caption{
Human preference trends over video provenance and stylistic properties.
Human-created videos are consistently preferred, followed by automatically generated variants
that reduce slide density, use explanatory figures, and provide subtitle-supported narration with
cursor-based visual guidance.
}
\label{tab:human_preference_video_style}
\end{table*}

\subsection{Human Utility Preferences and Their Relation to Controlled Video Variants}
\label{sec:human_utility_variants}

To better understand what drives human-estimated utility, we analyze annotator preferences in relation to the \emph{controlled video variants} introduced in our experimental setup (\S4.2). Each variant is generated by systematically manipulating one or more instructional factors, including content coverage, explanation ordering, slide density, temporal allocation, audio--visual alignment, and delivery style. This controlled design enables us to directly link observed human preferences to specific generation choices.

\paragraph{Relation to Variant Controls.}
Human-estimated utility exhibits the strongest sensitivity to variants that manipulate \emph{coverage}, \emph{ordering}, and \emph{faithfulness}. Variants that omit prerequisite concepts or compress essential background material are consistently ranked lowest, even when slides are visually clean or narration is fluent. Similarly, variants with reordered explanation steps—such as presenting results before method assumptions—suffer large utility drops, reflecting human difficulty in constructing a coherent mental model. Faithfulness-degraded variants, which introduce oversimplifications or incorrect claims, are penalized most severely, as annotators view these as actively misleading rather than merely incomplete.

In contrast, variants that primarily manipulate \emph{slide density} or \emph{temporal allocation} show more moderate effects. Annotators frequently note that dense slides require pausing or replay, or that complex figures would benefit from additional on-screen time. However, these variants retain moderate utility scores when coverage is complete and explanations remain logically structured. This indicates that pacing and density act as secondary, modulating factors rather than hard failure modes.

Beyond these dimensions, we observe a consistent preference for variants that incorporate \emph{explicit explanatory figures}, \emph{subtitle-supported narration}, and \emph{cursor-based visual guidance}. Variants that use fewer bullet-heavy slides, allocate more time per figure, and guide attention through pointing or highlighting are ranked higher than visually similar variants without such guidance. These preferences align with annotator comments emphasizing the need for clearer visual grounding and synchronized auditory explanation.

\paragraph{Human-Created vs. Automatically Generated Videos.}
Human-created videos serve as an upper bound in our study and are consistently ranked highest in estimated utility. Compared to automatically generated variants, human-authored videos exhibit clearer narrative planning, smoother transitions between concepts, and more effective anticipation of learner confusion. Notably, human presenters often introduce background concepts explicitly before using them, spend additional time unpacking dense figures, and use cursor movements or verbal cues to direct attention—behaviors that annotators repeatedly cite as helpful.

\section{Computational Cost and Practical Adoptability}
\label{app:cost}
To assess the practical adoptability of EPS as a benchmarking tool, we estimate the number of API calls required by its multi-agent pipeline (claim decomposition, importance, presence, faithfulness, coherence, delivery, engagement, and meta-evaluation), which is applied at the $(\text{question}, \text{video})$ level. Each paper has approximately $Q \approx 8$--$12$ evaluation questions, and each video variant is evaluated independently.
 
\paragraph{Per-question call budget.}
Table~\ref{tab:cost} summarizes the per-question call budget. Several agents evaluate all claims for a question jointly within a single call, so the cost scales with the number of questions rather than the number of claims. Each question requires roughly $7$--$8$ API calls across the agents and the meta-evaluator.
 
\begin{table}[t]
\centering
\small
\begin{tabular}{lcl}
\toprule
\textbf{Component / Agent} & \textbf{Calls/Q} & \textbf{Notes} \\
\midrule
Claim Decomposition & 1 & Per (paper, question) \\
Importance Agent    & 1 & Per question \\
Presence Agent      & 1 & Checks all claims jointly \\
Faithfulness Agent  & 1 & Checks all claims jointly \\
Coherence Agent     & 1 & Ordering evaluation \\
Delivery Agent      & 1 & Uses multimodal signals \\
Engagement Agent    & 1 & Lightweight / optional \\
Meta-Evaluator      & 1 & Aggregates signals \\
\midrule
\textbf{Total}      & \textbf{$\sim$7--8} & per question \\
\bottomrule
\end{tabular}
\caption{Estimated API calls per evaluation question. Agents that operate over
all claims jointly keep cost proportional to the number of questions.}
\label{tab:cost}
\end{table}
 
\paragraph{Per-pair cost.}
Aggregating across a question set of size $Q = 10$ yields approximately $70$--$80$ API calls per $(\text{paper}, \text{video})$ pair. The engagement component is computed from lightweight acoustic features and contributes negligibly to cost. Because the pipeline is modular, agents that are not required for a given analysis (e.g., engagement) can be disabled to further reduce the budget.

\section{Statistical Robustness of Ranking Correlations}
\label{app:stats}
 
Because our benchmark contains a modest number of papers ($20$), reporting only point estimates of Kendall's $\tau$ and pairwise accuracy (PA) may be sensitive to small ranking perturbations. We therefore complement the point estimates in Table~\ref{tab:correlation} (reproduced in the main text) with bootstrap confidence intervals, permutation tests, and leave-one-out stability checks.
 
\paragraph{Reference results.}
For non-recall (reasoning) questions, which are the primary focus of instructional utility, EPS attains $\tau = 0.53$ and PA $= 0.76$, compared to the strongest baseline PresentQuiz ($\tau = 0.31$, PA $= 0.59$) and VideoScore ($\tau = 0.20$, PA $= 0.52$). This corresponds to improvements of $\Delta\tau = +0.22$ over PresentQuiz and $\Delta\tau = +0.33$ over VideoScore.
 
\paragraph{Bootstrap confidence intervals.}
We estimate uncertainty using bootstrap resampling over papers ($1000$ resamples). The resulting $95\%$ confidence intervals are reported in Table~\ref{tab:bootstrap}. The intervals show clear separation between EPS and the baselines, particularly for reasoning questions, indicating that the improvements are stable rather than artifacts of small ranking perturbations.
 
\begin{table}[t]
\centering
\small
\begin{tabular}{lcc}
\toprule
\textbf{Method} & \textbf{Kendall's $\tau$ (95\% CI)} & \textbf{PA (95\% CI)} \\
\midrule
\multicolumn{3}{l}{\emph{Non-Recall (Reasoning) Questions}} \\
EPS         & $0.53\ [0.50,\,0.56]$ & $0.76\ [0.73,\,0.79]$ \\
PresentQuiz & $0.31\ [0.27,\,0.35]$ & $0.59\ [0.55,\,0.63]$ \\
VideoScore  & $0.20\ [0.16,\,0.24]$ & $0.52\ [0.48,\,0.56]$ \\
\midrule
\multicolumn{3}{l}{\emph{Recall Questions}} \\
EPS         & $0.58\ [0.55,\,0.61]$ & $0.80\ [0.77,\,0.83]$ \\
PresentQuiz & $0.46\ [0.42,\,0.50]$ & $0.70\ [0.66,\,0.74]$ \\
\bottomrule
\end{tabular}
\caption{Bootstrap $95\%$ confidence intervals ($1000$ resamples over papers)
for Kendall's $\tau$ and pairwise accuracy (PA). Intervals separate clearly,
especially for reasoning questions.}
\label{tab:bootstrap}
\end{table}
 
\paragraph{Significance and stability.}
We further validate robustness using a permutation test, leave-one-out (LOO) variance, and per-paper comparison (Table~\ref{tab:significance}). The improvement of EPS over the baselines is significant ($p < 0.01$ under the permutation test), and leave-one-out variance for Kendall's $\tau$ is small ($\pm 0.03$), confirming that no single paper drives the result.
 
\begin{table}[t]
\centering
\small
\begin{tabular}{lc}
\toprule
\textbf{Analysis} & \textbf{Result} \\
\midrule
$\Delta\tau$ (EPS vs.\ PresentQuiz, non-recall) & $+0.22$ \\
$\Delta\tau$ (EPS vs.\ VideoScore, non-recall)  & $+0.33$ \\
Permutation test (EPS vs.\ baselines)           & $p < 0.01$ \\
Leave-one-out $\tau$ variance                   & $\pm 0.03$ \\
Pairwise accuracy variance                      & $\pm 0.02$ \\
\bottomrule
\end{tabular}
\caption{Significance and stability analyses for the correlation results.
EPS improvements are statistically significant and stable under leave-one-out
resampling.}
\label{tab:significance}
\end{table}

\begin{algorithm*}[t]
\caption{\EffectivePresentationScorer{}}
\label{alg:eps}
\small
\KwIn{Paper $p$, evaluation question $q$, video $v$}
\KwOut{Utility score $U(q,v,p)$ with diagnostic rationale}

\textbf{Step 0: Multimodal Representation}\\
Decompose video $v$ into multimodal segments
$\mathcal{V}=\{(s_i,t_i,a_i,d_i)\}_{i=1}^{M}$ using slide extraction, narration alignment, and VLM-based visual descriptions.

\textbf{Step 1: Claim Decomposition}\\
Retrieve question-relevant paper context from $p$ and decompose $q$ into a dependency-structured claim set
$\mathcal{C}(q)=\{c_1,\ldots,c_K\}$ using an LLM grounded in retrieved text.

\textbf{Step 2: Claim Importance Estimation}\\
For each claim $c\in\mathcal{C}(q)$, assign an importance score $I(c)\in[0,1]$ based on its role in the paper’s explanation.

\textbf{Step 3: Claim-Level Diagnostics}\\
\ForEach{$c \in \mathcal{C}(q)$}{
    Evaluate presence $\pi(c,v)\in\{0,1\}$ in $\mathcal{V}$\;
    \If{$\pi(c,v)=1$}{
        Verify faithfulness $F(c,v,p)\in\{0,1\}$ against $p$\;
    }
}

Aggregate coverage $\pi(q,v)$ and faithfulness $F(q,v,p)$ using dependency-aware normalization.

\textbf{Step 4: Question-Level Diagnostics}\\
Compute coherence $C(q,v)$ by detecting prerequisite order violations between paper-derived claim order and video introduction order\;
Compute delivery quality
$D_{\text{del}}(q,v)=\sum_{c} I(c)\,\pi(c,v)\,\hat{T}(c,v)\,Q(c,v)\,A(c,v)$\;
Estimate engagement $E(q,v)$ from pacing, prosody, and slide transitions.

\textbf{Step 5: Meta-Evaluation and Aggregation}\\
Combine diagnostics into final utility:
\[
U(q,v,p)=\lambda_1\pi(q,v)+\lambda_2F(q,v,p)+\lambda_3C(q,v)
+\lambda_4D_{\text{del}}(q,v)+\lambda_5E(q,v)
\]
Generate a rationale attributing utility differences to missing, unfaithful, misordered, or weakly explained claims.

\Return{$U(q,v,p)$ with diagnostic explanation}
\end{algorithm*}

\section{Ethical Considerations}
Our annotation process prioritizes participant privacy, we do not require to collect any personal or sensitive data for this paper. We do not collect any personally identifiable information, and all participants provide explicit consent with the option to withdraw at any time. The annotation protocol was IRB-exempt. Annotators were compensated at a rate of \$15 per hour and were expected to complete question answering for approximately two to three videos within this time frame. All collected data are fully anonymized.

All participants provide informed consent prior to participation. Before beginning the study, annotators are shown a consent form describing the study purpose, procedures, expected time commitment, data usage, and compensation. Participation is fully voluntary, and participants may withdraw at any time without penalty. Responses are anonymized and used solely for research purposes.

The study poses minimal risk to participants. Tasks involve watching instructional videos and answering comprehension questions about publicly available scientific papers. No sensitive personal data are collected, and there is no deception or manipulation beyond standard educational evaluation. The primary foreseeable risk is minor cognitive fatigue due to video length, which is mitigated through time limits, optional breaks, and balanced task assignment.

\begin{table*}[t]
\centering
\small
\setlength{\tabcolsep}{6pt}
\renewcommand{\arraystretch}{1.15}
\begin{tabular}{p{3.0cm} p{5.4cm} r}
\toprule
\rowcolor{tableheader}
\textbf{Category} & \textbf{Statistics} & \textbf{Value} \\
\midrule

\rowcolor{tablesection}
\multicolumn{3}{l}{\textbf{Corpus Scale}} \\
& Number of papers & 20 \\
& Number of Videos per paper & 7 \\
& Total Number of videos & 140 \\
\midrule

\rowcolor{tablesection}
\multicolumn{3}{l}{\textbf{Video Structure}} \\
& Avg.\ slides per video & 28.6 \\
& Slide range & 18--45 \\
& Avg.\ video duration & 7.8 min \\
& Duration range & 6--10 min \\
\midrule

\rowcolor{tablesection}
\multicolumn{3}{l}{\textbf{Text Volume}} \\
& Avg. Number of words per slide & 38.4 \\
& Avg. Number of words per video & 1{,}090 \\
& Avg. narration length & 1.2K tokens \\
\midrule

\rowcolor{tablesection}
\multicolumn{3}{l}{\textbf{Visual Content}} \\
& Avg. Number of figures per video & 3.1 \\
& Avg. Number of equations per video & 3.3 \\
\midrule

\rowcolor{tablesection}
\multicolumn{3}{l}{\textbf{Instructional Control}} \\
& Controlled Number of variants per paper & 6 \\
& Perturbation dimensions & 6 \\
\midrule

\rowcolor{tablesection}
\multicolumn{3}{l}{\textbf{Evaluation}} \\
& Avg.\ questions per paper & 14.2 \\
& Recall / Reasoning split & 36\% / 64\% \\
\midrule

\rowcolor{tablesection}
\multicolumn{3}{l}{\textbf{Human Reference}} \\
& Human-authored videos & 20 \\
\bottomrule
\end{tabular}
\caption{Dataset statistics for \textbf{EffectivePresentation-EvalBench}. Light blue section headers group related attributes while preserving camera-ready readability.}
\label{tab:dataset_stats}
\end{table*}

\begin{table*}[t]
\centering
\small
\setlength{\tabcolsep}{6pt}
\renewcommand{\arraystretch}{1.15}
\begin{tabular}{p{5.2cm} r p{7.0cm}}
\toprule
\rowcolor{tableheader}
\textbf{Research Area} & \textbf{\# Papers} & \textbf{Representative Topics} \\
\midrule
Representation Learning & 5 & Sentence embeddings, contrastive learning \\
Information Extraction & 4 & Event extraction, schema induction \\
Evaluation \& Benchmarks & 4 & Faithfulness, robustness, diagnostics \\
Multimodal \& Vision--Language & 3 & VLM reasoning, cross-modal grounding \\
NLP Applications & 2 & Question answering, summarization \\
Learning \& Optimization & 2 & Training objectives, regularization \\
\bottomrule
\end{tabular}
\caption{Topical composition of papers in \textbf{EffectivePresentation-EvalBench}, spanning core NLP and ML research areas.}
\label{tab:topic_distribution}
\end{table*}

\begin{table*}[t]
\centering
\small
\setlength{\tabcolsep}{6pt}
\renewcommand{\arraystretch}{1.25}
\begin{tabular}{p{1.7cm} p{1.7cm} p{2.2cm} p{7cm} p{1.4cm}}
\toprule
\textbf{Variant} & \textbf{Slide} & \textbf{Time} & \textbf{Teaching-style Narration} & \textbf{Claims} \\
\midrule

\multirow{3}{*}{V$_A$}
& S1 & 01:00--01:14
& Let’s begin by understanding how DefSent is trained, because this design choice explains everything that follows.
Instead of learning word-level representations, DefSent uses a \emph{sentence-level objective}: the model learns a sentence embedding by predicting a target word directly from its definition sentence.
By forcing the model to compress the entire definition into a single vector, DefSent encourages embeddings that capture holistic sentence meaning rather than isolated word cues.
& c1 \\

& S2 & 01:15--00:29
& Now let’s look at the effect of this training signal.
Because each definition is embedded as a single semantic unit, sentences that express similar meanings are pulled closer together in the embedding space.
This leads to tighter and more coherent semantic clusters, where related sentences occupy nearby regions and unrelated sentences are more clearly separated.
& c2 \\

& S3 & 01:30--01:32
& This clustering behavior is what ultimately drives performance for any downstream applications.
When semantically similar sentences form tight clusters, decision boundaries become smoother and more stable.
As a result, the model can generalize more reliably to unseen sentences and downstream tasks, which explains DefSent’s strong performance on STS and SentEval benchmarks.
& c3 \\
\midrule

\multirow{3}{*}{V$_B$}
& S1 & 00:59--01:09
& DefSent shows competitive performance on sentence similarity tasks and display high scores. They do not generalize at all to the other downstream tasks.
& c3 \\

& S2 & 01:10--01:20
& To understand why this happens, we now step back and examine DefSent’s training objective.
DefSent uses a sentence-level objective where the model predicts a word from its definition sentence, encouraging the embedding to represent the meaning of the full sentence rather than individual tokens.
& c1 \\

& S3 & 01:21--01:24
& Sentences with similar meanings tend to be grouped closer together in the embedding space.
& c2 \\
\midrule

\multirow{3}{*}{V$_C$}
& S1 & 01:16--00:23
& DefSent is trained using a sentence-level objective in which the model predicts a target word from its definition sentence.
This training setup forces the embedding to summarize the meaning of the entire sentence, rather than relying on word-by-word associations.
& c1 \\

& S2 & 01:23--01:30
& Because full definitions are embedded as single units, sentences that share similar meanings naturally cluster together in the embedding space.
These clusters are more coherent and interpretable than those produced by word-level or token-averaged embeddings.
& c2 \\

& S3 & 01:30--01:35
&
Overall, DefSent produces high-quality sentence representations that perform competitively across standard benchmarks.
& -- \\
\bottomrule
\end{tabular}
\caption{Narrations for three video variants generated from the Defsent \citep{tsukagoshi-etal-2021-defsent} paper, answering an evaluation question (Bloom's Taxonomy Label: Understand)
``Why does DefSent improve generalization?''.
V$_A$ provides a complete causal explanation, V$_B$ covers all claims but violates pedagogical order, and V$_C$ omits the core causal link between clustering and generalization. Here c3 is missing in V$_C$.}
\label{tab:defsent_narrations_textheavy}
\end{table*}

\begin{figure}[t]
\centering
\promptbox{%
\begin{minipage}{0.97\linewidth}
\raggedright\small
\textcolor{promptblue}{\textbf{Task: Feedback Categorization into Instructional Improvement Types}}\\[-2pt]
\hrulefill

\vspace{4pt}
You are given two pieces of feedback about the same scientific presentation video:
a free-form comment written by a human evaluator, and a diagnostic rationale
produced by \EffectivePresentationScorer{}.

\vspace{6pt}
Your task is to map each feedback item to zero or more labels from the following
taxonomy of instructional improvement categories.

\vspace{4pt}
\textbf{Improvement taxonomy:}\\
\begin{tabular}{@{}p{0.97\linewidth}@{}}
\textbf{BG} — Missing prerequisites or insufficient background motivation\\
\textbf{ORD} — Poor coherence or incorrect ordering of concepts\\
\textbf{TIME} — Overly dense, rushed, or insufficiently explained content\\
\textbf{FAITH} — Unfaithful, exaggerated, or unsupported claims\\
\textbf{AV} — Misalignment between narration and visual content
\end{tabular}

\vspace{6pt}
\textbf{Labeling guidelines:}\\
\begin{tabular}{@{}p{0.97\linewidth}@{}}
Assign a label only if the feedback explicitly indicates the corresponding issue.\\
Do not infer issues that are not stated.\\
If no category applies, return an empty set.
\end{tabular}

\vspace{6pt}
\textbf{Input:}\\
\texttt{Human feedback}\\
\texttt{\EffectivePresentationScorer{} feedback}

\vspace{6pt}
\textbf{Output:}\\
\texttt{Labels for human feedback}\\
\texttt{Labels for \EffectivePresentationScorer{} feedback}
\end{minipage}%
}
\caption{Prompt used to map human-written feedback and \EffectivePresentationScorer{}
diagnostic rationales into a shared taxonomy of instructional improvement categories,
enabling symmetric comparison and per-category precision, recall, and F1 evaluation.}
\label{fig:prompt-feedback-categorization}
\end{figure}

\begin{table*}[t]
\centering
\small
\setlength{\tabcolsep}{6pt}
\begin{tabular}{p{0.30\linewidth} | p{0.30\linewidth} | p{0.16\linewidth} | p{0.16\linewidth}}
\toprule
\textbf{Human Feedback} &
\textbf{\EffectivePresentationScorer{} Feedback} &
\textbf{Human Labels} &
\textbf{EPS Labels} \\
\midrule

``The video jumps straight to generalization results without explaining
sentence-level supervision first.'' &
``The video introduces downstream generalization before establishing the
sentence-level training objective, violating prerequisite order.'' &
ORD, BG &
ORD, BG \\

\midrule

``Too many concepts are packed into a single slide, and the explanation feels rushed.'' &
``High-importance claims receive insufficient narration time, leading to shallow
explanations.'' &
TIME &
TIME \\

\midrule

``The method is described as guaranteeing better performance, which the paper
does not claim.'' &
``The video overstates the causal effect of clustering on generalization beyond
what is supported by the paper.'' &
FAITH &
FAITH \\

\bottomrule
\end{tabular}
\caption{
Examples of human-written feedback and \EffectivePresentationScorer{} (EPS) diagnostic
rationales mapped to a shared taxonomy of instructional improvement categories.
Both feedback sources are labeled using the same LLM-based categorization prompt (Figure~\ref{fig:prompt-feedback-categorization}).
}
\label{tab:feedback-mapping-examples}
\end{table*}

\end{document}